\documentclass{emulateapj}
\usepackage{float}
\slugcomment{}
\begin{document}

\title{Internal Energy Dissipation of Gamma-Ray Bursts Observed with {\em Swift}: Precursors, Prompt Gamma-rays, Extended emission and Late X-ray Flares}

\author{You-Dong Hu\altaffilmark{1}, En-Wei Liang\altaffilmark{1,2}, Shao-Qiang Xi\altaffilmark{1}, Fang-Kun Peng $^{1}$, Rui-Jing Lu\altaffilmark{1}, Lian-Zhong L\"u\altaffilmark{1}, Bing Zhang\altaffilmark{3}
}
\altaffiltext{1}{Department of Physics and GXU-NAOC Center for Astrophysics
and Space Sciences, Guangxi University, Nanning 530004,
China; lew@gxu.edu.cn}
\altaffiltext{2}{Key Laboratory for the Structure and Evolution of Celestial Objects, CAS, Kuming 650011}
\altaffiltext{3}{Department of Physics and Astronomy, University of
Nevada, Las Vegas, NV 89154; Zhang@physics.unlv.edu}

\begin{abstract}
We jointly analyze the gamma-ray burst (GRB) data observed with BAT and XRT on board the {\em Swift} mission to present a global view on the internal energy dissipation processes in GRBs, including precursors, prompt gamma-ray emission, extended soft gamma-ray emission, and late X-ray flares. The Bayesian block method is utilized to analyze the BAT lightcurves to identify various emission episodes. Our results suggest that these emission components likely share a same physical origin, which is repeated activation of the GRB central engine. What we observe in the gamma-ray band may be the tip-of-iceberg of more extended underlying activities. The precursor emission, which is detected in about 10\% of {\em Swift} GRBs, is preferably detected in those GRBs that have a massive star core-collapse origin. The soft extended emission (EE) tail, on the other hand, is preferably detected in those GRBs that have a compact star merger origin. Bright X-ray emission is detected during the BAT quiescent phases prior to subsequent gamma-ray peaks, implying that X-ray emission may be detectable prior the BAT trigger time. Future GRB alert instruments with soft X-ray capability would be essential to reveal the early stage of GRB central engine activities, sheding light into jet composition and jet launching mechanism in GRBs.
\end{abstract}
\section{Introduction\label{sec:intro}}
{\em Swift} (Gehrels et al. 2004) has greatly improved our understanding of the nature of the gamma-ray burst (GRB) phenomenon since it was successfully launched in 2004. The Burst Alert Telescope (BAT, Barthelmy et al. 2005b) and X-ray Telescope (XRT, Burrows et al. 2005b) onboard {\em Swift} observe the GRB prompt emission and afterglow up to days and even months (Zhang et al. 2006; OBrien et al. 2006; Liang et al. 2006; Zhang et al. 2007; Sakamoto et al. 2007; Chincarini et al. 2007; Falcone et al. 2007; Liang et al. 2007; Margutti et al. 2010). It is known that the conventional long-short GRB classification scheme, which was based on the bimodal burst duration ($T_{90}$) distribution discovered from the GRB survey observations with Burst And Transient Source Experiment (BATSE) on board Compton Gamma-Ray Observatory (CGRO) (Kouveliotou et al. 1993), meets the theoretical speculations of two types of GRBs from deaths of massive stars (Colgate 1974; Woosley 1993) and mergers of two compact objects (e.g., Paczy\'nski 1986, 1991; Eichler et al. 1989; Narayan et al. 1992; Bloom et al. 1999; see Berger 2013 for a review). However, the erratic, bright X-ray flares and long, soft extended emission tails (EE tails) observed with BAT and XRT challenge the simple classification using the $T_{90}$ criterion. It is also interesting that some short-duration GRBs are likely originated from collapse of massive stars (Virgili et al. 2011; Zhang et al. 2009; Levesque et al. 2010a, b; Xin et al. 2011; Belczynski et al. 2010). Qin et al. (2013) showed that the bimodal $T_{90}$ distribution is significantly affected by instrumental selection effects. A physically motivated classification scheme, i.e. Type II (massive star origin) vs. Type I (compact star origin), was proposed (Zhang 2006; Zhang et al. 2007), which demands multiple observational criteria (Zhang et al. 2009). Other classification parameters were also proposed (e.g. L\"u et al. 2010; Goldstein et al. 2010).

The duration $T_{\rm 90}$ is not a good representative of the duration of internal energy dissipation processes driven by central engine activities. Observationally, the GRB emission we observe starts from the detector trigger time in the gamma-ray band. It is unclear what may happen prior to the GRB trigger. The most extensively discussed GRB model is the internal shock model (Rees \& M\'esz\'aros 1994; Kobayashi et al. 1997; Daigne \& Mochkovitch 1998; Maxham \& Zhang 2009; Daigne et al. 2011), which suggests that the observed highly variable emission of GRBs are from internal shocks due to the collisions of baryon-dominated shells ejected from the central engines. Alternatively, a magnetically dominated outflow may be launched from the central engine (Zhang \& Pe'er 2009), which dissipates magnetic fields either continuously (e.g. Drenkhahn \& Spruit 2002) or abruptly at large radii through internal collision-induced magnetic reconnection and turbulence (ICMART, Zhang \& Yan 2011). In general, the internal dissipation site should be above the photosphere and below the deceleration radius, and should have two emission components: a quasi-thermal emission component from the photosphere and a non-thermal component from the energy dissipation site (internal shocks for a baryon-dominated outflow and magnetic dissipation site for a magnetically dominated flow), and the observed emission is generally believed to be a superposition of these components (e.g. M\'esz\'aros \& Rees 2000; Zhang \& M\'{e}sz\'{a}ros 2002; Toma et al. 2011; Pe'er et al. 2012). Indeed, photosphere emission is observed to be superposed on the non-thermal component in the spectra of some GRBs (e.g., Ryde 2005; Ryde \& Pe'er 2009; Ryde et al. 2010; Guiriec et al. 2011; Axelsson et al. 2012), and some GRBs even have a dominant thermal component, e.g. GRB 090902B (Ryde et al. 2010; Zhang et al. 2011). For GRBs with a massive star progenitor, there could be an additional emission site where the jet first breaks out from the star. The long-lasting soft emission from low-luminosity GRBs has been attributed to the shock breakout emission (e.g. GRB 060218, Campana et al. 2006; and X-ray outburst 080109, Soderberg et al. 2008). Some authors have suspected that shock breakout or photosphere emission may give rise to precursor emission for long-duration GRBs (e.g. Lyutikov \& Usov 2000; Ramirez-Ruiz et al. 2002). It would be interesting to check whether there is any observational evidence in support of these claims.

Another interesting issue regarding the lifetime of GRB internal energy dissipation processes is the residual emission post the episodes of the main burst. {\em Swift} detected an EE tail in the BAT band for some ``short'' GRBs and bright X-ray flares in the XRT band for about half of {\em Swift} GRBs. Both the EE tails and X-ray flares may signal the late central engine activities of GRBs (Burrows et al. 2005a; Zhang et al. 2006; King et al. 2005; Fan \& Wei 2005; Dai et al. 2006; Perna et al. 2006; Proga \& Zhang 2006; Liang et al. 2006; Lazzati \& Perna 2007; Lazzati et al. 2008; Lee et al. 2009; Maxham \& Zhang 2009). The EE tails (Norris \& Bonnell 2006) have attracted great attention of the community, especially after the detection of GRB 060614, which is a nearby long GRBs ($T_{90}=108.7$ seconds at $z=0.125$) without an accompanied SN, being different from other known nearby long GRBs, such as GRB 980425/SN 1998bw (Galama et al. 1998; Kulkarni et al. 1998), GRB 030329/SN 2003dh (Stanek et al. 2003; Hjorth et al. 2003), GRB 031203/SN 2003lw (Malesani et al. 2004), GRB 060218/SN 2006aj (Modjaz et al. 2006; Pian et al. 2006; Sollerman et al. 2006; Mirabal et al. 2006; Cobb et al. 2006), and 100316D/SN 2010bh (Starling et al. 2011; Fan et al. 2011). It was proposed that GRB 060614 may be similar to those of some nearby ``short'' GRBs that may have a compact star merger origin (e.g. Gehrels et al. 2006; Zhang et al. 2007; Gal-Yam et al. 2006). A handful of Type I GRBs show such a component in their lightcurves (Norris et al. 2008; Lin et al. 2008; Zhang et al. 2009), such as GRB050724 (Barthelmy et al. 2005a; Tanvir et al. 2005; Berger et al. 2005) and GRB050709 (Hjorth et al. 2005). Metzger et al. (2008) suggested that late-time accretion from a remnant disk created during a compact object merger can in principle provide sufficient energy to power such a late time central engine activity. Liu et al. (2012) proposed that radial angular momentum transfer in a massive disk may significantly prolong the lifetime of the accretion process and multiple episodes may be switched by a magnetic barrier (e.g., Proga \& Zhang 2006). In general, it would be interesting to check observationally whether the EE tail and X-ray flares share similar or different observational properties with respect to prompt gamma-rays.

This paper presents a joint analysis of both BAT and XRT data of GRBs in order to give a global view of the emission from internal energy dissipation processes in GRBs. We systematically study all the emission components that are likely related internal dissipation processes, including precursor emission, main episodes of prompt emission, extended emission tail, and X-ray flares. We pay special attention of the following questions: Is there any preference for precursor emission or EE tail to exist in Type I and Type II GRBs? What is the physical origin of precursor emission? Is there X-ray emission prior the gamma-ray triggers?  Our data and analysis method are presented in \S 2. The analysis results are reported in \S 3. Conclusions and discussion are presented in \S 4.

\section{Data Analysis\label{sec:data}}
\subsection{BAT Data}
We analyze the BAT and XRT data of 613 GRBs, which are triggered with {\em Swift}/BAT by May 14, 2012. The data are downloaded from the NASA {\em Swift} Achieve. We extract the lightcurves and spectra for the BAT data with the standard {\em Swift} scientific tools. Because of the narrowness of the BAT band, the BAT spectra are usually fit with a single power-law model, which is adequate to fit the spectra in our sample (e.g., Sakamoto et al. 2011; Zhang et al. 2007). In order to search for low-significance signal before and after the duration covered by $T_{90}$, the extracted BAT lightcurves usually cover 200-300s prior to the BAT triggers and 200-300s post $T_{90}$ of the GRBs. The method of Bayesian Blocks (BB; Scargle 1998) is used to search for possible signals in the lightcurves. We find that the radiation episodes are poorly identified with the 64ms-binned lightcurves. Therefore, we re-bin the lightcurves with a bin size of 1s, 2s, 4s, and 8s, respectively, and pick up all blocks from the lightcurves based on our analysis with the BB method.  We outline our BAT data analysis procedure as follows.

\begin{itemize}
\item Select background time intervals, which are normally taken from $100\sim 200$ seconds prior to the BAT trigger and $100\sim 200$ post $T_{90}$. The selection of the background intervals is essential to identify a weak signal. We set a background time window and move the window to search for a proper background interval in which no block identified with the BB method has a signal-to-noise ratio $S/N$ greater than $2\sigma$ in an 8-second binned lightcurve, where $\sigma$ is the standard deviation of the data in the window.

\item Utilize the BB method to pick up emission episodes by progressively re-bin the lightcurves with a bin size of 1s, 2s, 4s, and 8s. The criterion of an radiation episode is taken as $S/N>3\sigma$. The increase of bin size can reduce the level of background noises. We calculate the corresponding $\sigma$ of the background noises, and then analyze the temporal structure of the lightcurves with the BB method. Note that a proper bin size is critical to identify a weak signal. The increase of bin size is helpful to identify a weak signal with a long duration, but may smear out an extremely short duration signal. Therefore, we progressively re-bin the lightcurves in size of 1s, 2s, 4s, and 8s, until the possible weak signals can be identified.
\item Measure the duration and separation of each episode. The duration of an episode is estimated with $t_{\rm d}=T_{\rm e}-T_{\rm b}$, where $T_{\rm e}$ and $T_{\rm b}$ are the starting and ending times at $S/N=3$.
\end{itemize}

\begin{figure}
\centering
\includegraphics[angle=0,scale=.35]{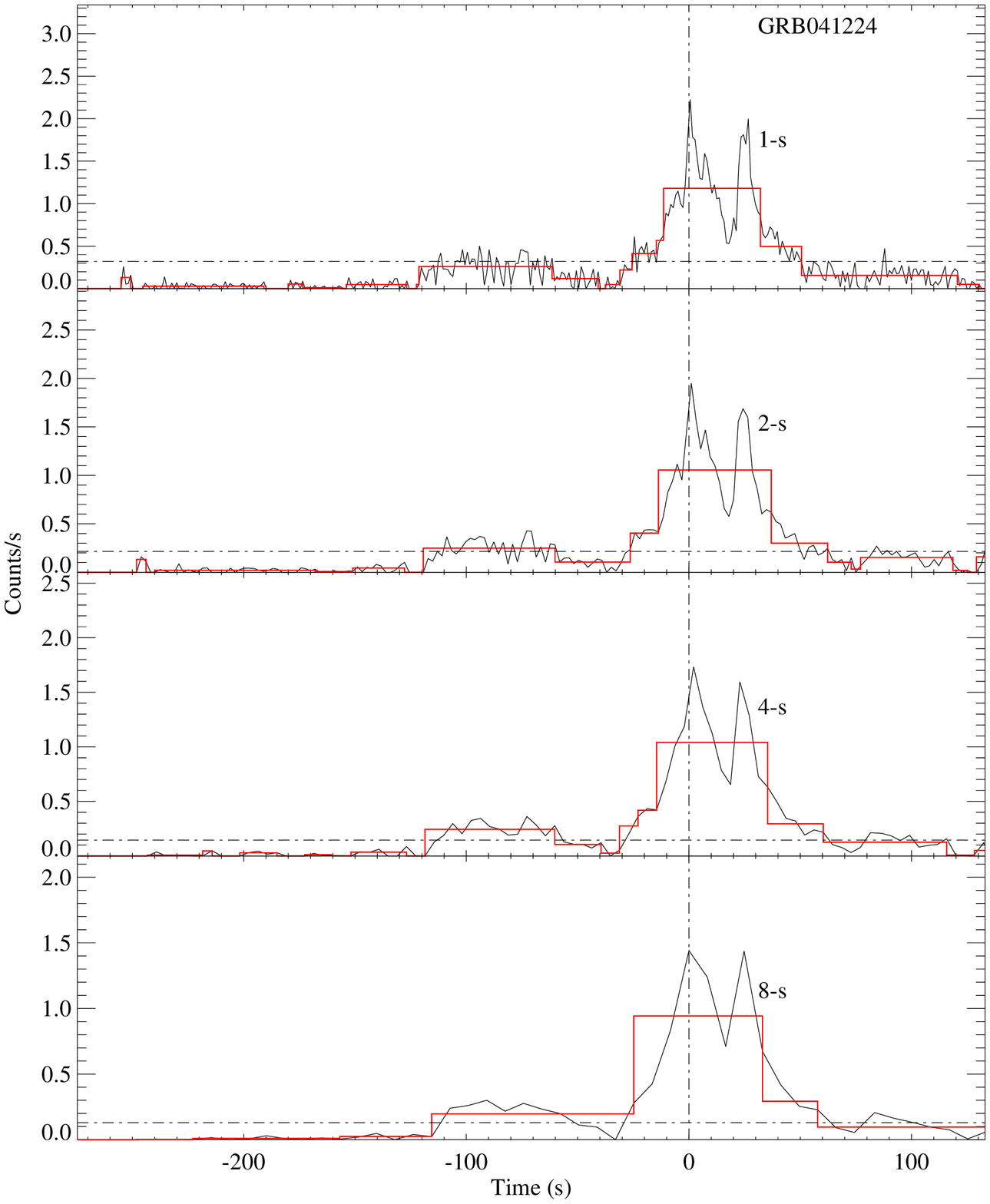}
\includegraphics[angle=0,scale=0.35]{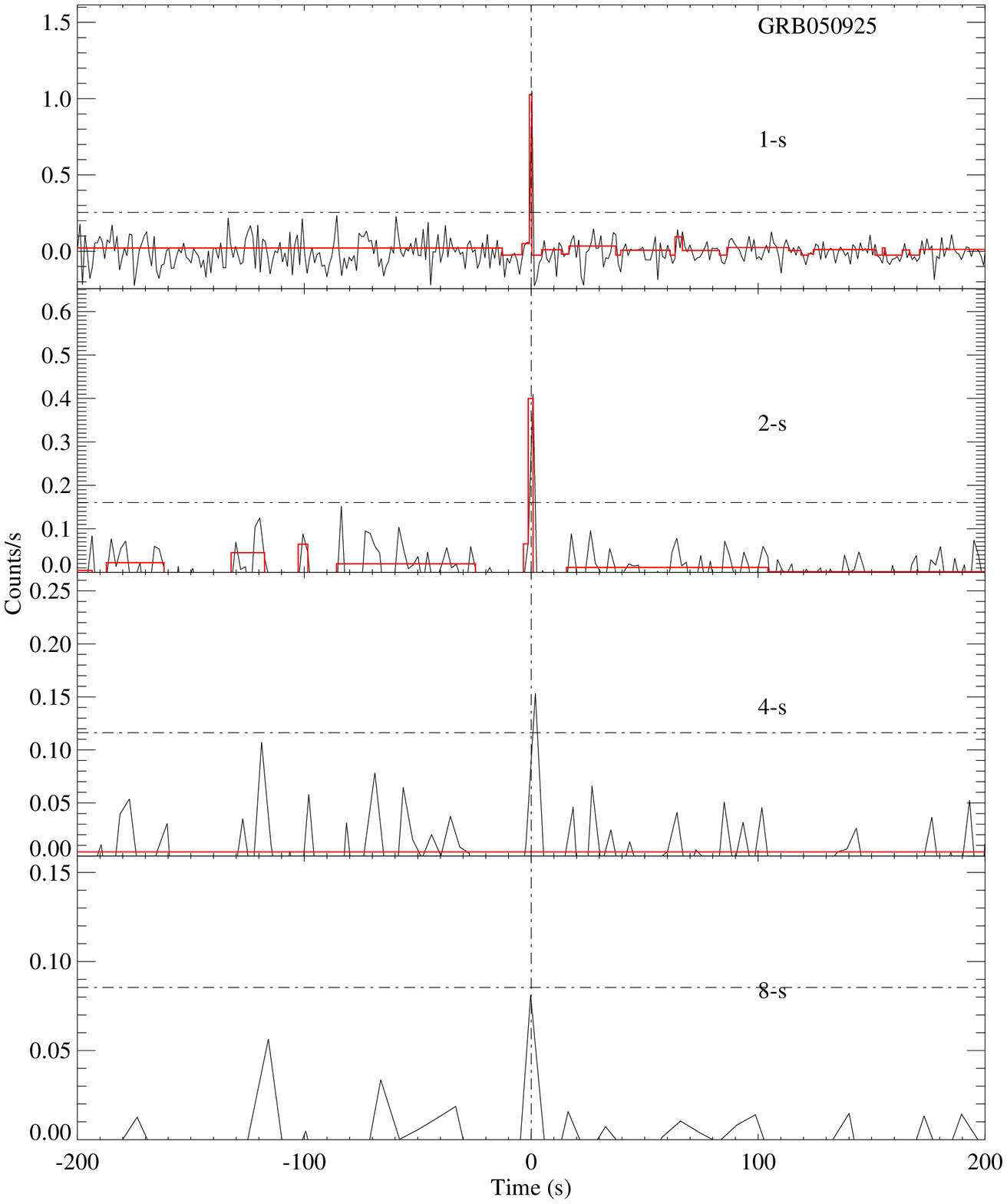}
\caption{Examples of our Beyasian block analysis (step lines) for lightcurves of long ({\em left panel}) and short ({\em right panel}) GRBs with different bin sizes (connected lines). The vertical lines mark the BAT trigger time.}
\label{Example}
\end{figure}

Examples of our analysis with the BB method are shown in Figure \ref{Example} with a typical long GRB 041224 and a typical short GRB 050925. Two separated emission episodes are identified for GRB 041224. One is the main episode in the time interval from $T_0-25$ to $T_{0}+60$ seconds, which is well identified from the 1s-binned lightcurve. It contains some overlapped pulses. It triggered BAT at nearly the peak time interval of the lightcurve. A long, weak signal is also seen prior to the main episode, but its $S/N$ is smaller than 3 $\sigma$  in the 1s-binned lightcurve. Increasing the bin size to 2s, this episode stands out with $S/N>3 \sigma$. We therefore pick up this episode from the 2s-binned lightcurve. Different from the main episode, this episode is composed of some significant flickerings and the emission level keeps almost constant. In the 4s and 8s binned lightcurves this episode can be recognized with a higher confidence level. The main episode is also followed by a weak, extended emission component, but its $S/N$ value is smaller than 3 in these lightcurves. We do not identify it as an extended emission component from the lightcurves. For short GRBs, especially for those extremely short GRBs with duration less than 1 second, a large bin size may smear out its main episode, as shown in Figure \ref{Example} for GRB 050925. We therefore prefer to identify an emission episode from the lightcurves in a small bin size for these GRBs.

\subsection{XRT Data}
The rapid slewing capacity of XRT makes it possible to catch the X-ray emission from very early to very late episodes of GRBs (Burrows et al. 2005a; Chincarini et al. 2007; Margutti et al. 2010). We obtain the XRT lightcurves of all bursts in our sample from http://www.swift.ac.uk (Evans et al. 2007, 2009). The X-ray data are mixed with both flares and underlying power law afterglow segments, and are non-uniformly binned. we do not utilize the BB method to analysis the XRT data, but identify a flare with a criterion of  $\Delta$F/F $\geq 5$, where $\Delta F=F_{\rm p}-F$ is the peak flux ($F_{\rm p}$) over the underlying flux $F$.

\section{Results}
\subsection{The Main Emission Episodes}
\begin{figure}
\centering
\includegraphics[angle=0,scale=.4]{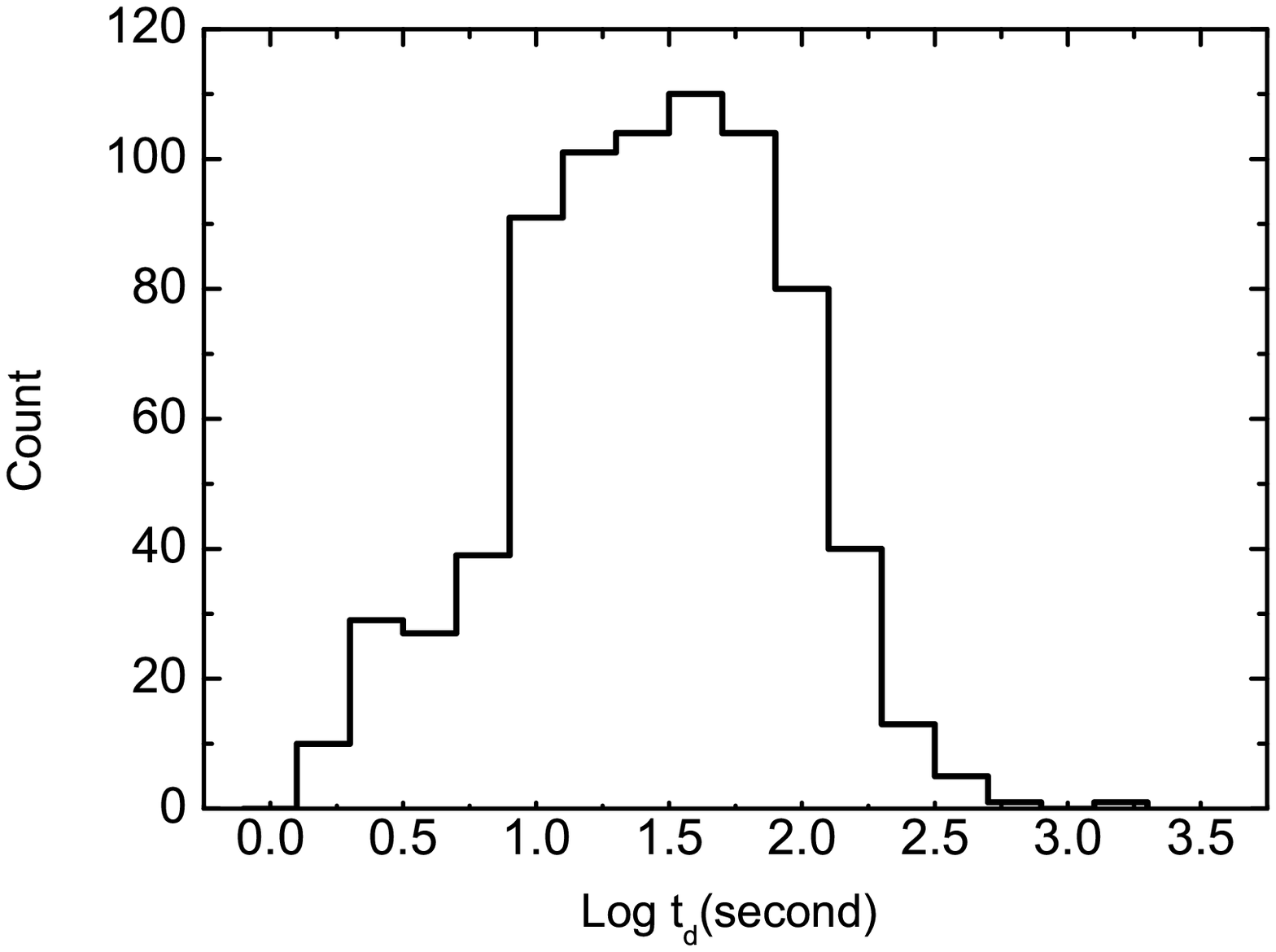}
\includegraphics[angle=0,scale=.4]{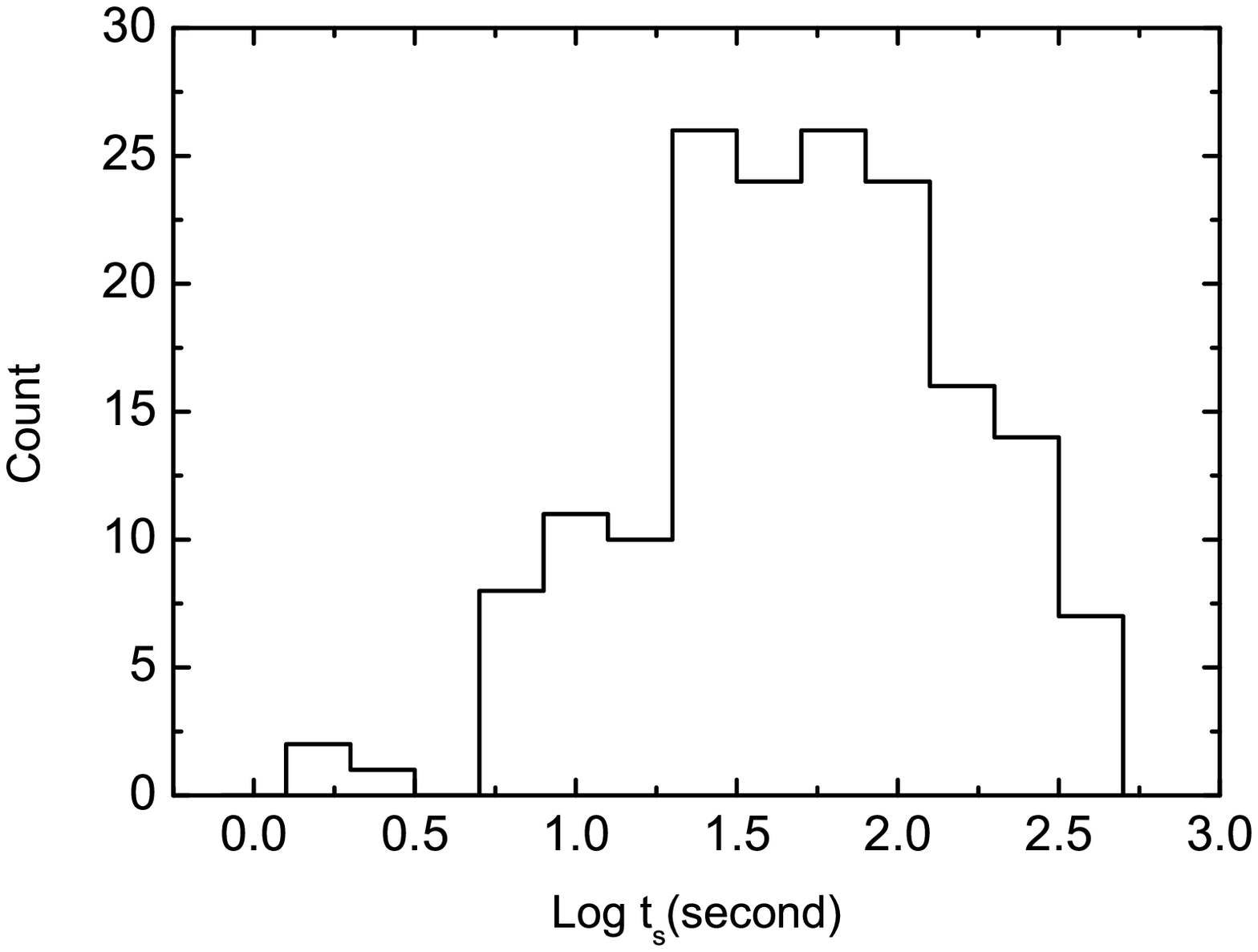}
\caption{Distributions of duration ($t_{\rm d}$, {\em left panel}) and separation interval ($t_{\rm s}$, {\em right panel}) of the episodes derived from our analysis with the BB method.}
\label{Distribution}
\end{figure}

Based on our analysis with the BB method, we identify emission episodes as connected blocks. One emission episode may contain one or several overlapped pulses. In view that X-ray flares are temporarily separated from prompt emission, we explore whether the main GRB component may sometimes include multiple separated episodes. We find that the BAT lightcurves of 73\% GRBs in our sample are composed of one episode only, and the others are composed of several episodes. We show the distributions of the durations ($t_d$) and separation time intervals ($t_s$) for the identified episodes in Figure \ref{Distribution}. It is found that the typical $t_d$ range from 10-100 seconds. The $t_s$ distribution is also in the same range.
\subsection{Precursors}
\begin{figure}
\centering
\includegraphics[angle=0,scale=.4]{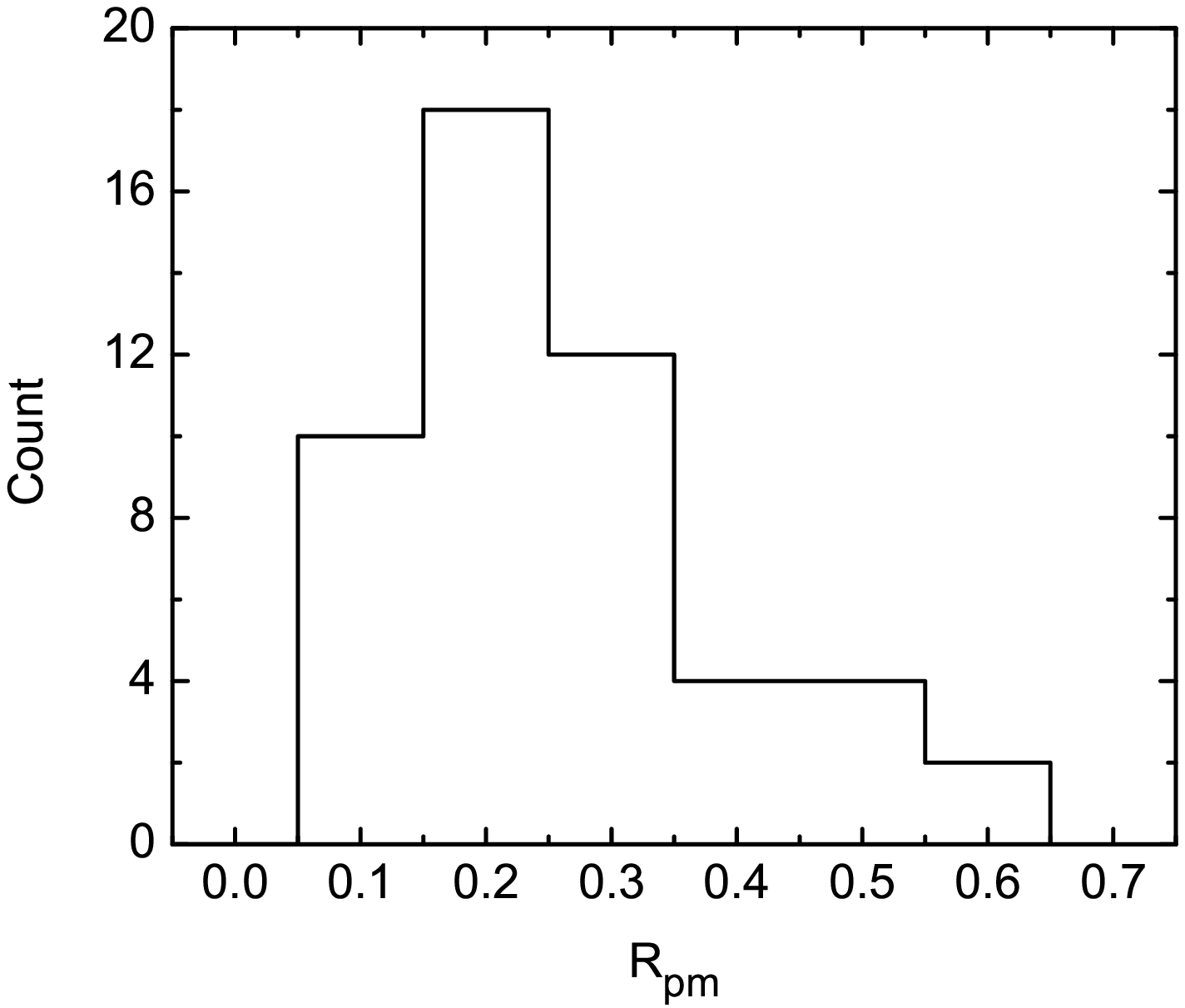}
\includegraphics[angle=0,scale=.4]{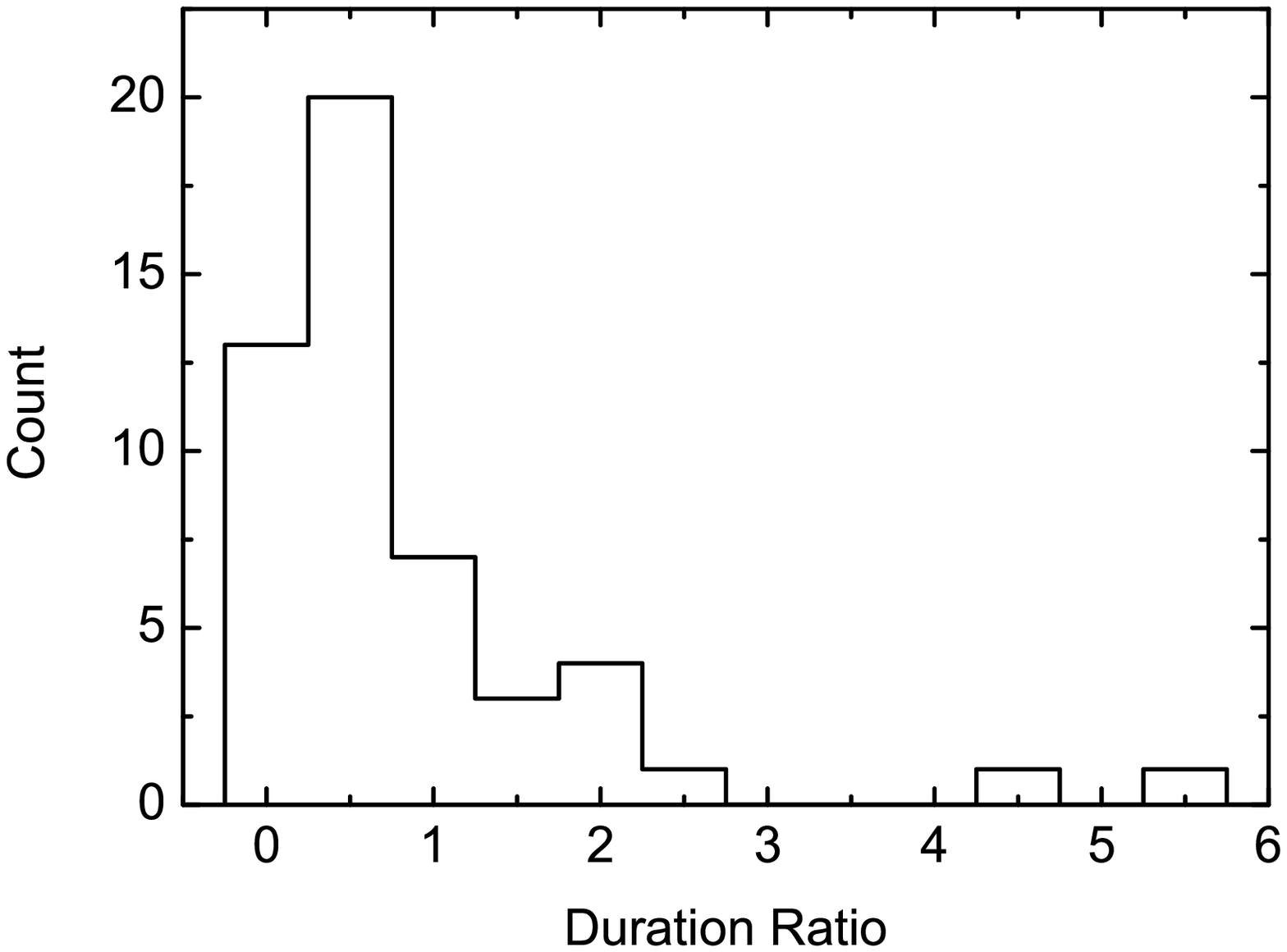}
\caption{Ratio Distributions of the average count rate ($R_{\rm pm}=C_{\rm p}/C_{\rm m}$, {\em left panel}) and duration ($t^{\rm p}_{\rm d}/t^{\rm m}_{\rm d}$, {\em right panel}) of the precursor to the main episode, where $t^{\rm p}_{\rm d}$ and $t^{\rm m}_{\rm d}$ are the durations of the precursor and main burst, respectively.}
\label{Precursor_Ratio_and_interval}
\end{figure}

\begin{figure}
\centering
\includegraphics[angle=0,scale=.4]{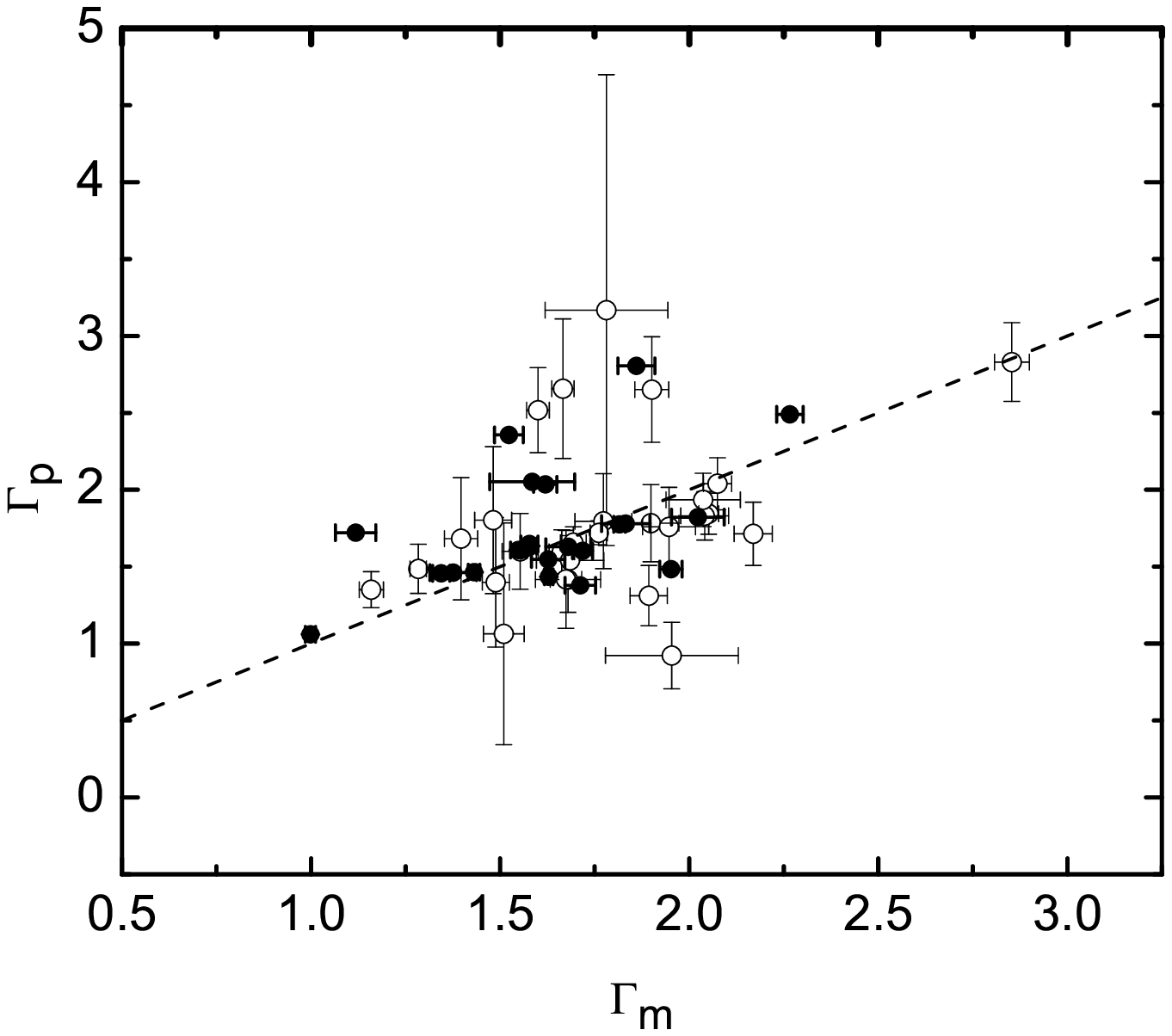}
\includegraphics[angle=0,scale=.4]{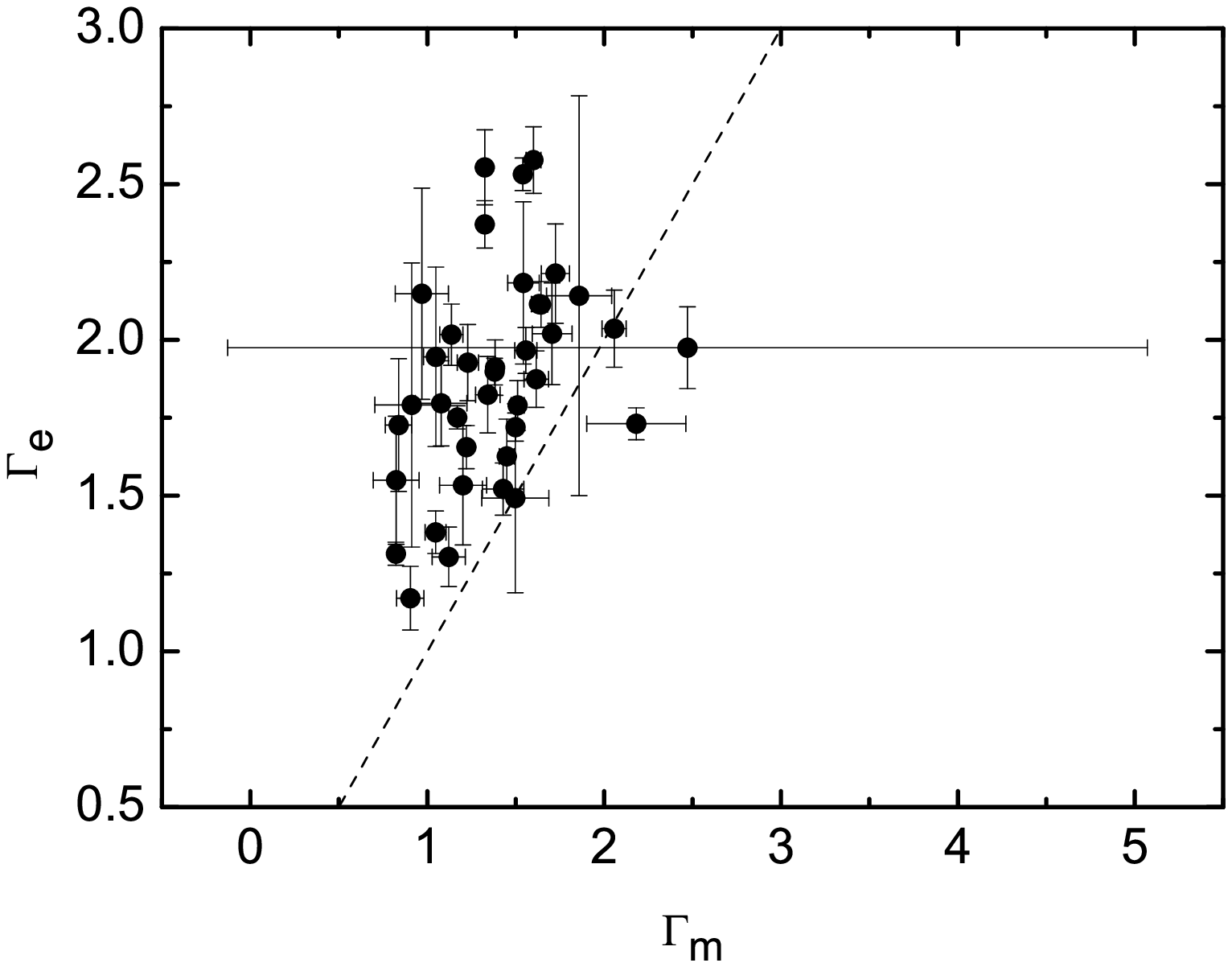}
\caption{Comparisons of the photon index of the main bursts to that of the precursor ({\em left panel}) and the EE tail ({\em right panel}). The solid and opened dots in the left panel are for the triggered and non-triggered precursors, respectively. The dashed lines are the equality lines.}
\label{Fig_Comp_Gamma_Main_Pre_and_EE}
\end{figure}

\begin{figure*}
\includegraphics[angle=0,scale=0.5]{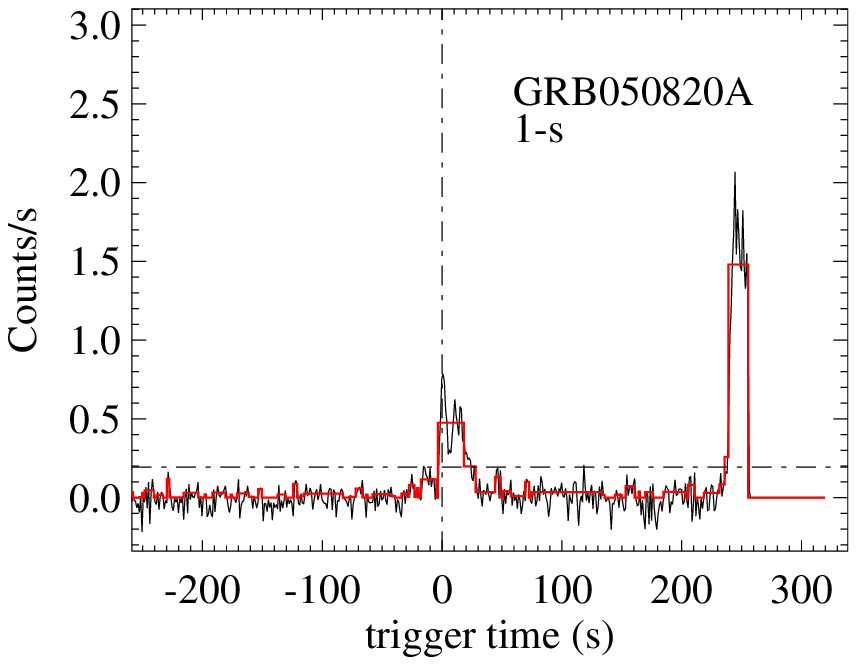}
\includegraphics[angle=0,scale=0.5]{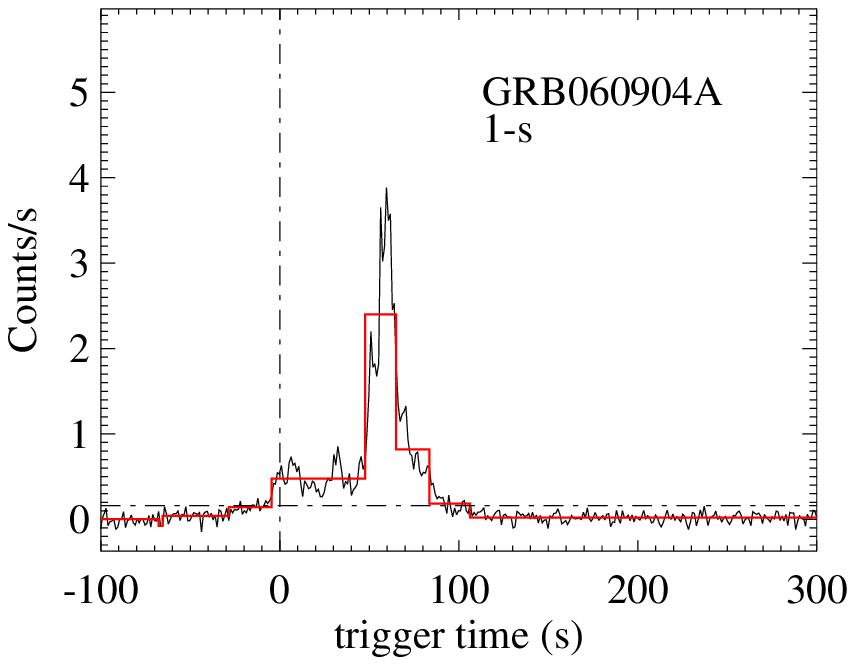}
\includegraphics[angle=0,scale=0.5]{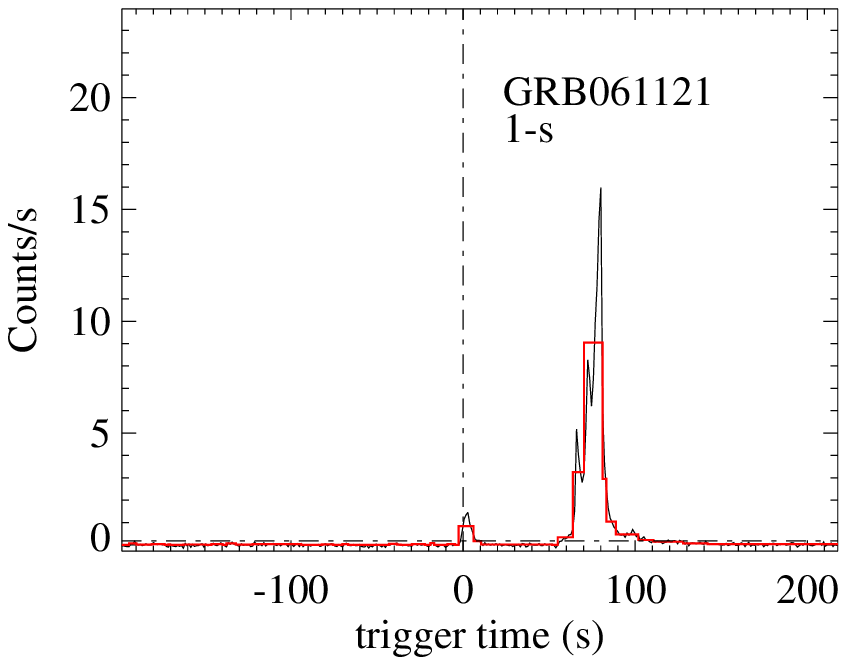}
\includegraphics[angle=0,scale=0.5]{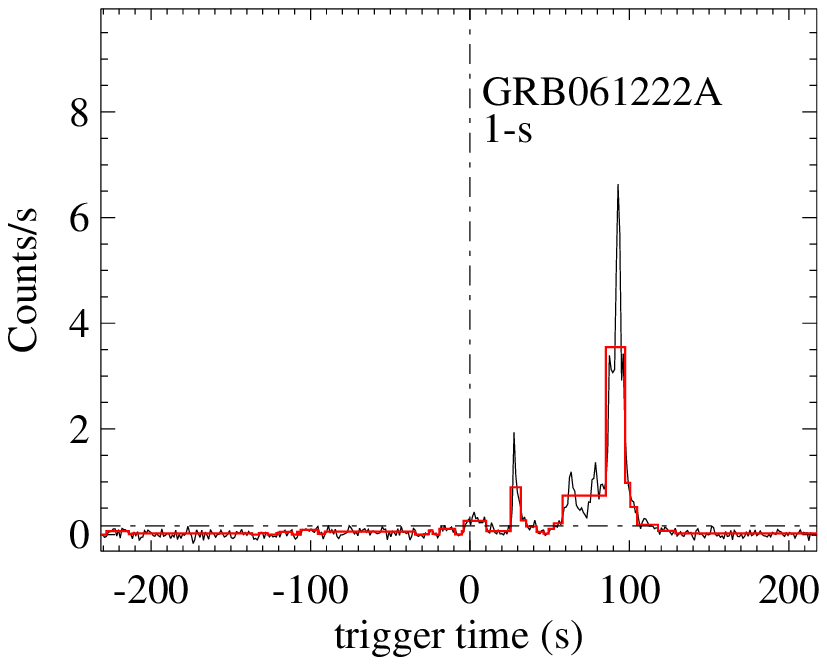}
\includegraphics[angle=0,scale=0.5]{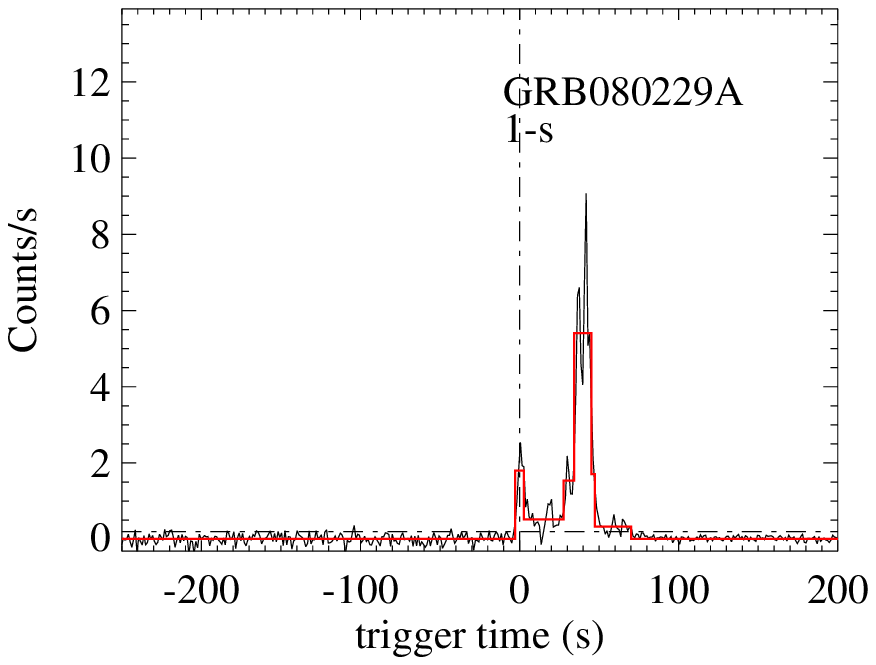}
\includegraphics[angle=0,scale=0.5]{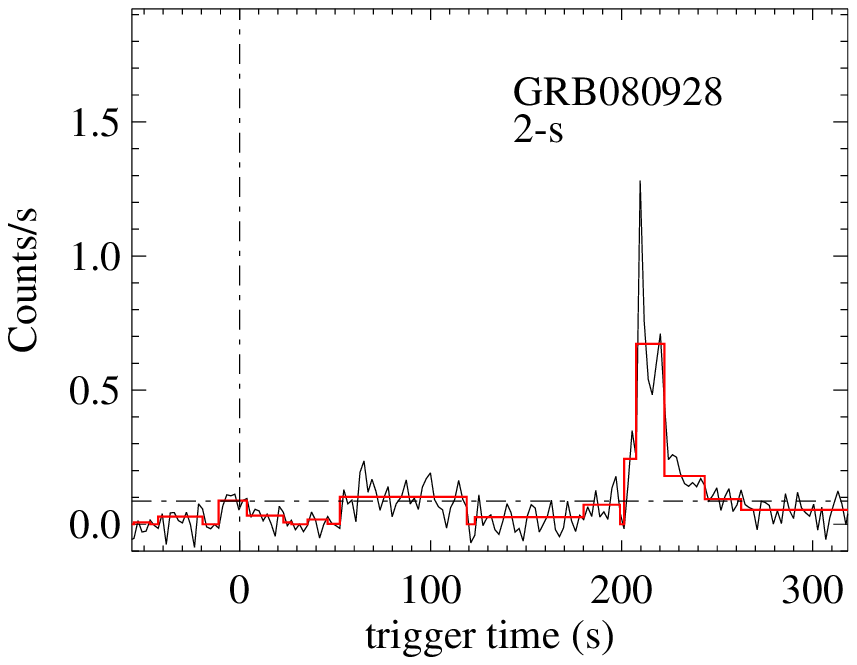}
\includegraphics[angle=0,scale=0.5]{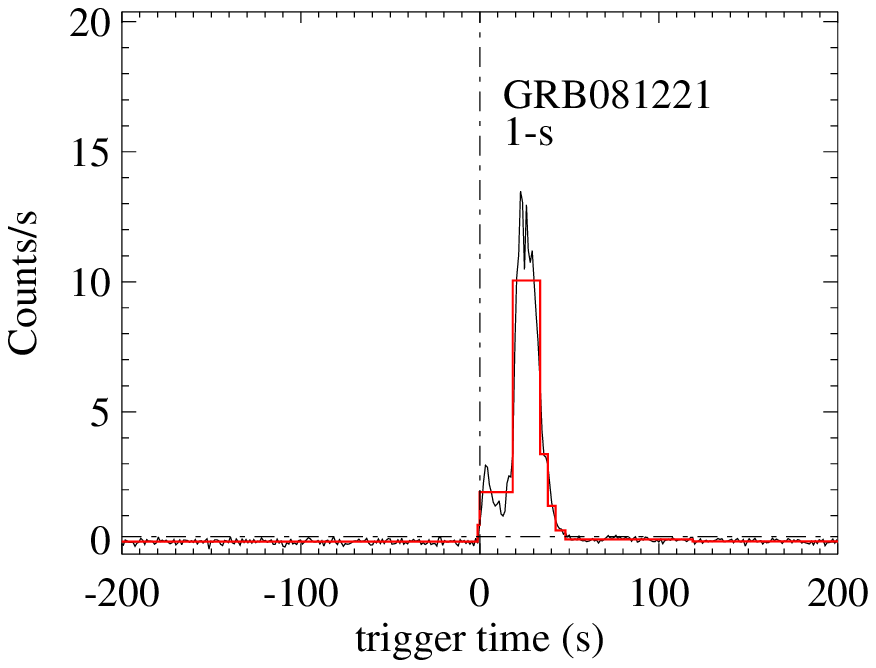}
\includegraphics[angle=0,scale=0.5]{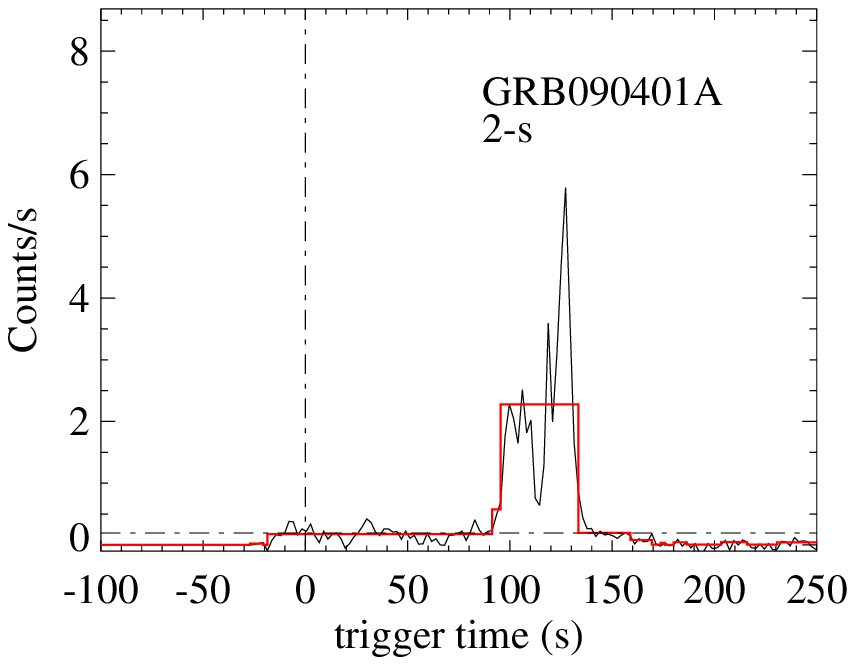}
\includegraphics[angle=0,scale=0.5]{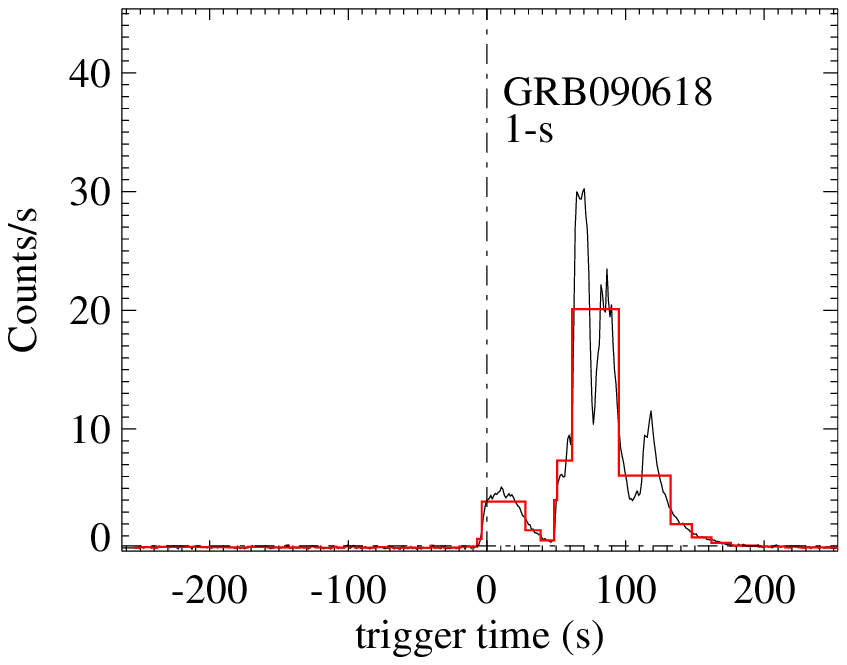}
\includegraphics[angle=0,scale=0.5]{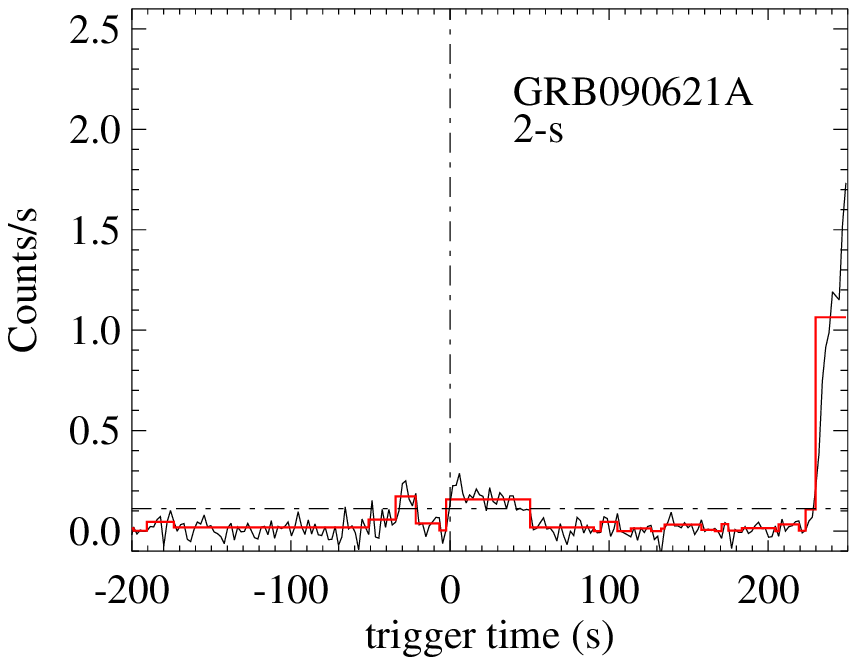}
\includegraphics[angle=0,scale=0.5]{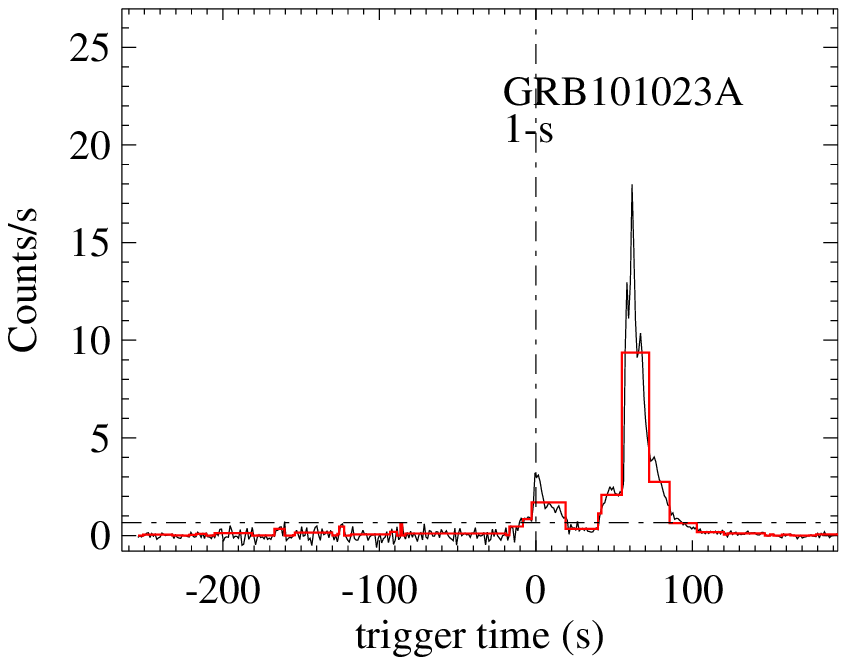}
\includegraphics[angle=0,scale=0.5]{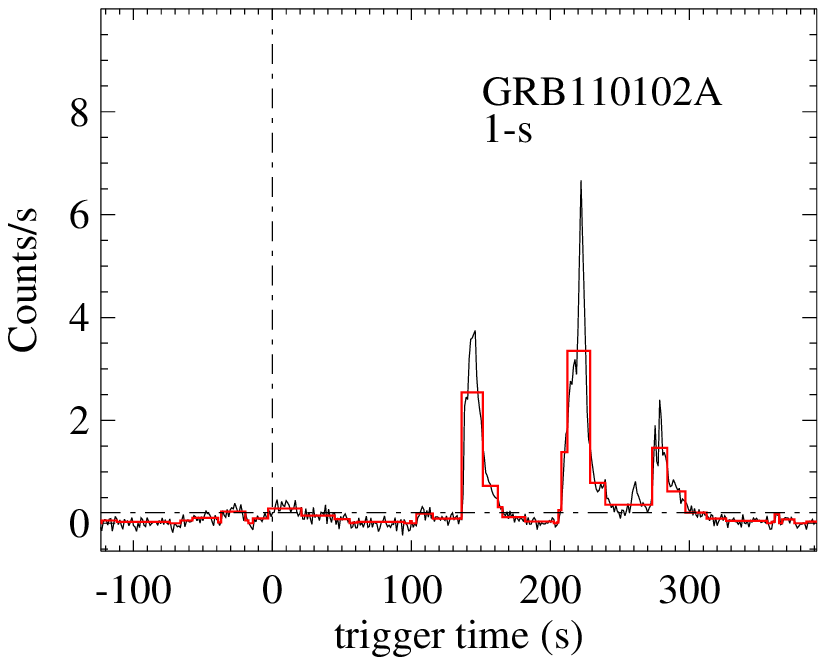}
\includegraphics[angle=0,scale=0.5]{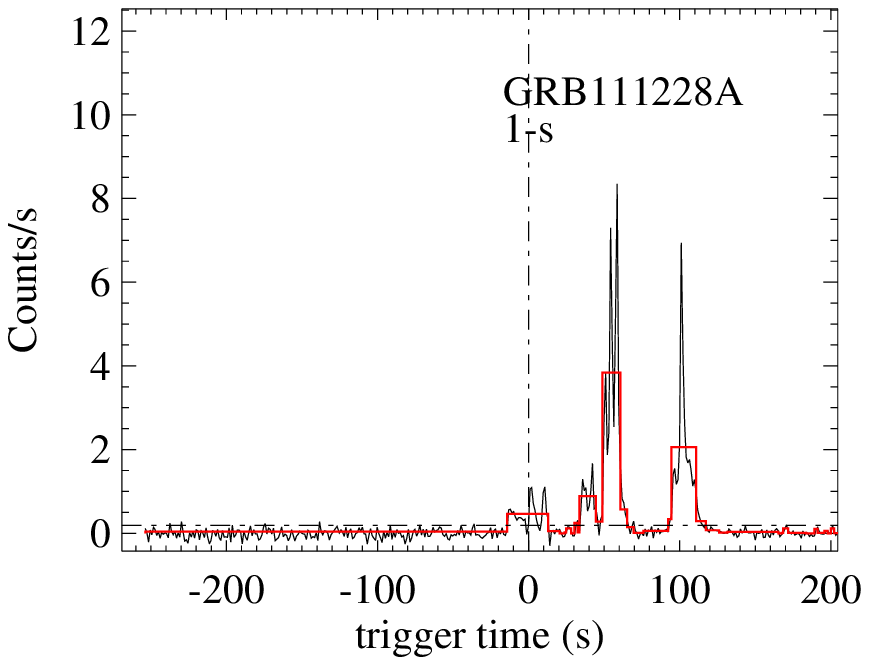}
\includegraphics[angle=0,scale=0.5]{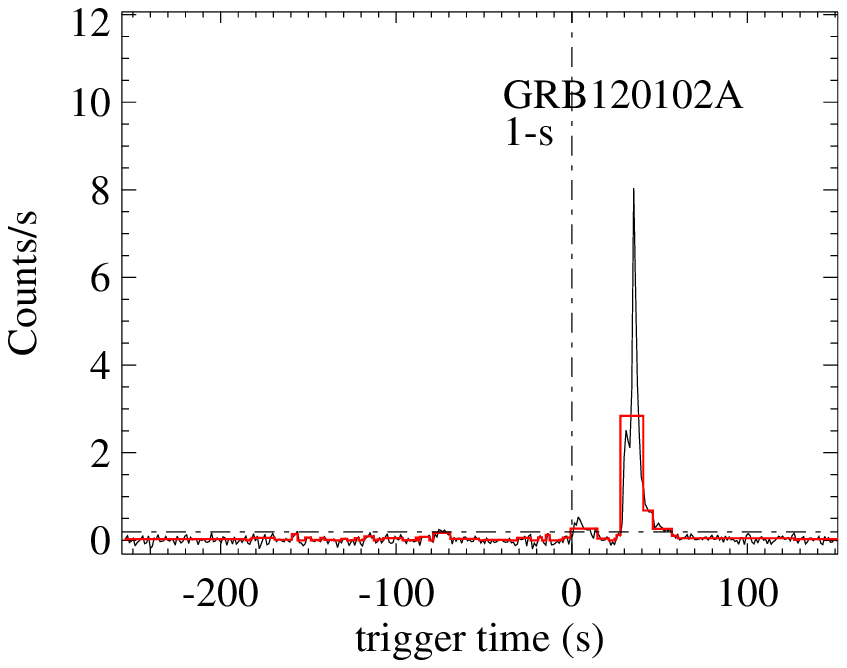}
\includegraphics[angle=0,scale=0.5]{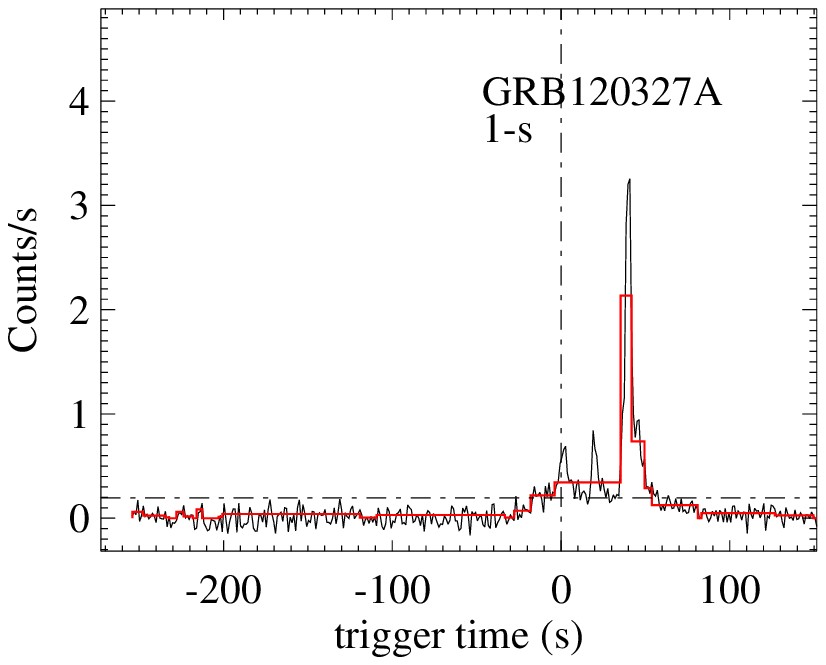}
\hfill
\caption{Examples of our analysis results (step lines) with the BB method for the BAT lightcurves (connected lines) of GRBs triggered with a precursor. The vertical dashed lines mark the BAT trigger time. The horizonal lines mark the $3\sigma$ of the background data in the given time interval as shown in Table 1. The selected bin sizes are also shown. }
\label{LC_Precursors_triggered}
\end{figure*}

\begin{figure*}
\includegraphics[angle=0,scale=0.5]{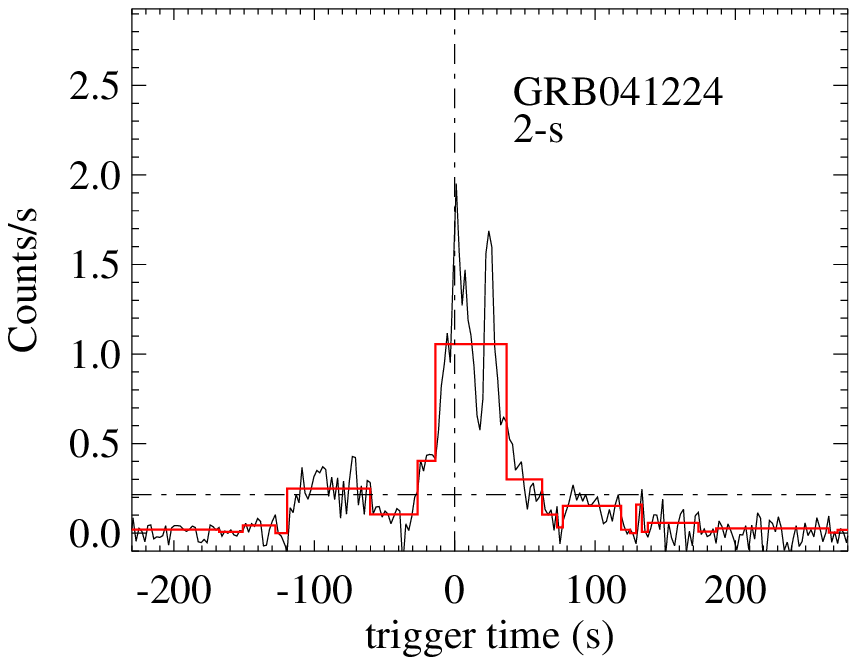}
\includegraphics[angle=0,scale=0.5]{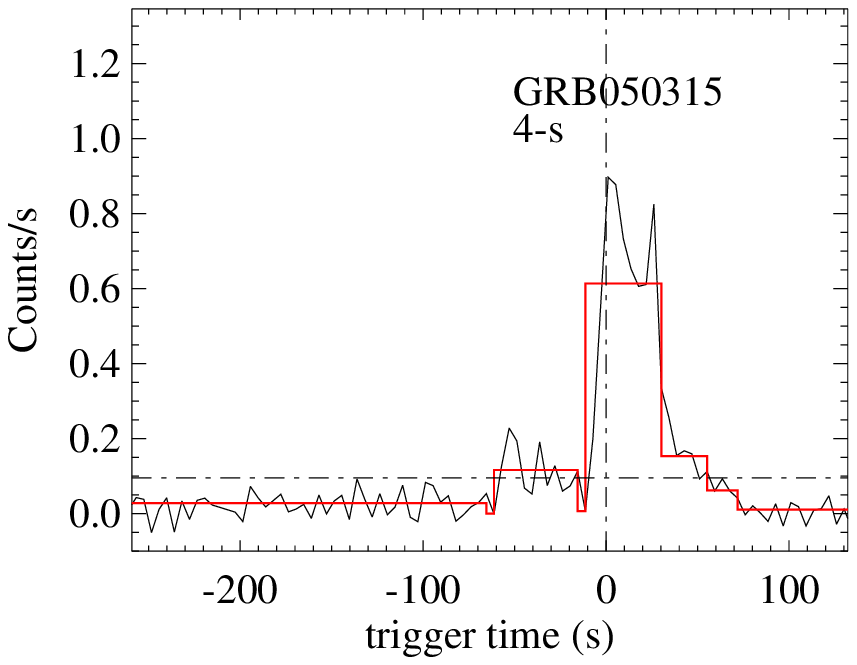}
\includegraphics[angle=0,scale=0.5]{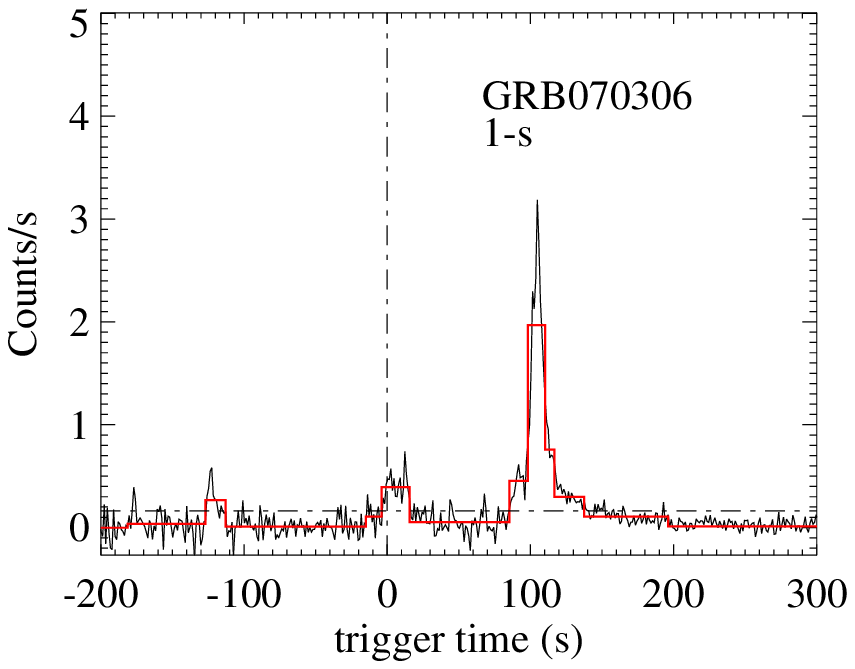}
\includegraphics[angle=0,scale=0.5]{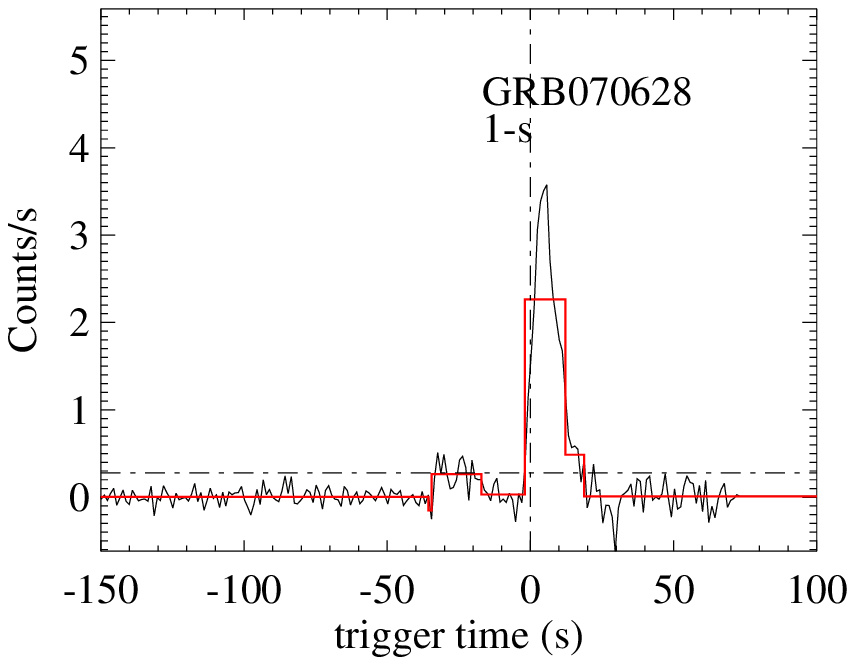}
\includegraphics[angle=0,scale=0.5]{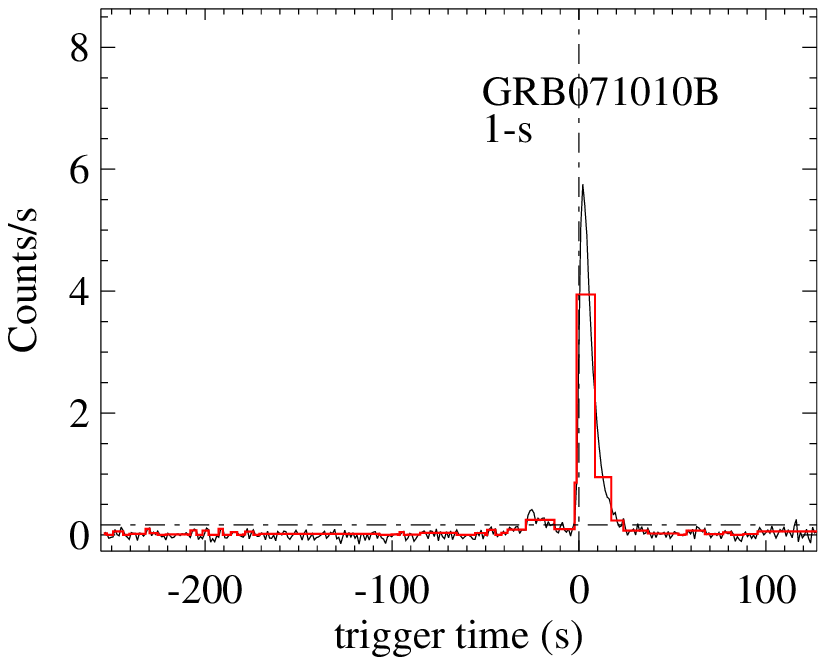}
\includegraphics[angle=0,scale=0.5]{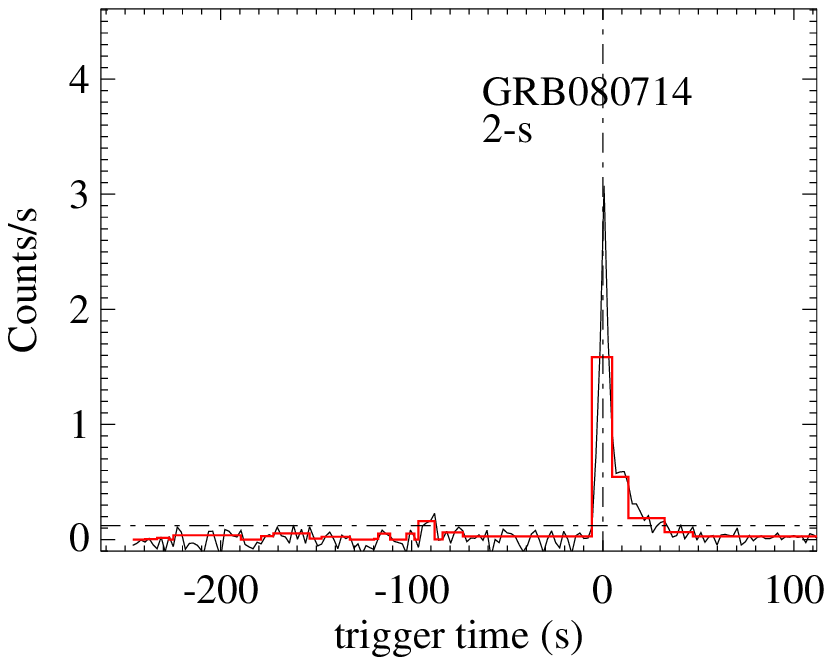}
\includegraphics[angle=0,scale=0.5]{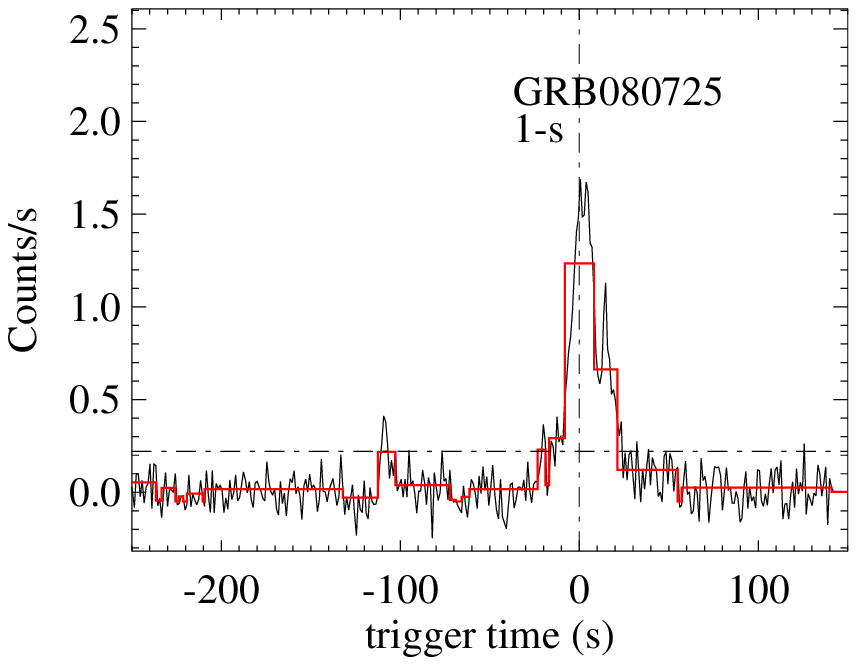}
\includegraphics[angle=0,scale=0.5]{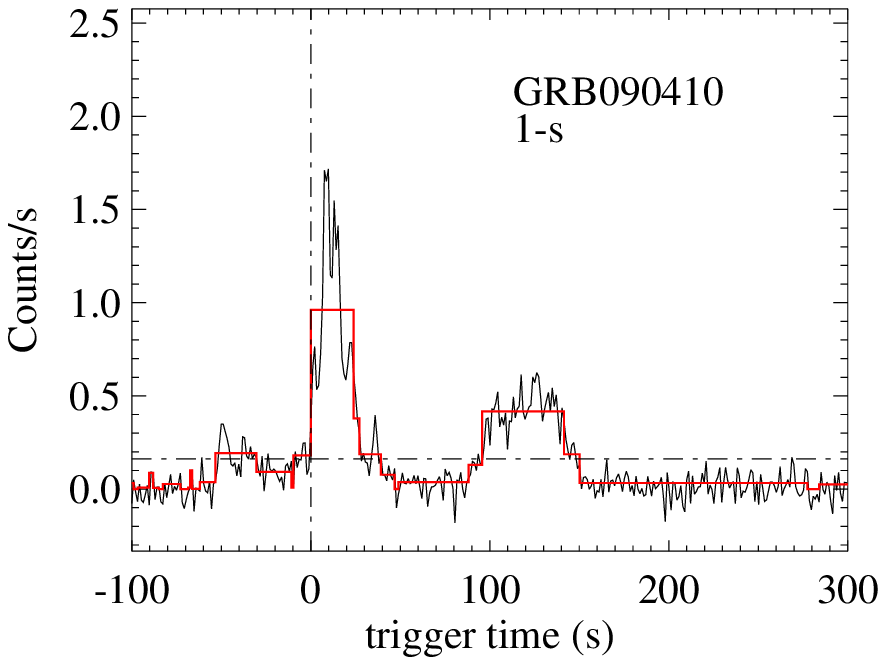}
\includegraphics[angle=0,scale=0.5]{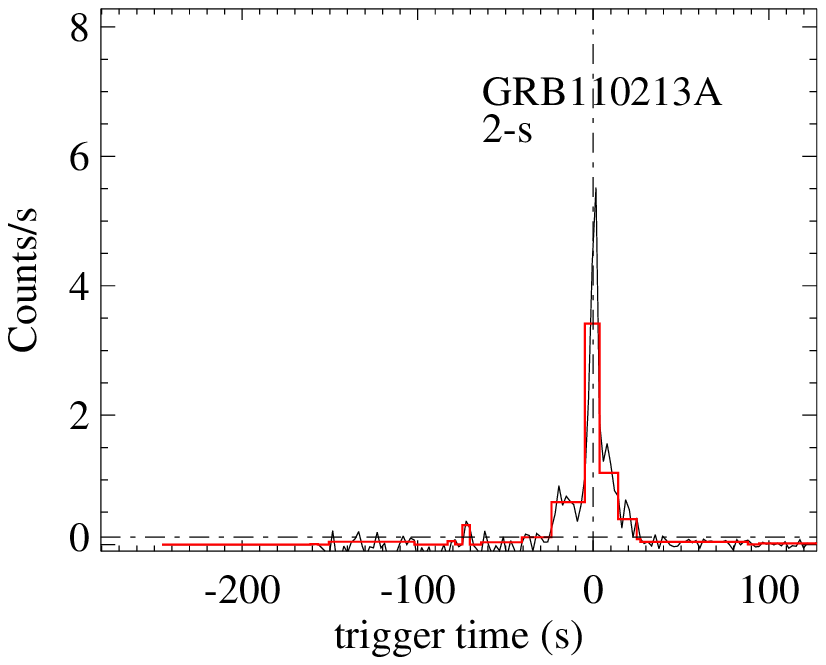}
\includegraphics[angle=0,scale=0.5]{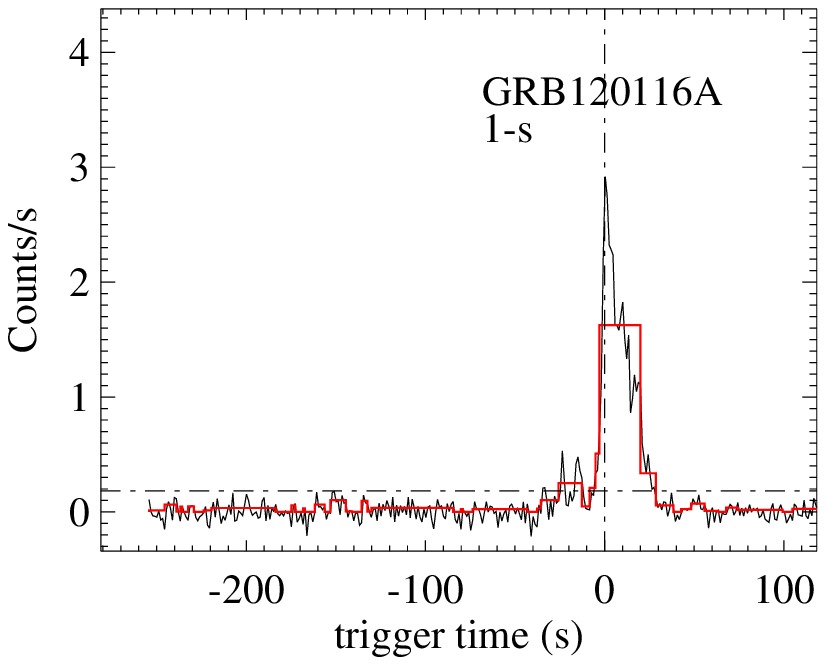}
\includegraphics[angle=0,scale=0.5]{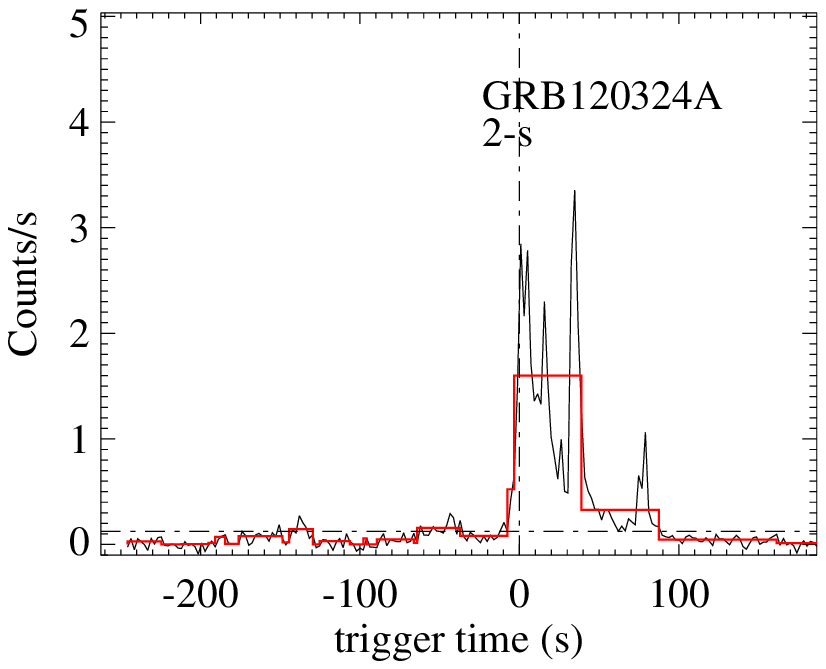}
\includegraphics[angle=0,scale=0.5]{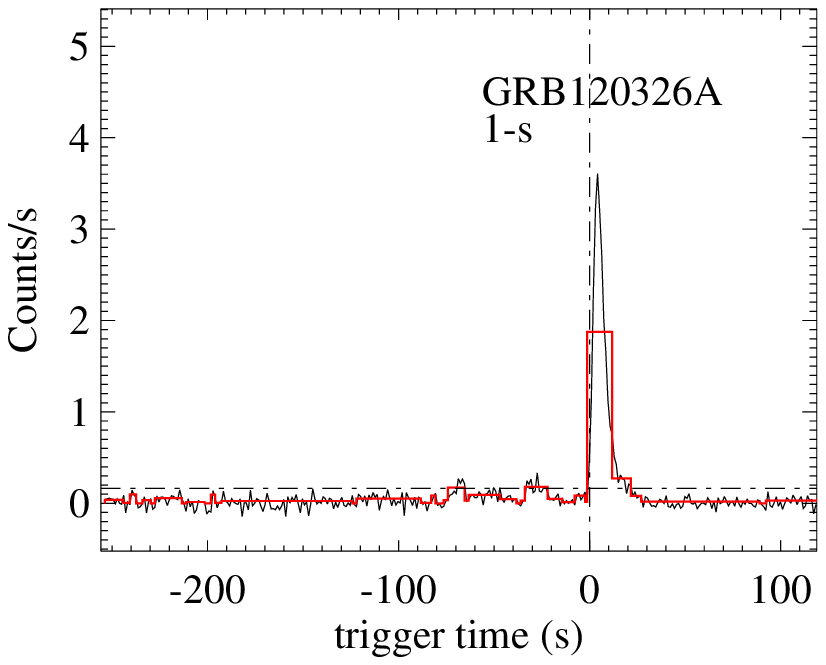}
\includegraphics[angle=0,scale=0.5]{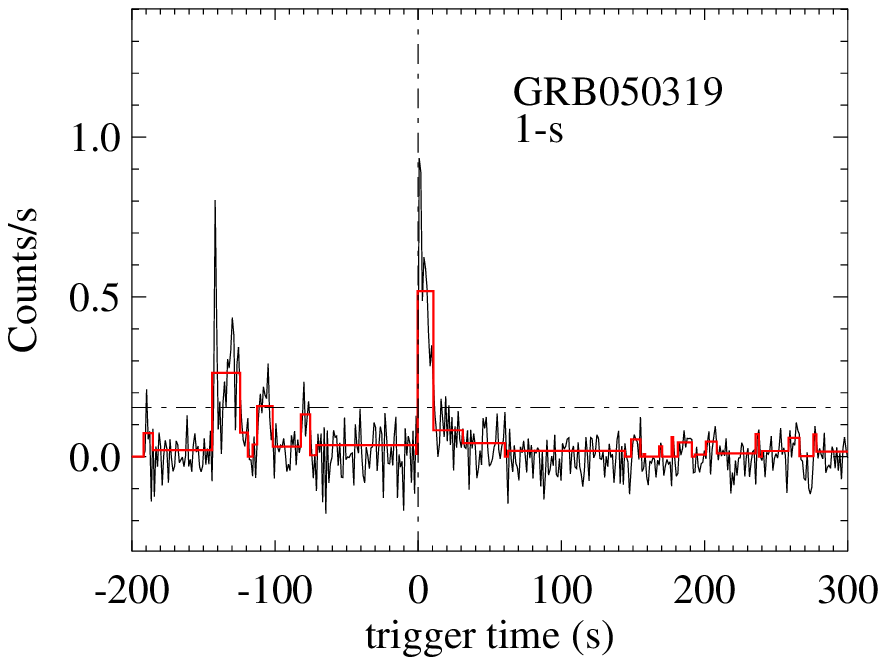}
\includegraphics[angle=0,scale=0.5]{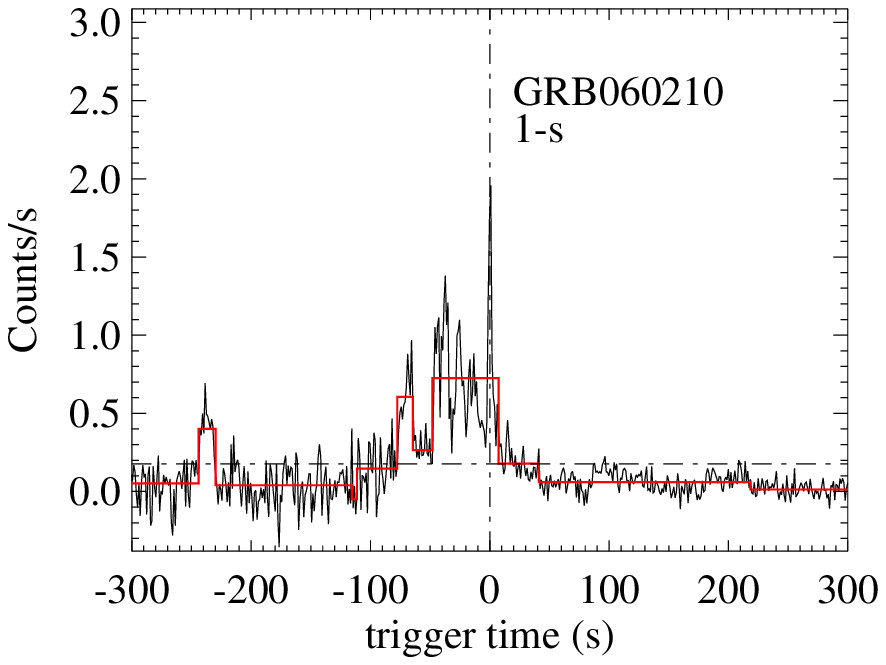}
\includegraphics[angle=0,scale=0.5]{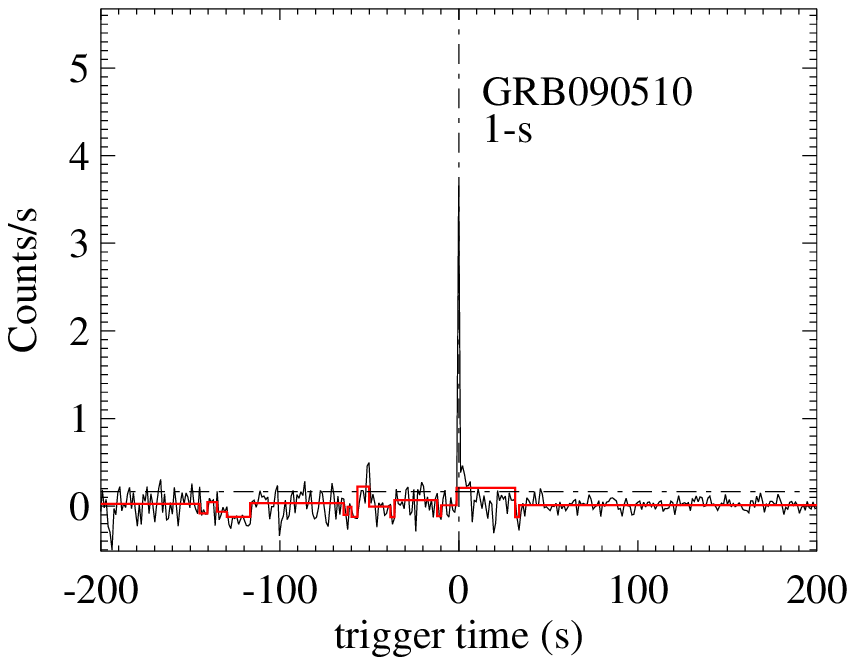}
\hfill
\caption{The same as Figure \ref{LC_Precursors_triggered}, but for non-triggered precursors.}
\label{LC_Precursors_non_triggered}
\end{figure*}

Precursors are of interest in our analysis. We find 50 GRBs whose lightcurves start with a low-$S/N$ episode followed by a bright emission episode. The low-$S/N$ episode may be separated from or closely connected to the bright episode. Traditionally, a precursor is defined if the early low-$S/N$ component triggers BAT.  We find that the low-S/N component in some GRBs did not trigger BAT. Here we adopt the term {\em precursor} to refer to an early low-S/N component leading the main emissioin episode regardless of whether it triggered BAT. Among the 50 GRBs, 22 were trigged with the precursor, whereas the other 28 GRBs were triggered with the main bursts. Figure \ref{LC_Precursors_triggered} and Figure \ref{LC_Precursors_non_triggered} show the examples for the triggered and non-triggered precursors, respectively. We measure the duration of the precursor and its separation to the main episode based on our BB analysis results for each burst. Our results are reported in Table 1. The average count rate ratio of the precursor to the main burst spans from 0.05 to 0.5, with a typical value of $\sim 0.2$, as shown in Figure \ref{Precursor_Ratio_and_interval} left. The distribution of the ratio between the durations of the precursor and the main episode is shown in Figure \ref{Precursor_Ratio_and_interval} right. It is found that the majority of precursors are shorter than the main burst. The duration ratio concentrates around 0.5. We also extract and fit the spectra of the precursors and the main bursts of these GRBs. Figure \ref{Fig_Comp_Gamma_Main_Pre_and_EE} left shows the comparison of the photon index of the precursor and main episode ($\Gamma_p$ and $\Gamma_m$). We find that photon indices of the precursors are comparable to those of the main episodes. We do not find any systematic differences between the triggered and non-triggered precursors. These results seems to indicate that the precursor is a weak emission episode that has the same origin as the main burst in a GRB.

Fifty seven GRBs with $T_{90}<2$ seconds are included in our sample. Except for GRB 090510, we do not find other strong cases that have a significant precursor\footnote {Short GRB 090510 detected by {\em Fermi}/GBM shows evidence of a precursor(Troja et al 2010). The S/N ratio of the precursor is only $\sim 3$. Since it is a high-luminosity short GRB, the possibility that it is a Type II GRB is not ruled out (Zhang et al. 2009; Panaitescu 2011; Bromberg et al. 2013).} This result likely indicates that the precursors preferably happen in GRBs associated with collapses of massive stars (Type II GRBs), and but not in GRBs produced by mergers of compact stars (Type I).

\subsection{The EE tails\label{sec:calculate}}
We recognize the weak episode following the main episode as an EE tail with a criterion that the EE tail has a count rate that is lower than 50\% of the mean count rate of the main burst (except for GRB 060614 whose EE tail is even brighter). Such an EE tail was detected for 36 GRBs in our sample. We extract the spectra of the EE tails in these GRBs and fit them with a simple power law model. Our analysis for the EE tails are reported in Table 2. Figure \ref{Fig_Comp_Gamma_Main_Pre_and_EE} right shows the comparison of the photon index of the EE component ($\Gamma_e$) to that of the main episode. It is found that the EE tails are softer than that of the main episode.

\begin{figure*}
\includegraphics[angle=0,scale=0.5]{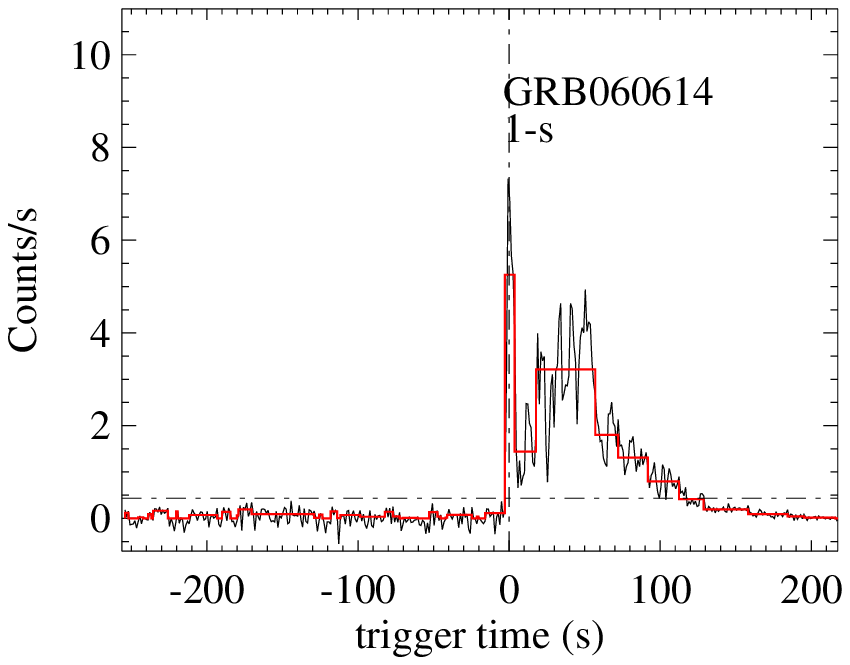}
\includegraphics[angle=0,scale=0.5]{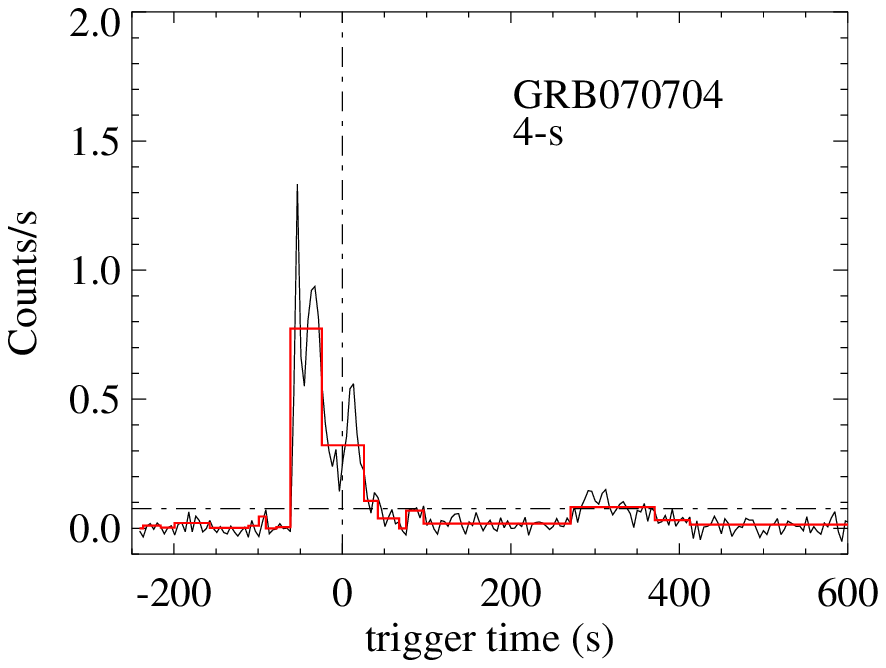}
\includegraphics[angle=0,scale=0.5]{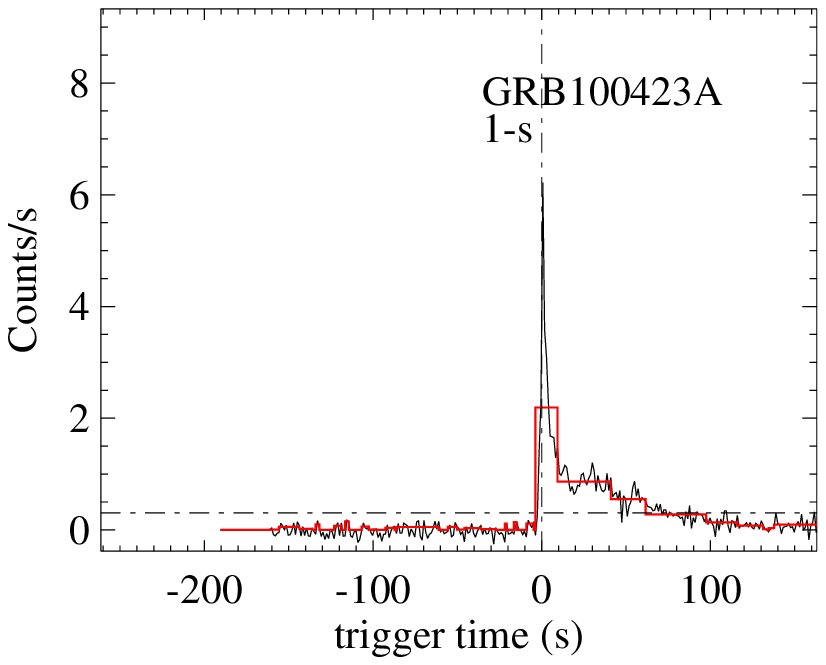}
\includegraphics[angle=0,scale=0.5]{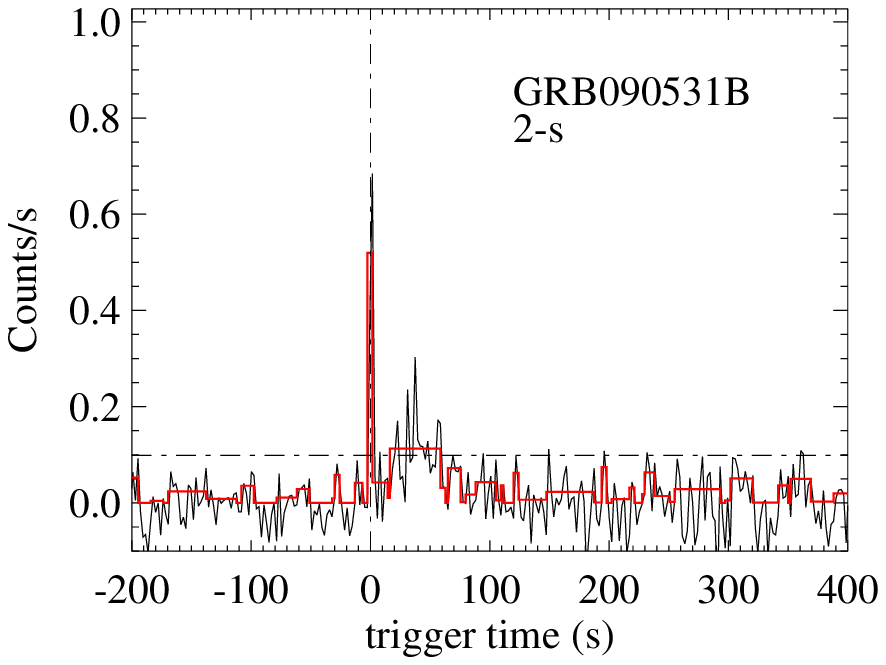}
\includegraphics[angle=0,scale=0.5]{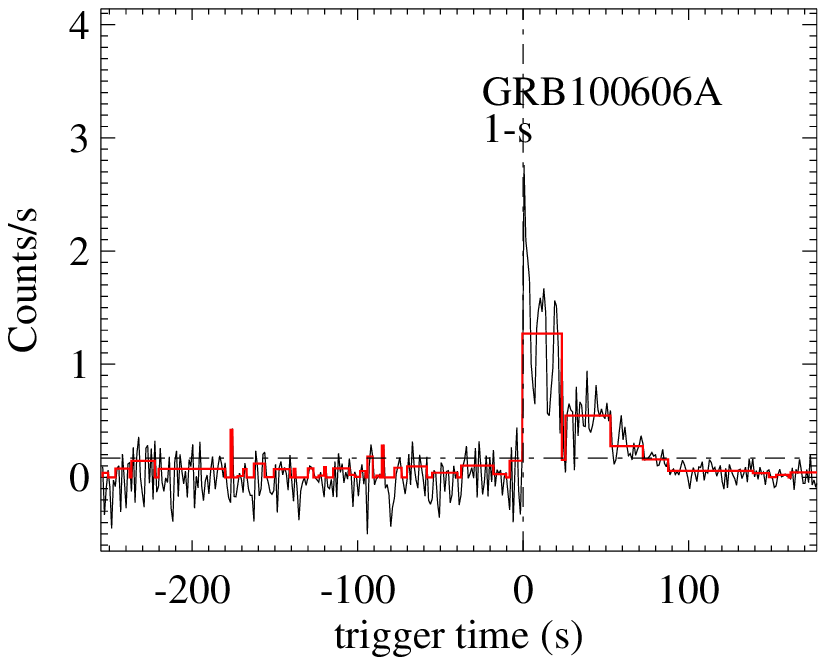}
\includegraphics[angle=0,scale=0.5]{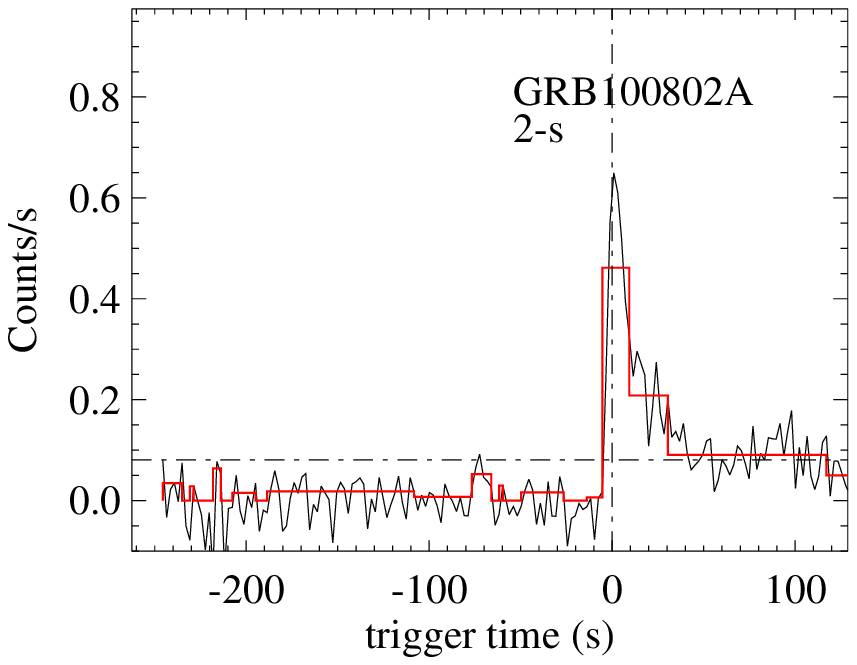}
\includegraphics[angle=0,scale=0.5]{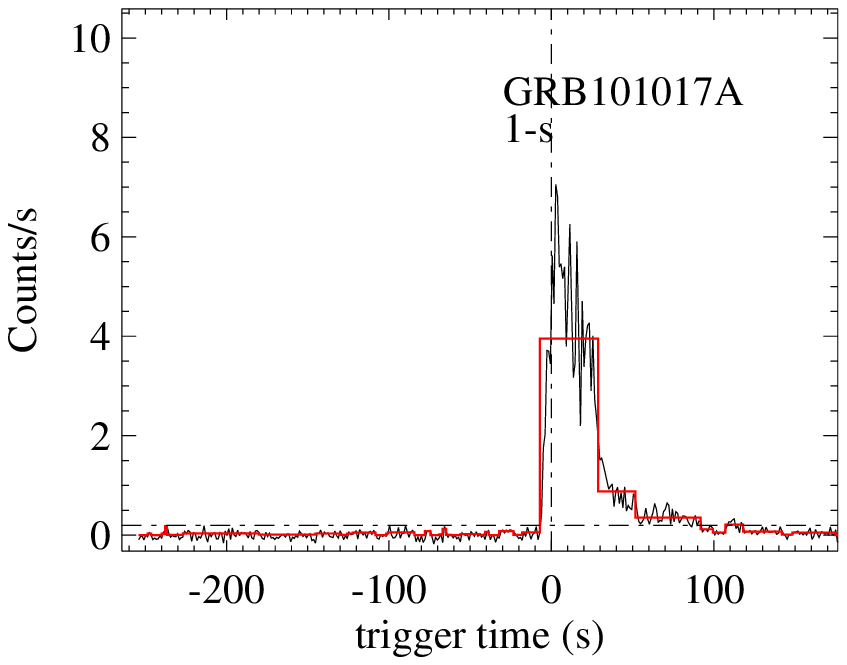}
\includegraphics[angle=0,scale=0.5]{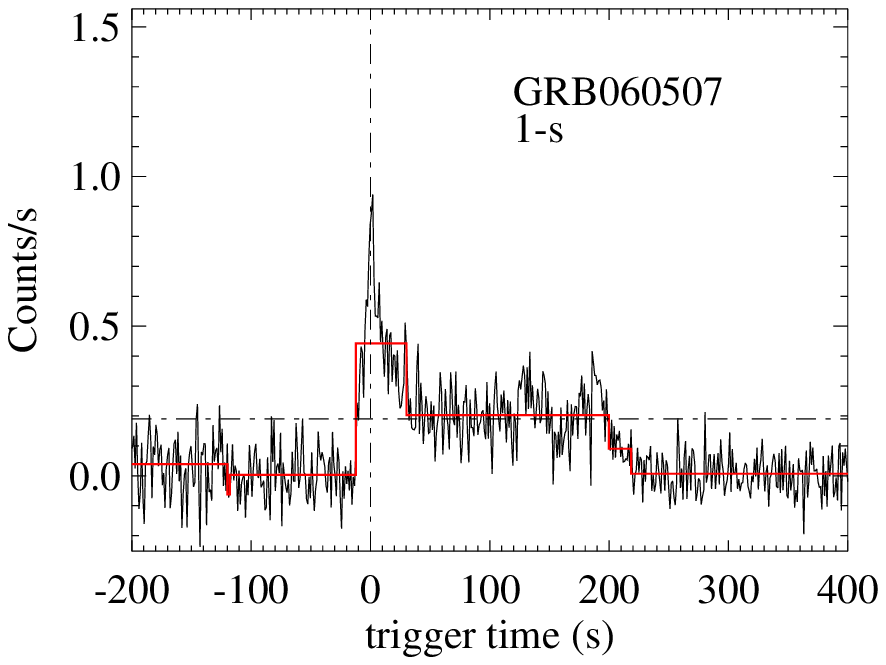}
\includegraphics[angle=0,scale=0.5]{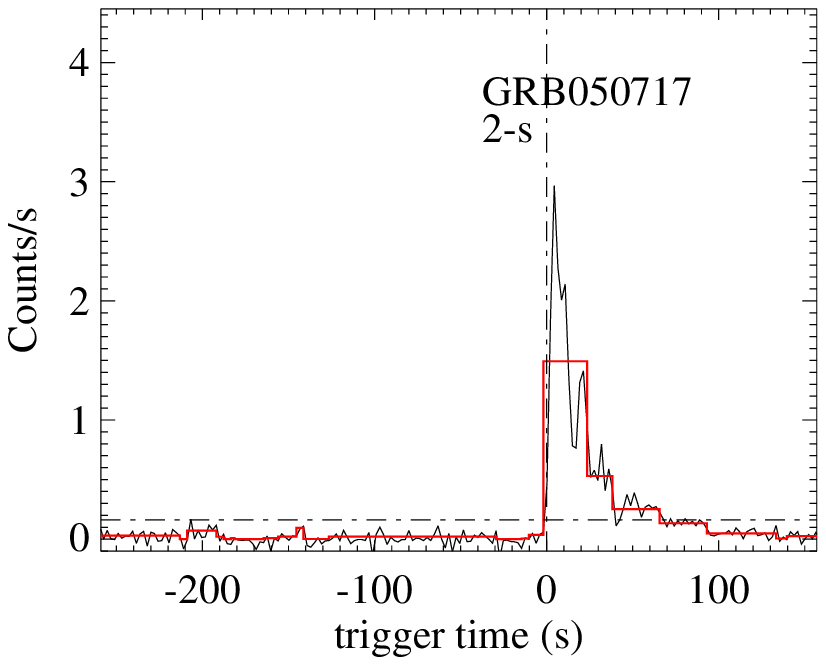}
\includegraphics[angle=0,scale=0.5]{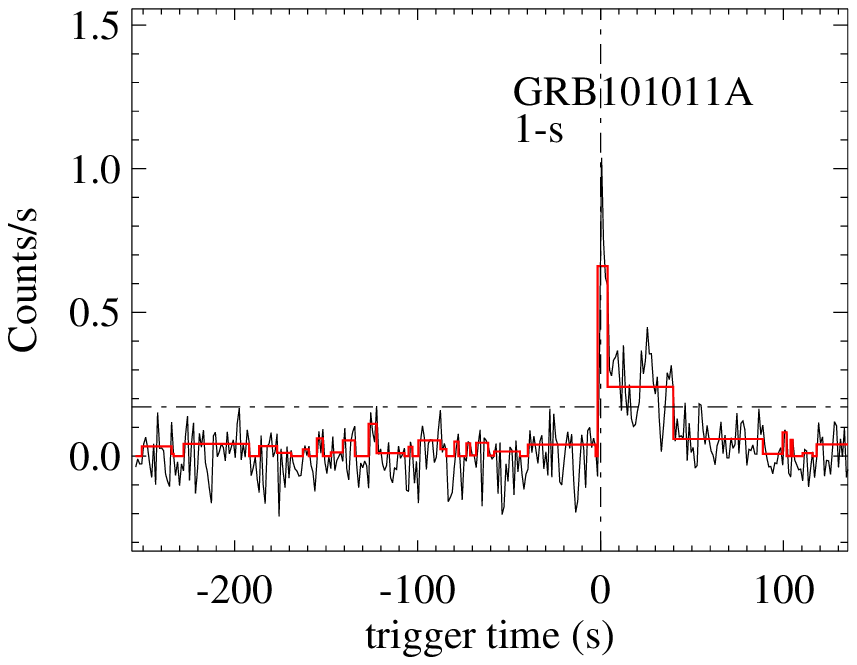}
\includegraphics[angle=0,scale=0.5]{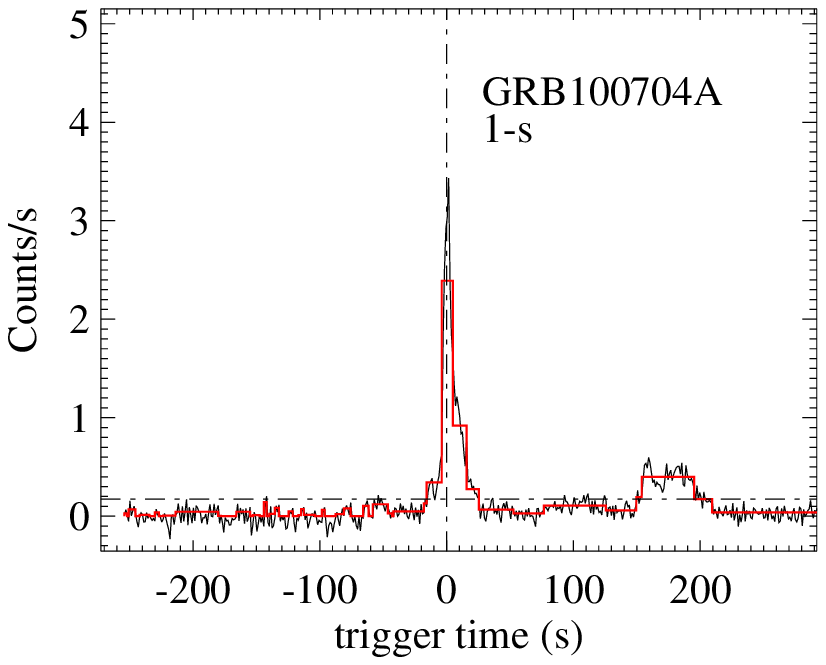}
\includegraphics[angle=0,scale=0.5]{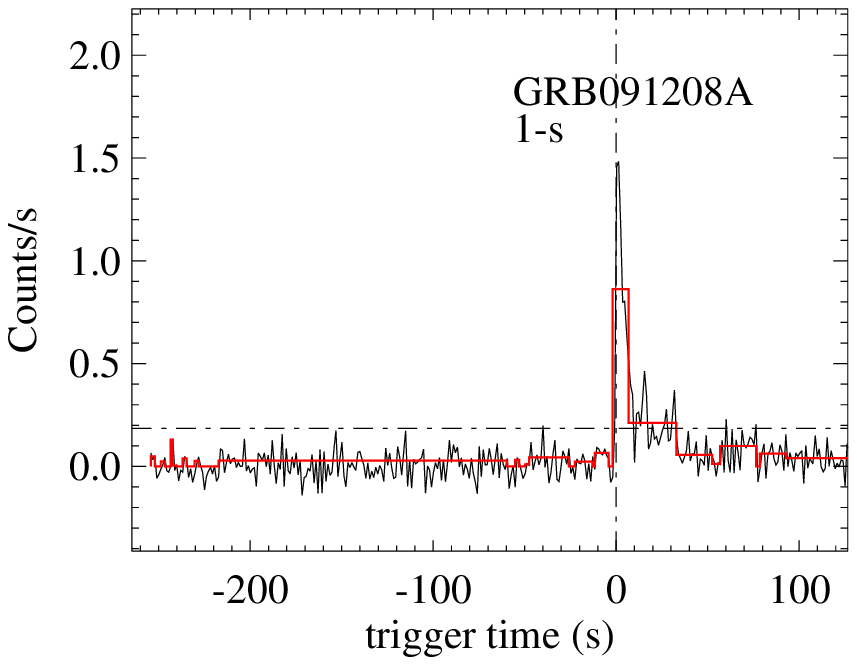}
\includegraphics[angle=0,scale=0.5]{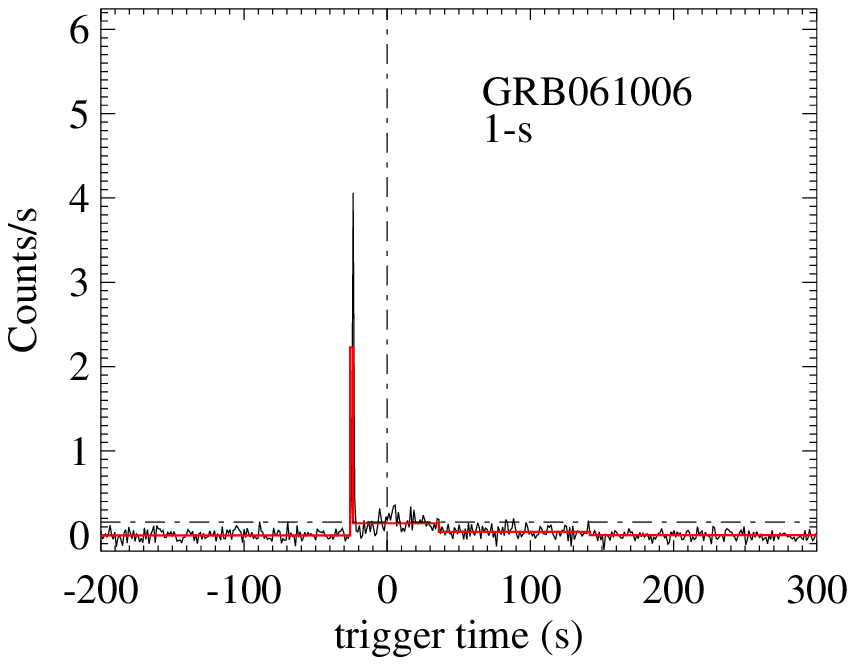}
\includegraphics[angle=0,scale=0.5]{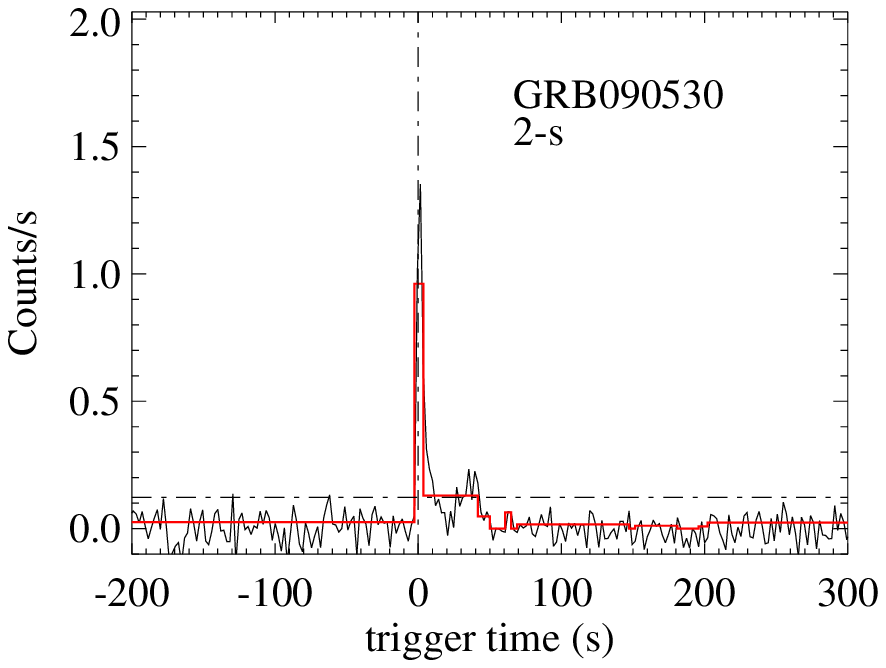}
\includegraphics[angle=0,scale=0.5]{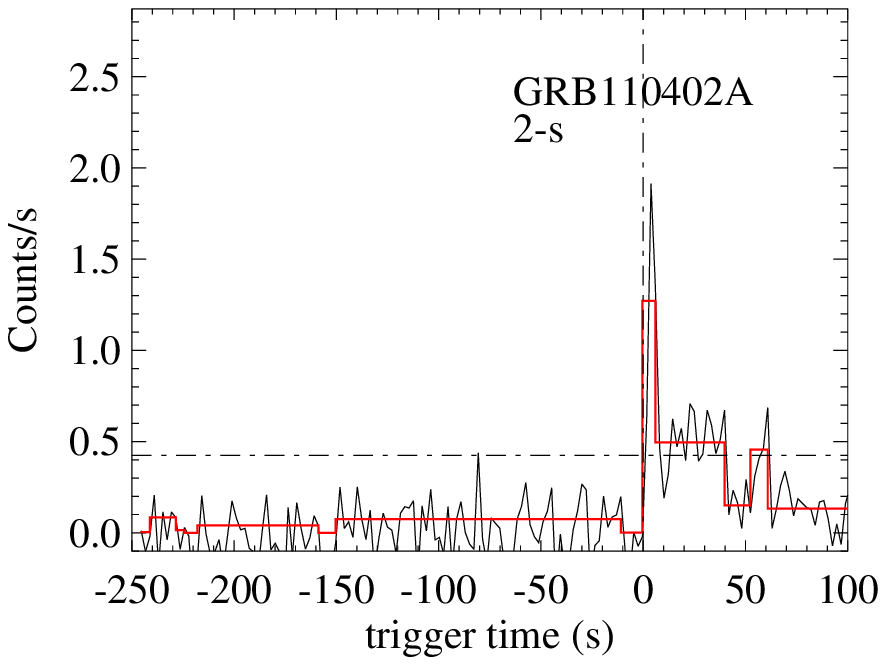}
\includegraphics[angle=0,scale=0.5]{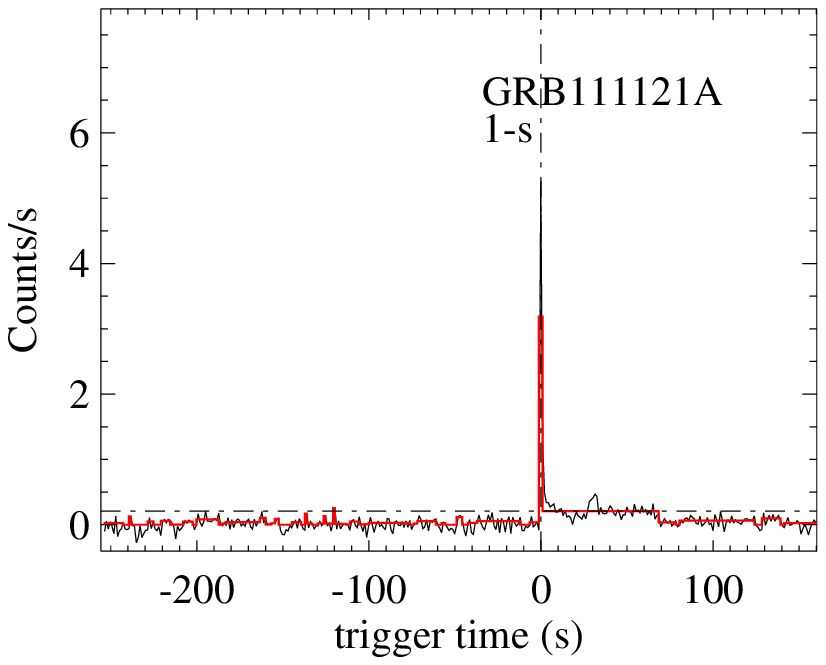}
\includegraphics[angle=0,scale=0.5]{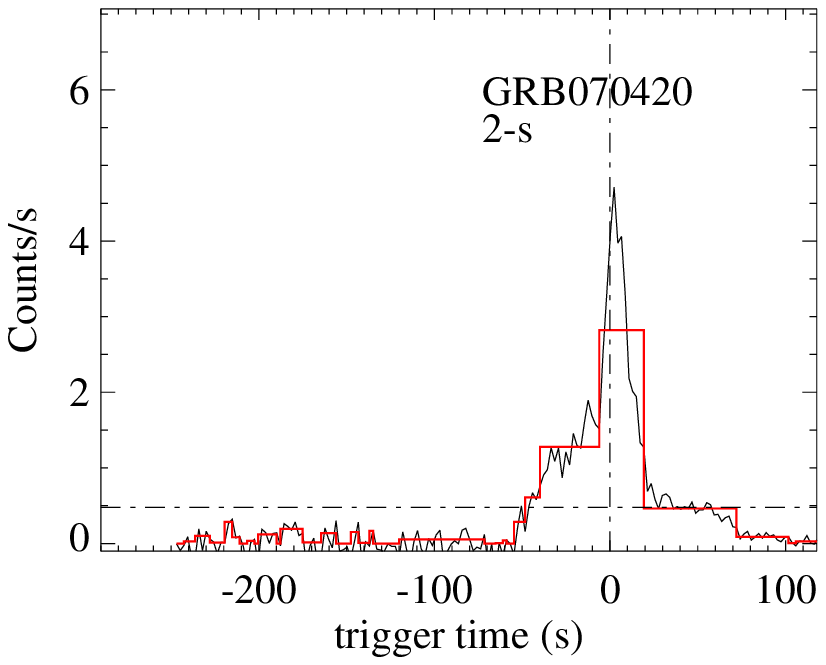}
\hfill
\caption{The same as Figure \ref{LC_Precursors_triggered}, but for the extended tails.}
\label{LC_EEs}
\end{figure*}

We show some typical lightcurves with such a component in Figure \ref{LC_EEs}. One of the most prominent feature of these lightcurves is that they usually start with a bright and sharp spikes. GRB 060614 is a prototype of the EE tails. Some GRBs whose lightcurves are similar to that of GRB 060614 are shown in the first three rows of Figure \ref{LC_EEs}. The EE tails are usually highly variable and their brightness gradually decreases. The bright EE tails detected in some GRBs may extend up to $\sim 100$ seconds, even hundreds of seconds, after the BAT trigger. Besides the GRB 060614-like events, the EE tails in some GRBs show as continuous fluctuations with amplitudes being much smaller than the initial spikes, such as those as observed in GRB 061006 and 111121A. According to our analysis, EE tails are usually detected in some ``short'' GRBs. These GRBs are usually classified as short since their EE tails are ignored in $T_{90}$ calculations\footnote{It was suggested that an EE tail presents in some short GRBs, such as 050724,070223, 0507242, 050911,051227, 061210, 070714B,071227,090715A, and 090916 (e.g., Barthelmy et al. 2007; Zhang et al. 2007; Zhang et al. 2009). Our analysis shows a low-S/N block after the main burst in the lightcurves of these GRBs, but the S/N ratio is lower than 3$\sigma$. We therefore do not include these GRBs in our analysis to keep a uniform criterion for both precursors and extended emission tails.}. The lightcurves of typical long GRBs are composed of several separated pulses. In case that the amplitude of these pulses decreases with time, the later pulses may be identified as EE tails. Excluding this kind of GRBs, we find a prominent EE tail only in GRB 070420. Therefore, if both GRB 060614-liked GRBs and GRB 061006-liked GRBs have the same origin (e.g., Zhang et al. 2007), the EE tails may preferably happen in GRBs from compact star mergers (Type I).

\subsection{X-Ray Emission from Internal Energy Dissipation}
\begin{figure*}
\includegraphics[angle=0,scale=0.350]{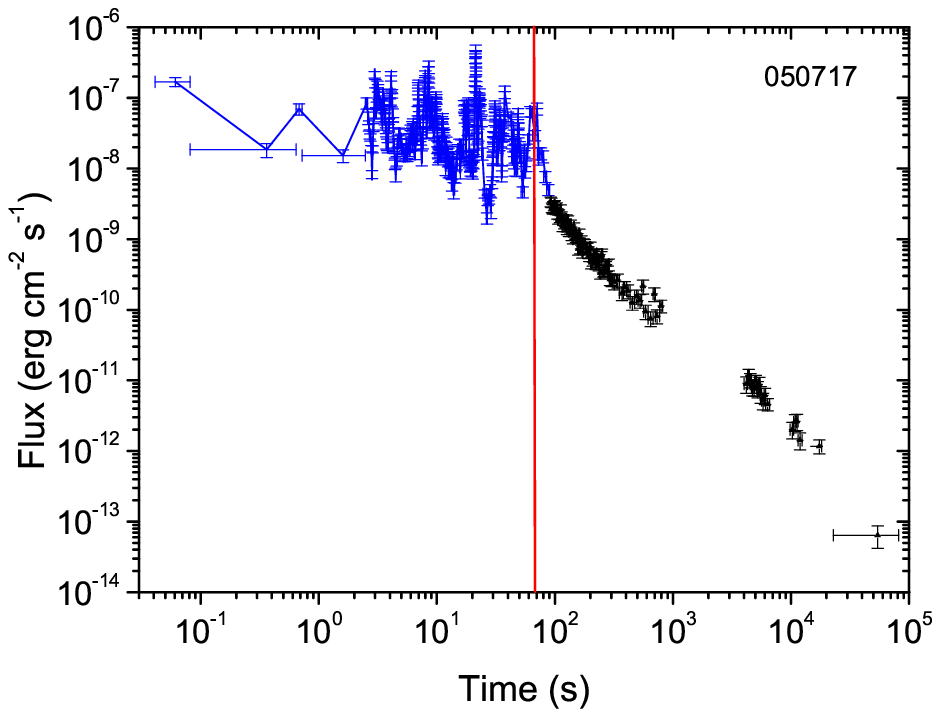}
\includegraphics[angle=0,scale=0.350]{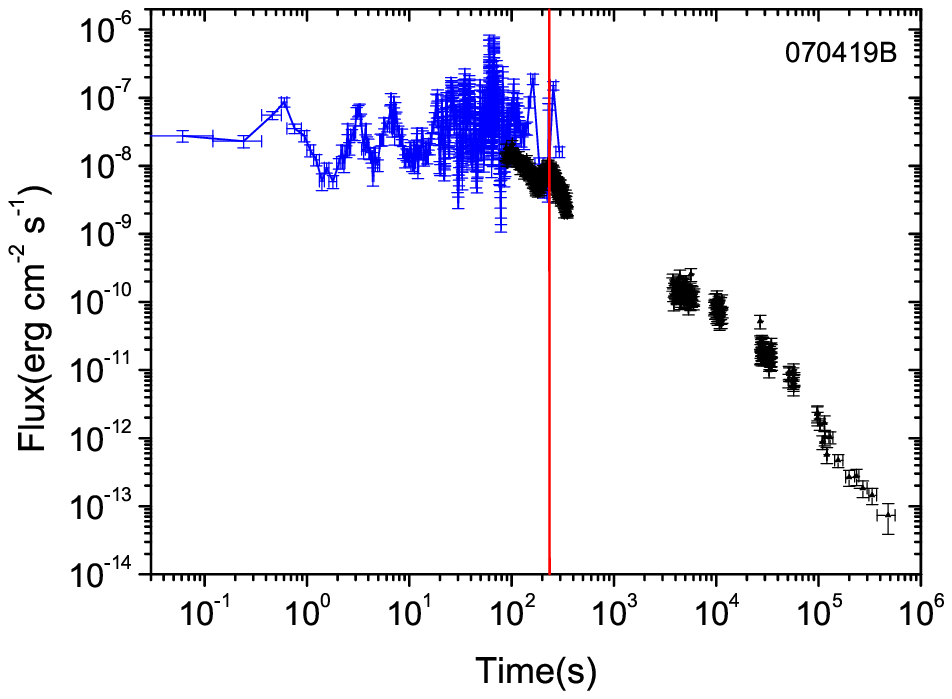}
\includegraphics[angle=0,scale=0.350]{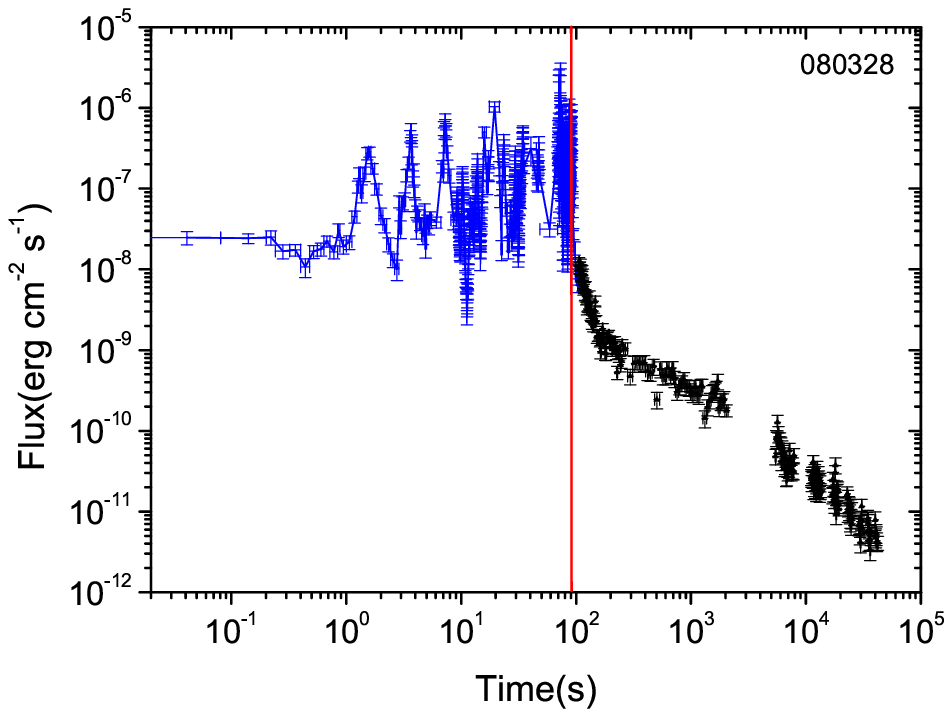}
\includegraphics[angle=0,scale=0.350]{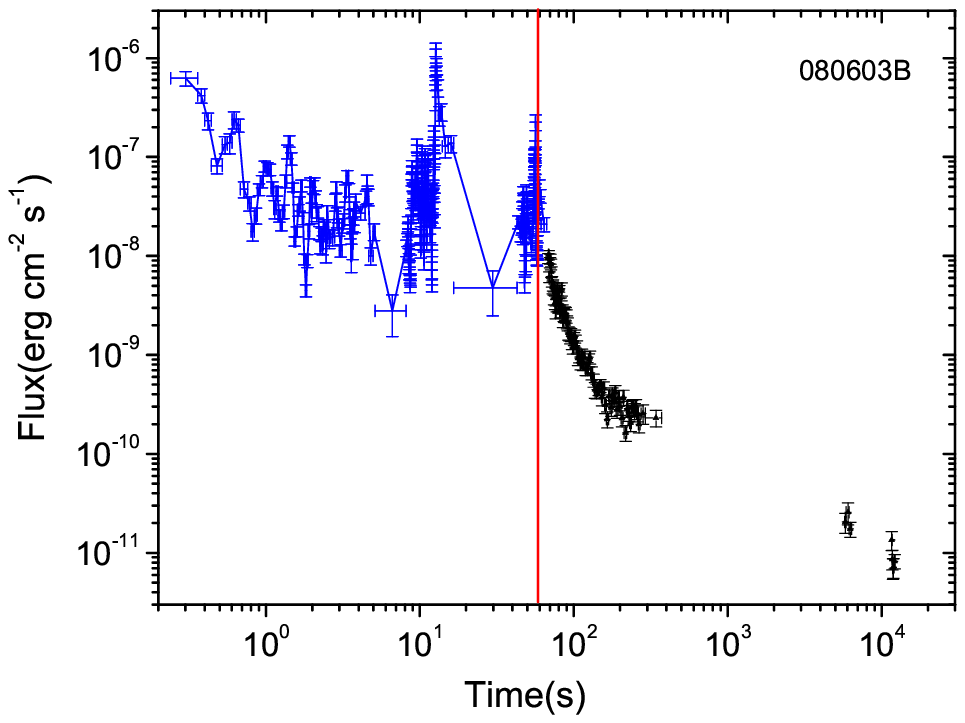}
\includegraphics[angle=0,scale=0.350]{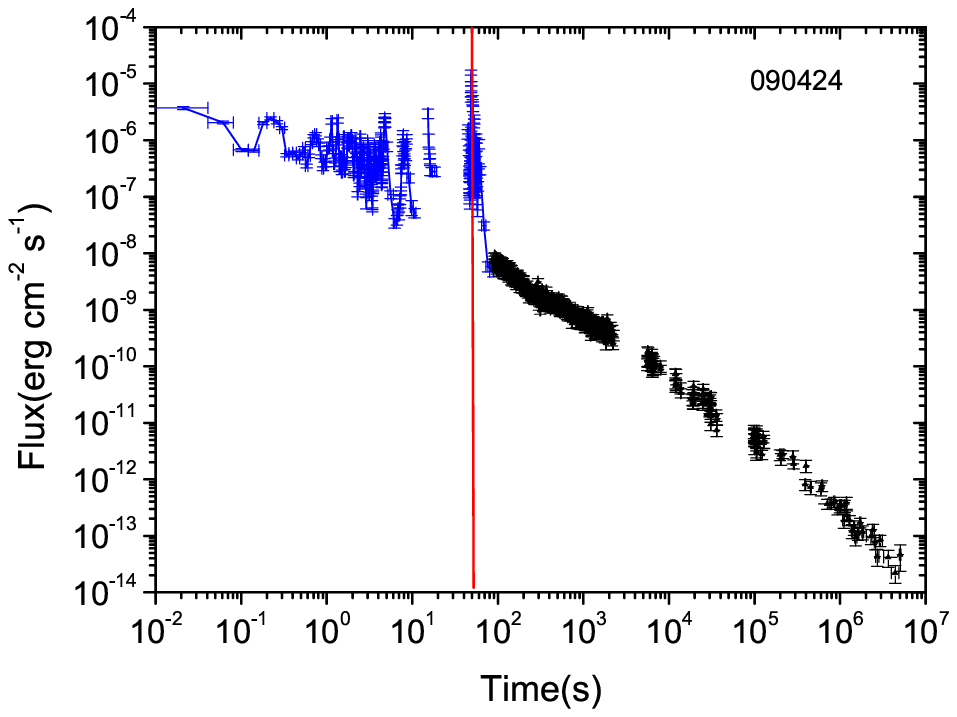}
\includegraphics[angle=0,scale=0.350]{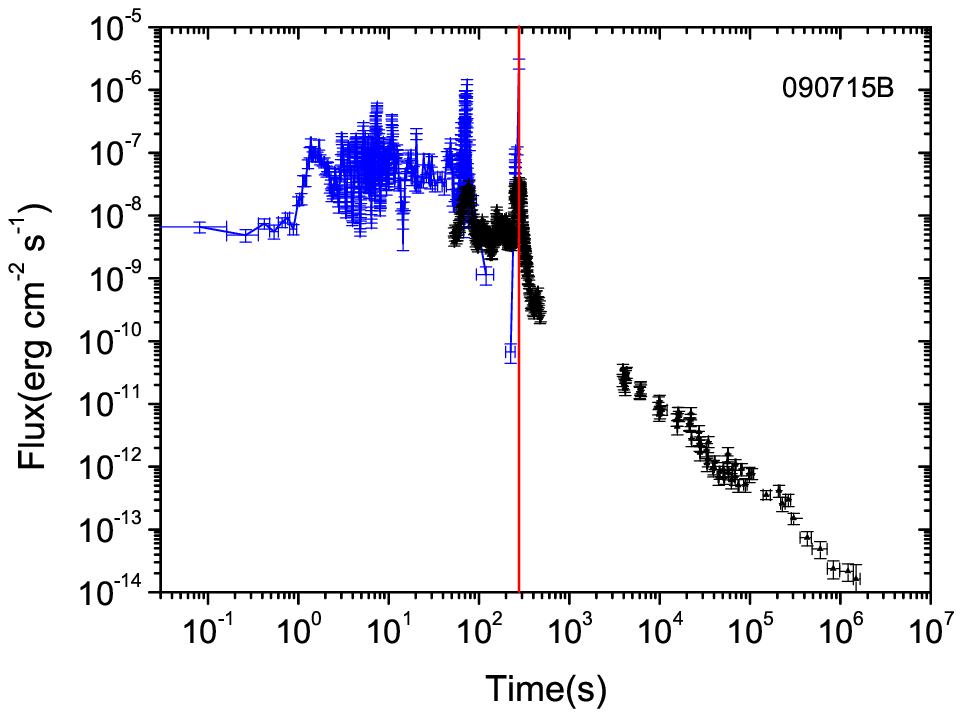}
\includegraphics[angle=0,scale=0.350]{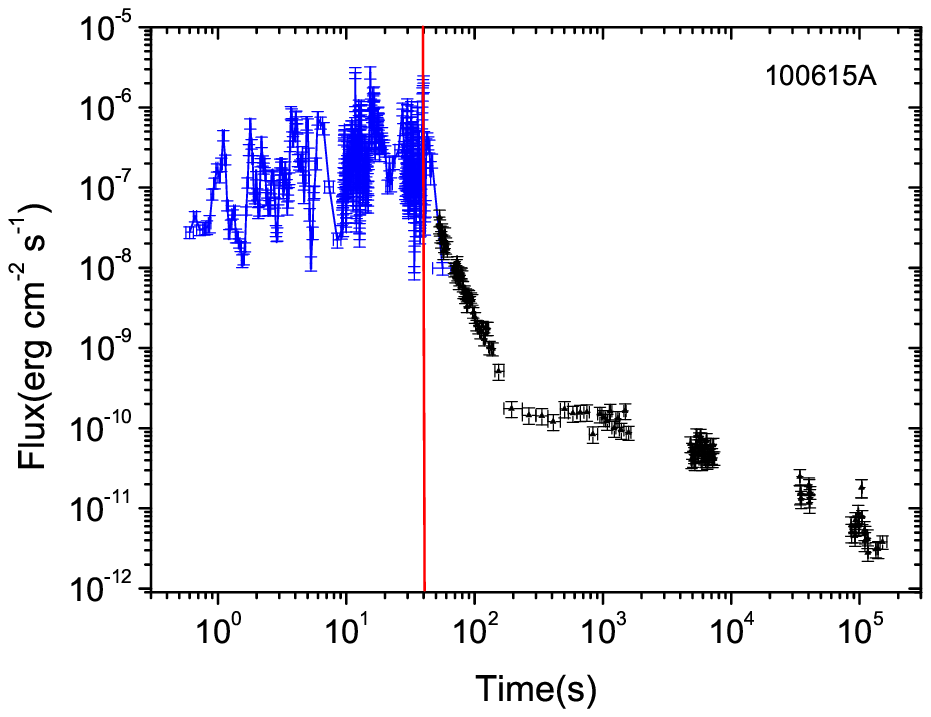}
\includegraphics[angle=0,scale=0.350]{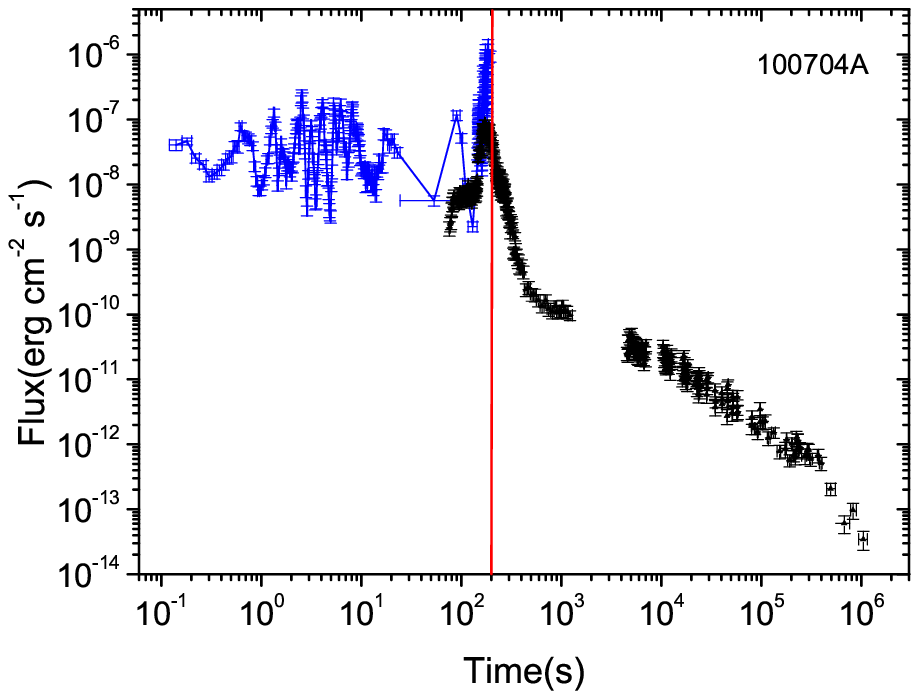}
\includegraphics[angle=0,scale=0.350]{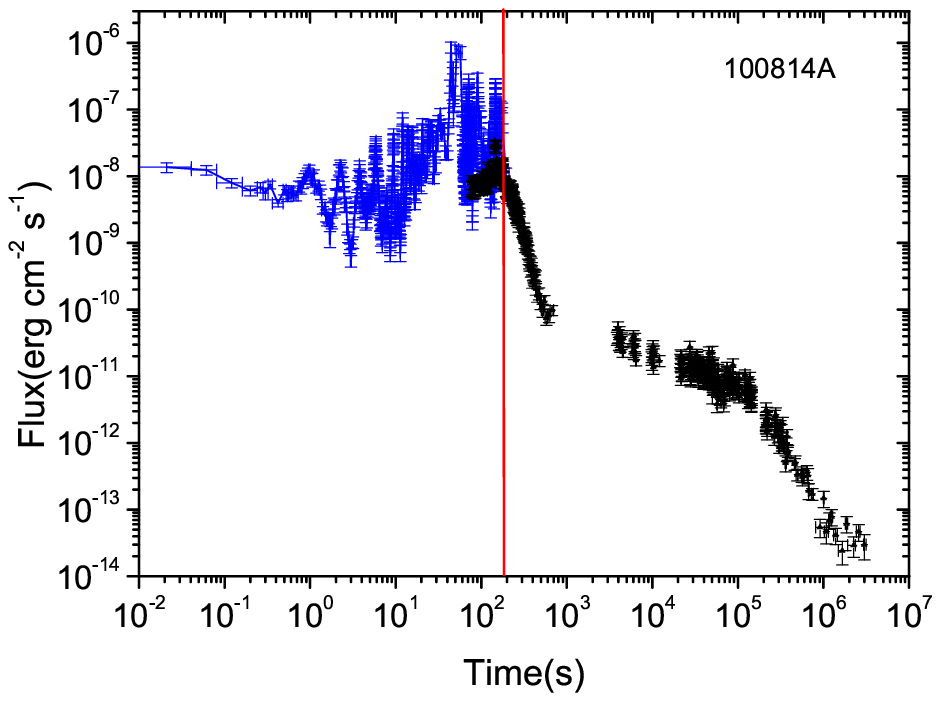}
\includegraphics[angle=0,scale=0.350]{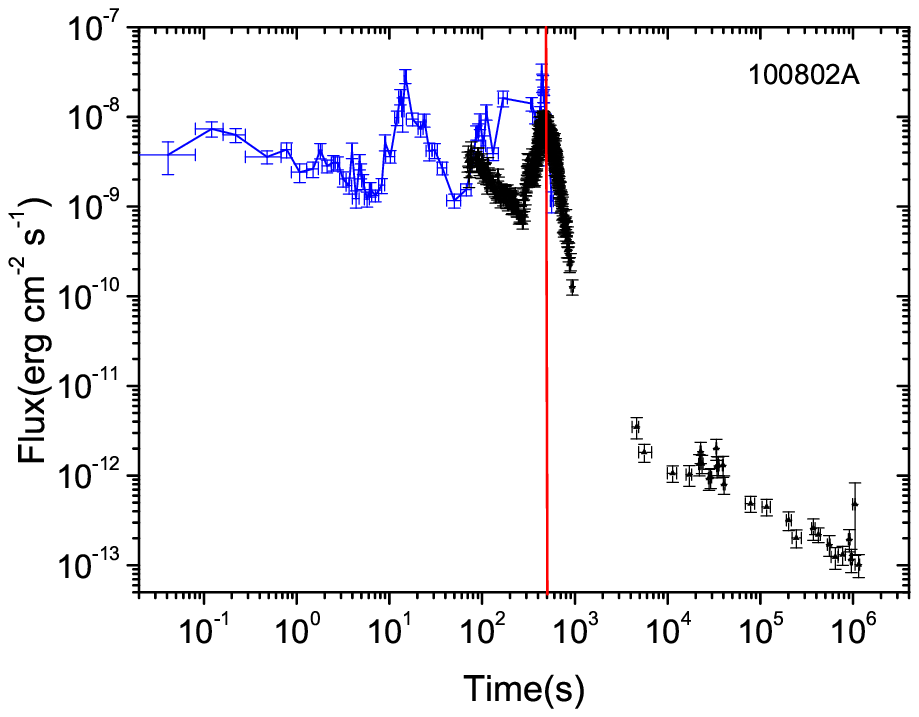}
\hfill
\caption{Examples of X-ray lightcurves from the BAT trigger to late XRT observational epochs for the GRBs that have no significant flares after $T_{90}$, where the prompt X-rays (the connected dots) are derived by extrapolating the BAT data to the XRT band. The vertical lines mark the end time of $T_{90}$.}
\label{BAT_XRT_1}
\end{figure*}

\begin{figure*}
\includegraphics[angle=0,scale=0.350]{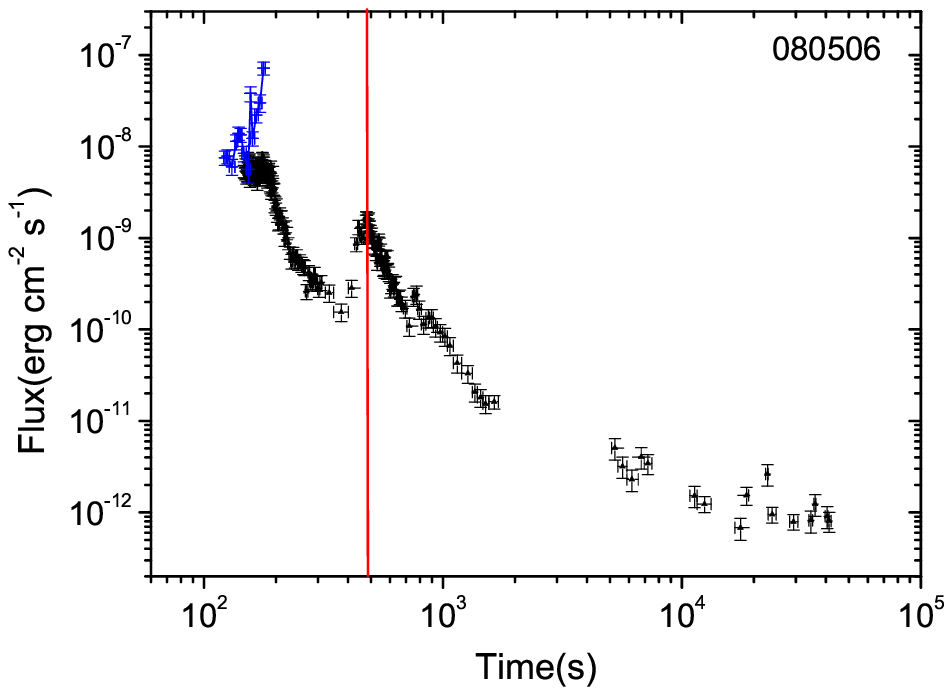}
\includegraphics[angle=0,scale=0.350]{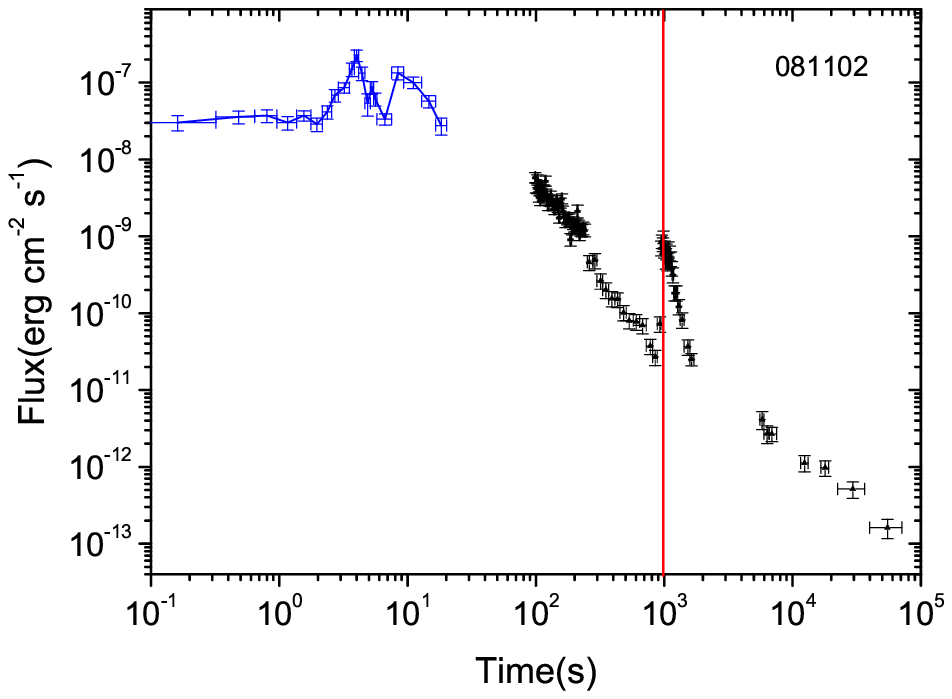}
\includegraphics[angle=0,scale=0.350]{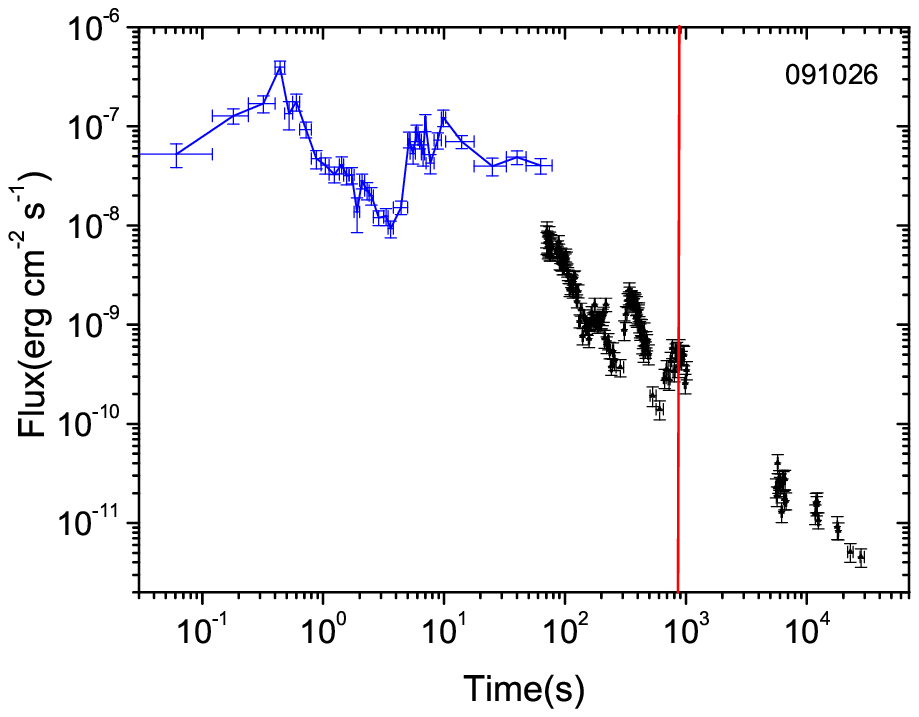}
\includegraphics[angle=0,scale=0.350]{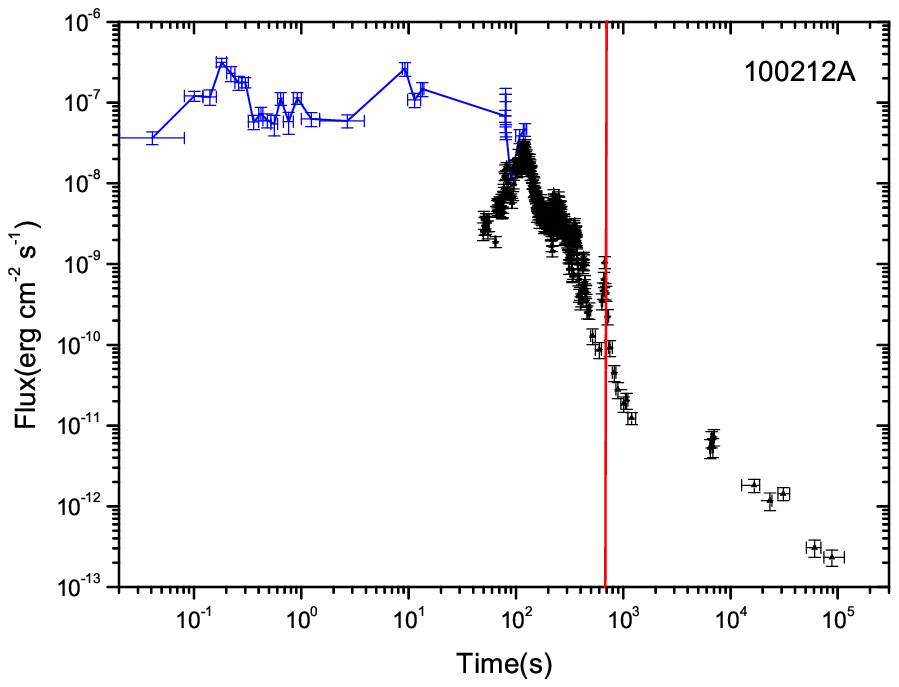}
\includegraphics[angle=0,scale=0.350]{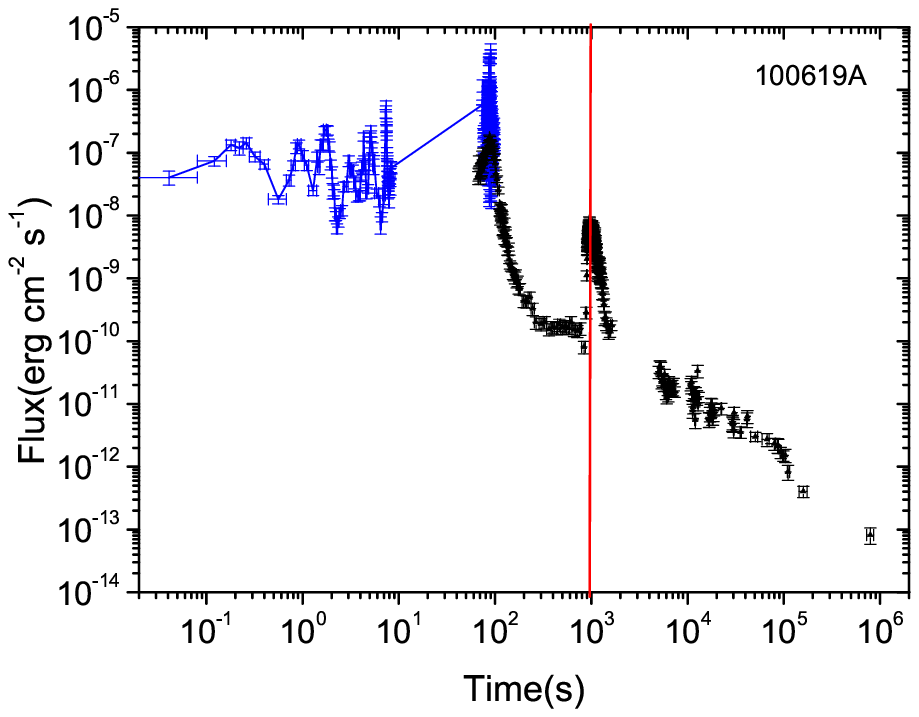}
\includegraphics[angle=0,scale=0.350]{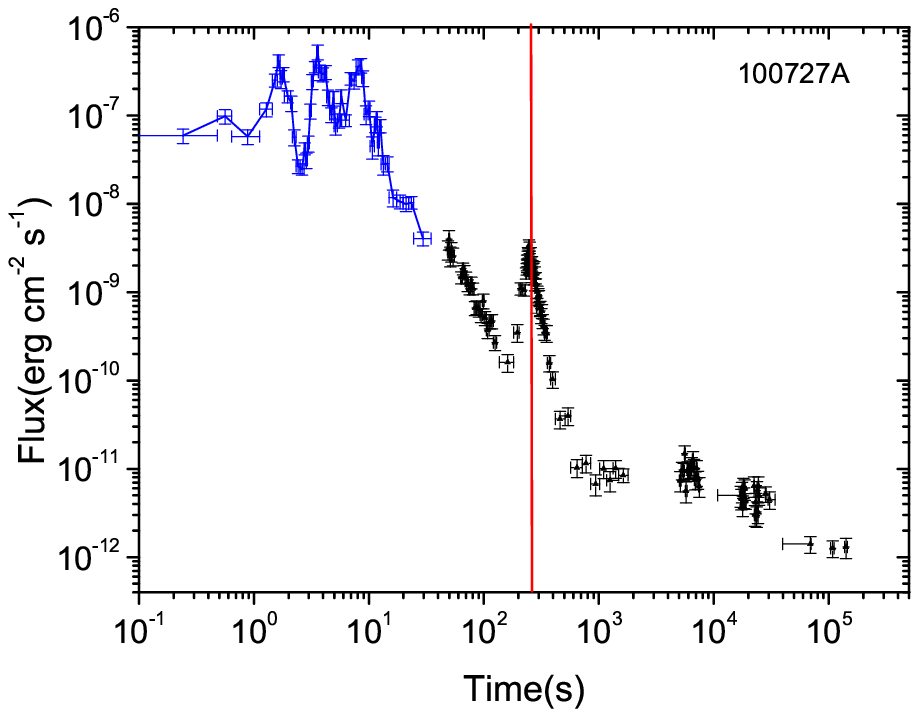}
\includegraphics[angle=0,scale=0.350]{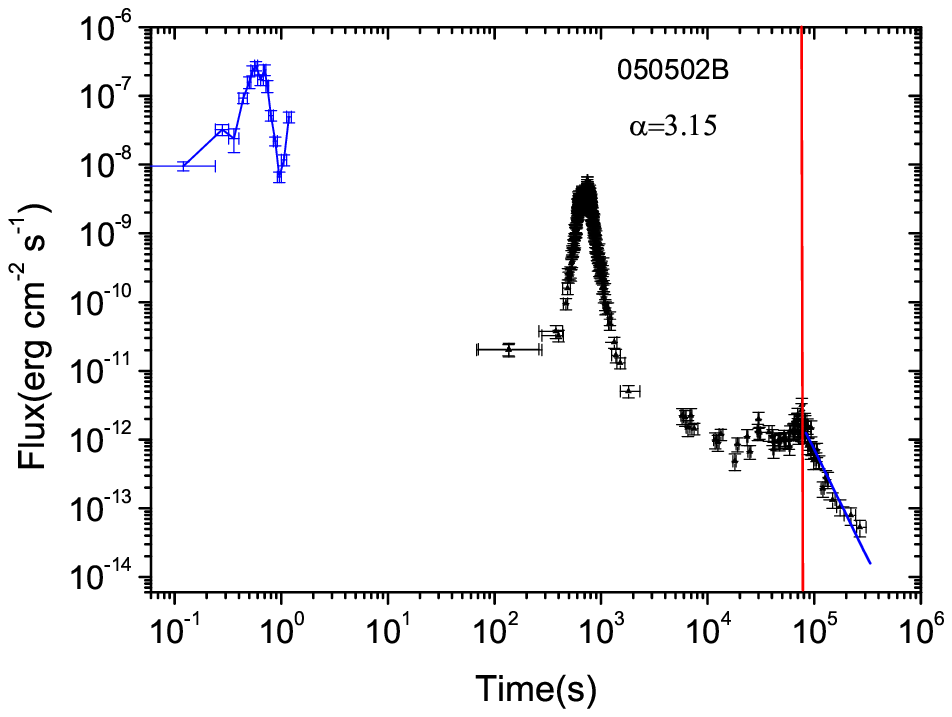}
\includegraphics[angle=0,scale=0.350]{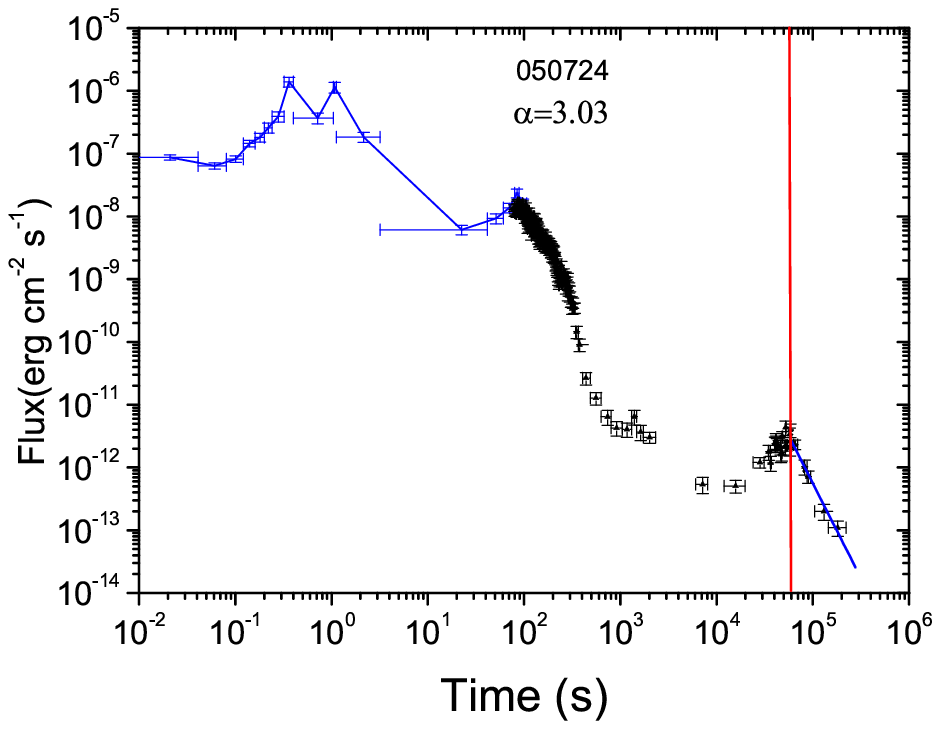}
\includegraphics[angle=0,scale=0.350]{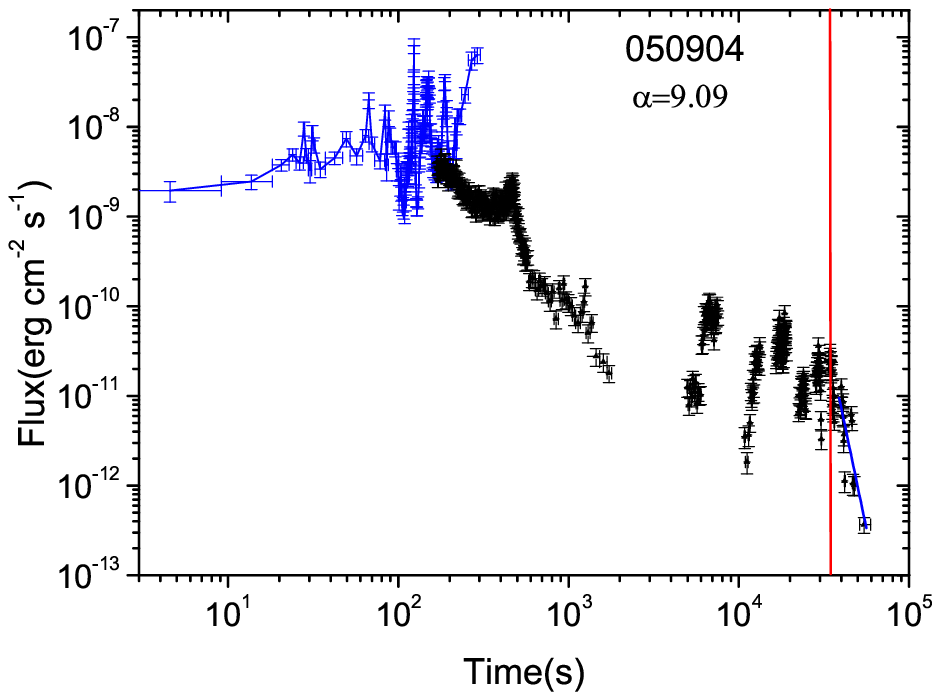}
\includegraphics[angle=0,scale=0.350]{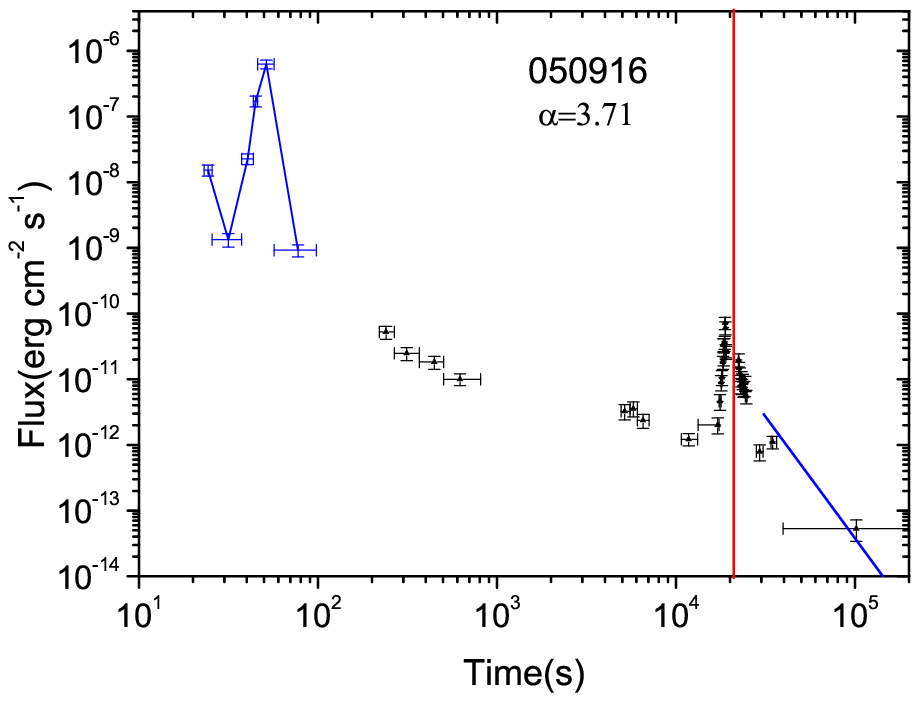}

\hfill
\caption{The same as Figure \ref{BAT_XRT_1}, but for those GRBs whose significant flares are observed after $T_{90}$. The lifetime of the internal energy dissipation process of these GRBs is much longer $T_{90}$.}
\label{BAT_XRT_2}
\end{figure*}

X-ray flares observed with XRT evidence that GRB phenomenon is also bright in the X-ray band. We derive the X-ray flux in the XRT band by extrapolating the BAT data to the XRT band and make a joint X-ray lightcurve from BAT trigger to late epochs as observed with XRT. We take the peak time ($T_{\rm f}$) of the last X-ray flare as an indicator of the end of the internal energy dissipation processes. In order to give a robust estimate of $T_{\rm f}$, we search for sub-samples of GRBs whose BAT and XRT data are well connected without a gap, or whose significant flares are observed after a gap indicating that the internal energy dissipation process is still alive at late times. We have such a sub-sample of 159 GRBs. We find that the $T_{90}$ values of 30\% GRBs are comparable to $T_{\rm f}$, indicating that $T_{90}$ would be a good measure of the duration of the the internal energy dissipation processes in these GRBs. However, the $T_{\rm f}$ values of 70\% GRBs are much larger than $T_{90}$, suggesting that the internal energy dissipation processes extend to much longer than $T_{90}$. This is in general agreement with Zhang et al. (2013), who defined the duration of a burst based on X-ray data. We show some typical x-ray lightcurves of the two type of GRBs in Figure \ref{BAT_XRT_1} and Figure \ref{BAT_XRT_2}, respectively.

A long quiescent phase in the BAT band is observed in some GRBs. The XRT observations during the BAT quiescent phase can reveal the feature of the internal energy dissipation processes in this phase. We show the joint BAT and XRT lightcurves of 8 GRBs that have a long BAT quiescent phase in Figure \ref{LC_quiescence}. Doubled triggered GRB 110709B (Zhang et al. 2012) and GRB 121217A are the most prominent cases. BAT was triggered by GRB 110709B again by the second bright episode after a quiescent phase of 11 minutes. XRT promptly slewed to catch the X-ray emission after the first trigger. One can find that the bright X-ray emission was detected during the quiescent phase, indicating that the burst events are not indeed quiescent. For GRB 121217A, a rapid decay segment of the XRT lightcurve before $T_{0}+100$ seconds should be the tail emission of the first gamma-ray peak being due to the curvature effect (e.g., Zhang et al. 2007). The X-rays since $T_{0}+200$ seconds increase with significant flares and peak at $T_{0}+750$ seconds. The second bright, short gamma-ray peak is just at the peak of the X-ray emission. The duration of the second peak of the gamma-ray emission is only a very short slice of the burst event in the time interval from $T_{0}-200$ s to $T_{0}+1300$ s.

\section{Conclusions and Discussion}
We have presented a joint analysis of 613 gamma-ray burst (GRB) data observed with {\em Swift} mission to investigate the internal energy dissipation processes as observed in the BAT and XRT band, including precursors, prompt gamma-ray emission, extended soft gamma-ray emission, and late X-ray flares. Based on our analysis with the BB method, we show that the BAT lightcurves of 73\% of GRBs in our sample are composed of one single emission episode only. A precursor is observed in about 10\% of the BAT GRBs, most of which are of long-duration GRBs. About half of the precursors triggered BAT to alert the GRB, but we do not find any statistical differences of the photon indices between the triggered and non-triggered precursors, and their photon indices are also roughly consistent with those of the main burst. The EE tails in the BAT band are usually observed in short duration GRBs. They are statistically softer than the emission in the main burst. We derive the X-ray flux in the XRT band by extrapolating the BAT data to the XRT band and make joint X-ray lightcurves from the BAT trigger to late observational epochs by XRT. Take the peak time of the last X-ray flare as an indicator of the end of the internal energy dissipation processes, we show that the duration of internal energy dissipation process of about $2/3$ GRBs are much longer than $T_{90}$ measured in the BAT band. Time intervals between bright episodes in the band range from tens of seconds to hundreds of seconds, and a long quiescent phase is shown for some GRBs. However, bright X-ray emission is detected during the BAT quiescent phase. These results suggest that what we have seen in the gamma-ray band may be only the tip of iceberg of more extended underlying activities. The duration of the gamma-ray emission may be only a short period of the life time of the entire internal energy dissipation processes.

The physical origin of the precursor is of great interest. Previously some authors suggested an early photosphere emission origin or shock break out origin (Lyutikov \& Usov 2000; Ramirez-Ruiz et al. 2002). Our analysis shows no significant difference between the precursor emission and the main emission episode, suggesting that very likely precursors share the same physical origin as the prompt episode, which is directly related to the central engine activities.

\begin{figure*}
\centering
\includegraphics[angle=0,scale=.25]{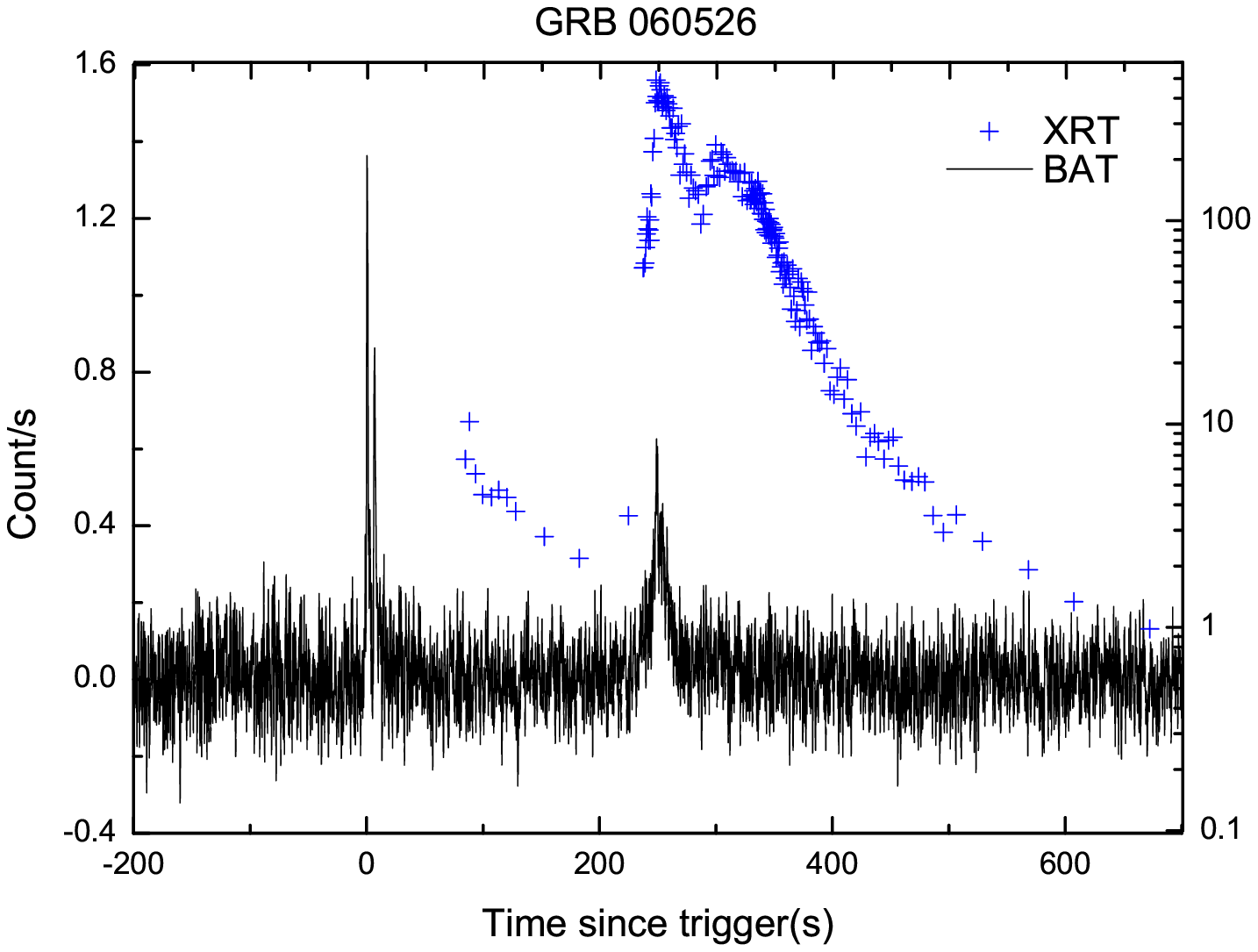}
\includegraphics[angle=0,scale=.25]{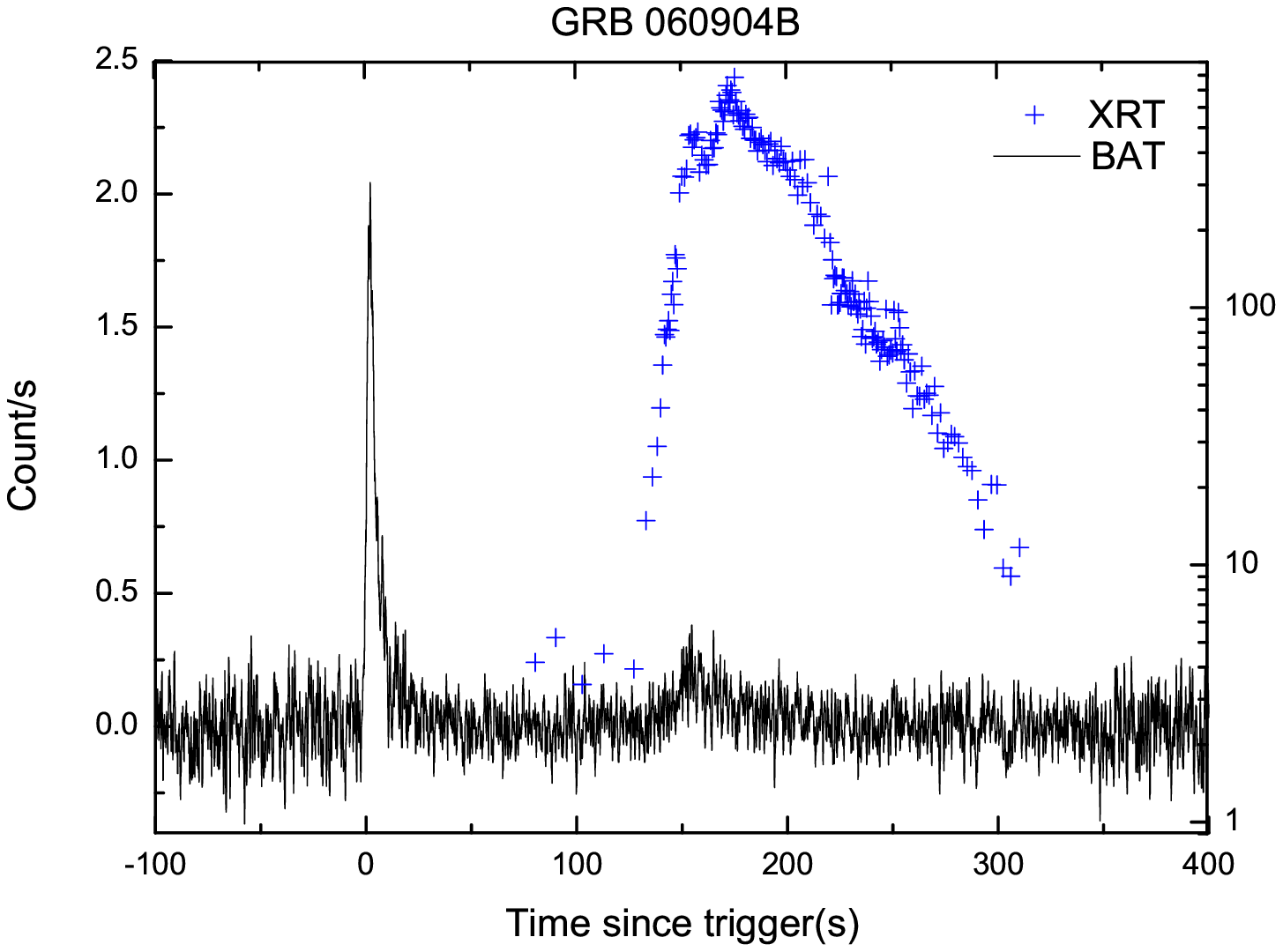}
\includegraphics[angle=0,scale=.25]{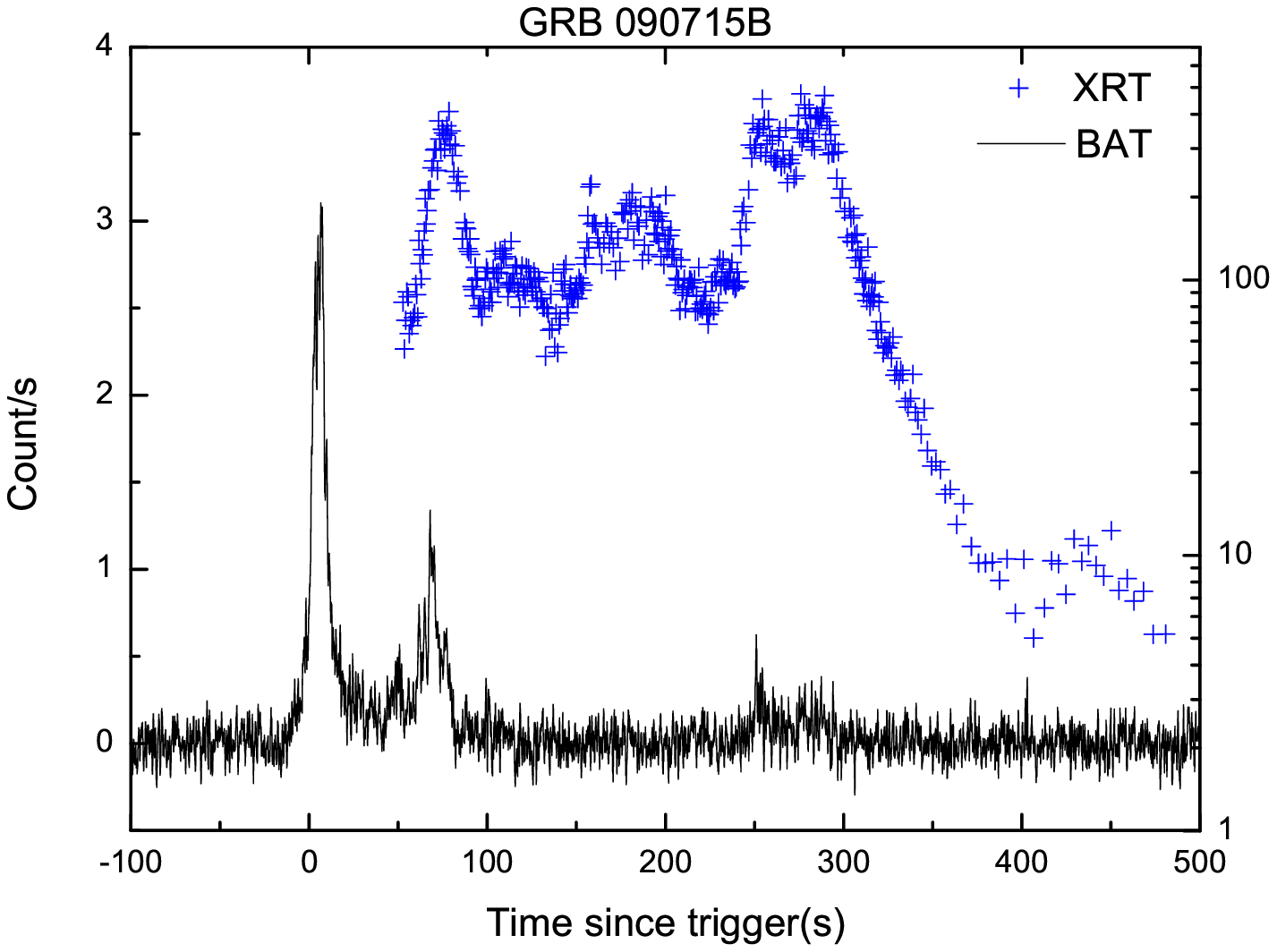}
\includegraphics[angle=0,scale=.25]{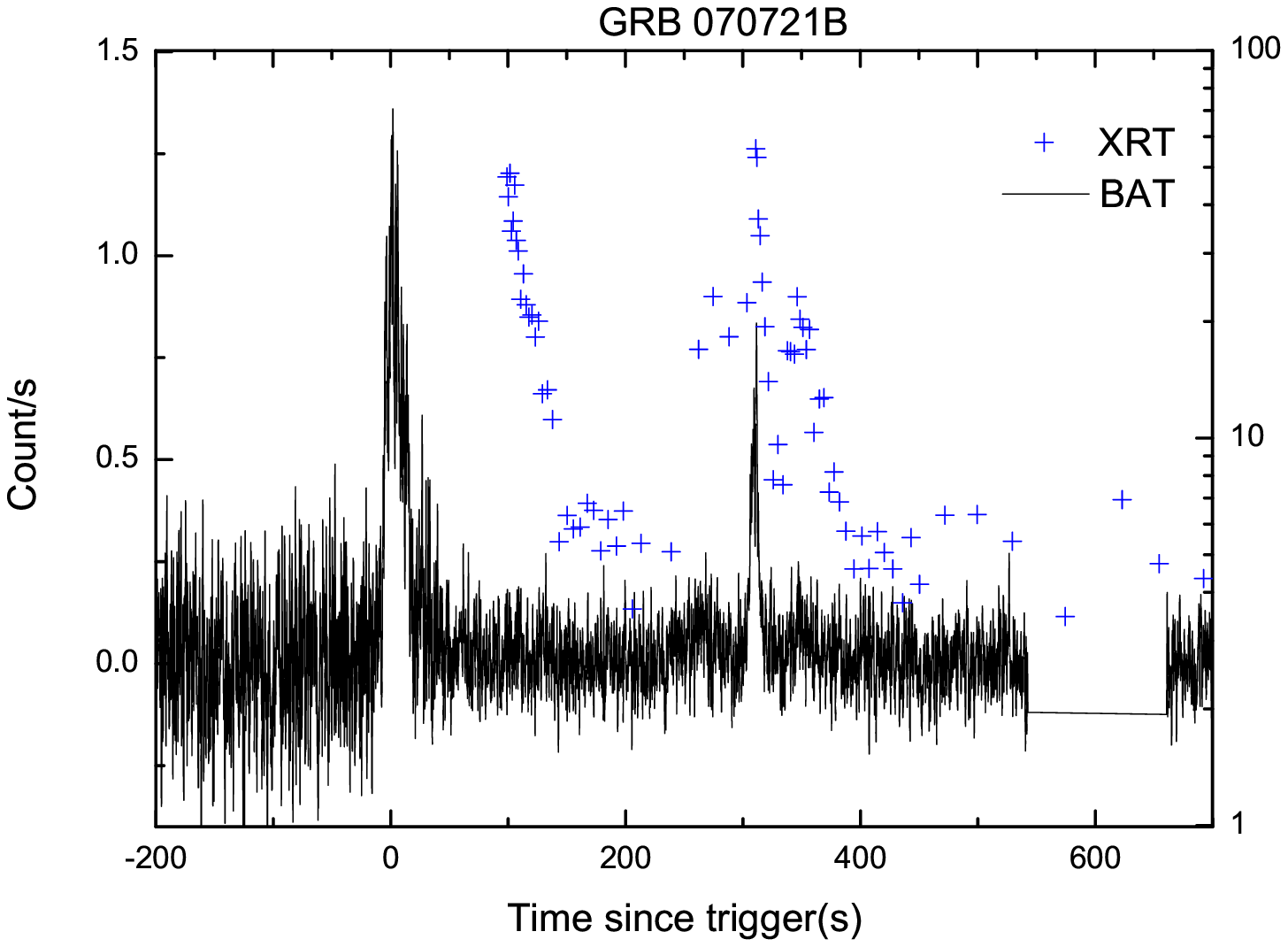}
\includegraphics[angle=0,scale=.25]{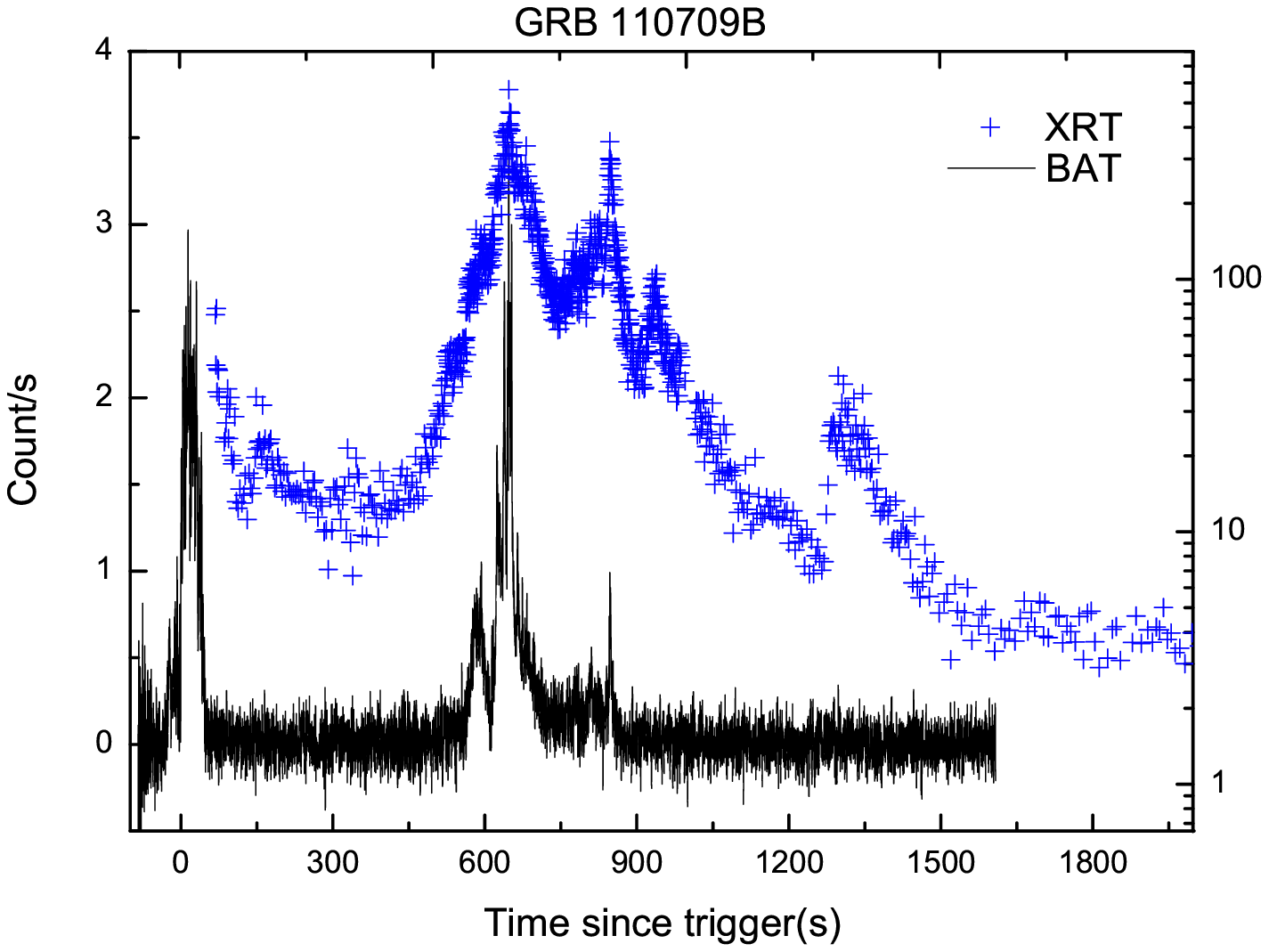}
\includegraphics[angle=0,scale=.25]{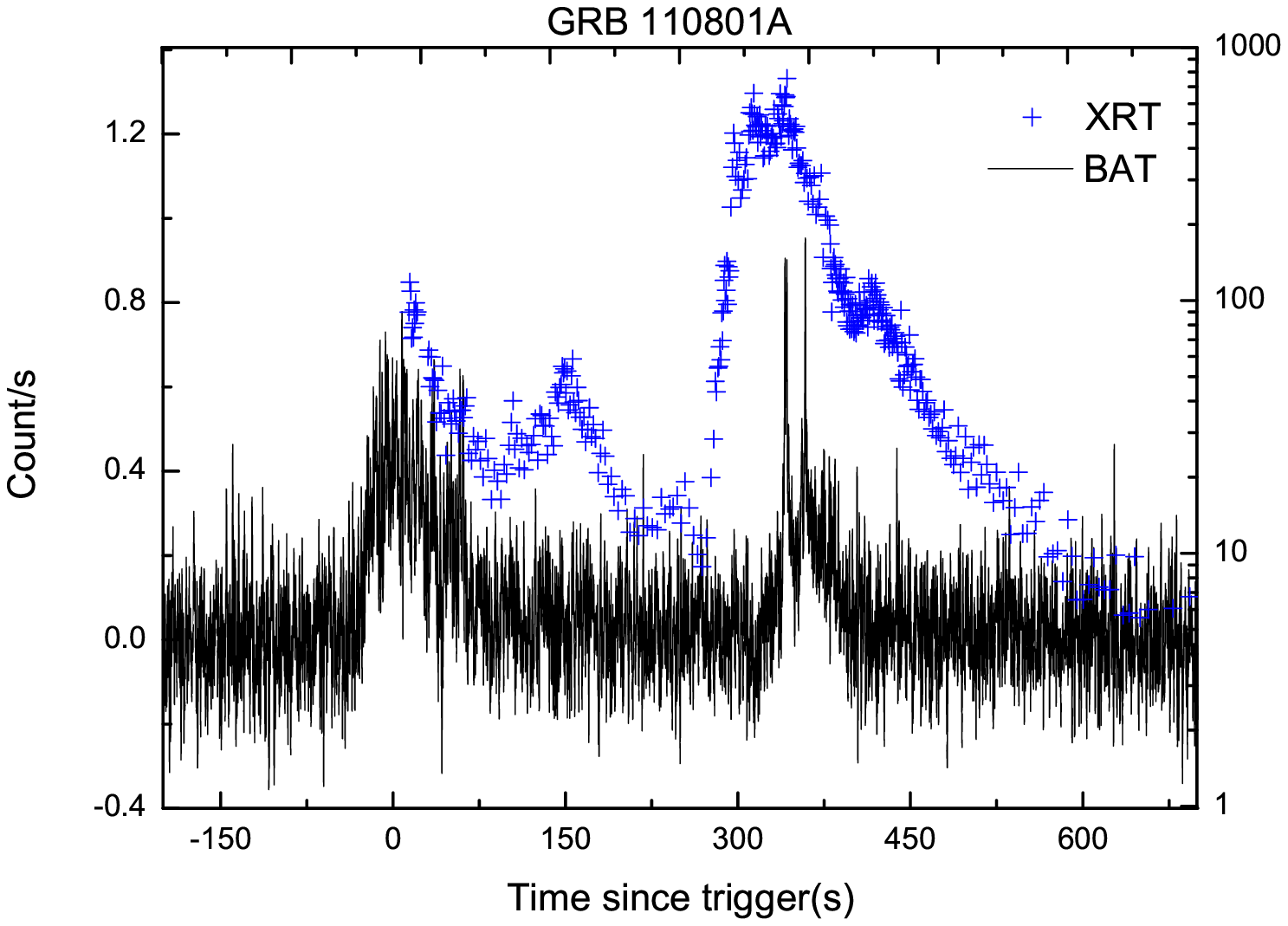}
\includegraphics[angle=0,scale=.25]{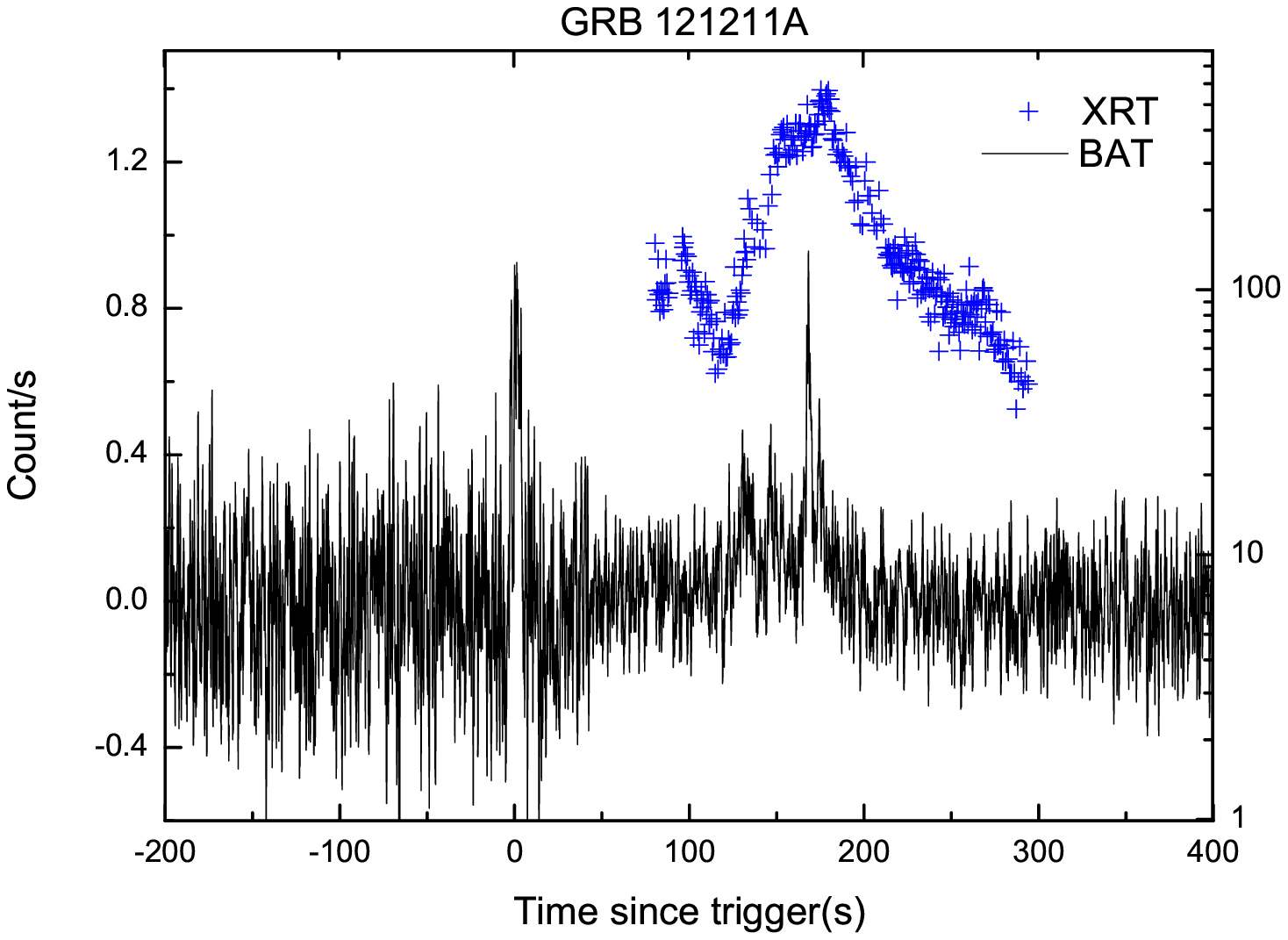}
\includegraphics[angle=0,scale=.25]{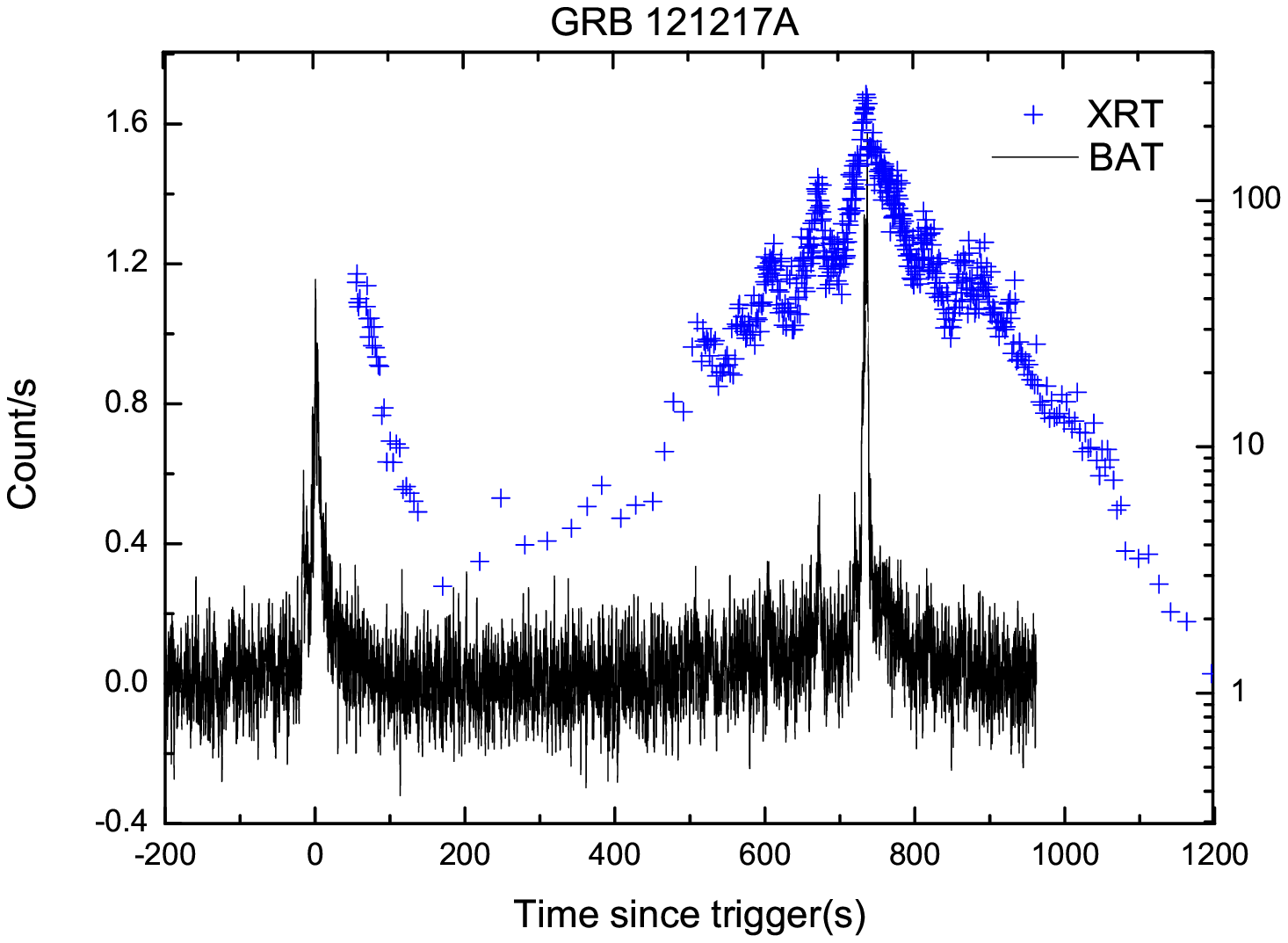}
\hfill
\caption{Examples of Joint BAT (connected lines) and XRT (crosses) lightcurves for GRBs with a long quiescence phase in the BAT band. The left and right vertical axes of each panel are for the count rate for the BAT data and XRT data, respectively.}
\label{LC_quiescence}
\end{figure*}

One interesting question in GRB studies is whether there is emission prior to the GRB trigger time. The detection of precursors prior to the main burst indicate that the burst activities can in principle start prior to the main explosion. A GRB trigger is instrumentally dependent. The trigger probability of low S/N signals drops rapidly (e.g., Qin et al. 2013). An off-line analysis of CGRO/BATSE data showed that about 1838 out of 3906 GRBs were detected but did not trigger BATSE (Stern et al. 2001). As shown in our analysis, only about $44\%$ of precursors triggered BAT. In addition, our joint BAT and XRT lightcurves shown in Figure \ref{LC_quiescence} indicate that bright X-ray emission may be detectable during the BAT quiescent phase. Currently, the XRT observations usually start at least tens of seconds post the GRB trigger. We do not have X-ray observations prior to the BAT triggers. On the other hand, based on XRT observations to the second peaks of GRBs 110709B and 121217A, one may suspect that bright X-ray emission may come even prior to the first episode that triggers BAT. The time interval of $T_{90}$ would be only a short, violent episode of the bursts. Therefore, the exact starting time of a burst event could be (much) prior to the BAT trigger. This effect may also significantly affect our understanding of the temporal behaviors of the early shallow-decay X-ray emission (e.g. Yamazaki 2009; Liang et al. 2009, but see Birnbaum et al. 2012).

It is interesting that precursors preferably happen in type II GRBs, which are produced by collapses of massive stars. As discussed above, the precursors, main burst, and late flares would all come from the internal energy dissipation of a relativistic outflow powered by the GRB central engine. Popular GRB central engine models are related to a hyper-accreting black hole (e.g., Popham et al. 1999; Narayan et al. 2001; Di Matteo et al. 2002; Kohri \& Mineshige 2002; Gu et al. 2006; Chen \& Beloborodov 2007; Liu et al. 2007; Lei et al. 2009; Lei et al. 2013). The detection of precursors in Type II GRBs likely imply that the GRB central engine may have some weak activities prior to the main explosion during the collapse of a massive star. The lack of detection of a precursor in Type I GRBs, together with the fact that the lightcurves of type I GRBs usually starts with extremely sharp and bright spikes, seem to suggest that compact star mergers may start with most violent activities of the central engine.

The EE tails of about $1/3$ GRBs in Table 2 are composed of some well separated pulses, such as GRBs 100906A, 060814, 100814, 111008A, 060306, 100522, 100704A, 080509B, 120514A. They seem to be part of the main bursts. However, the EE tails of about $2/3$ GRBs are not clearly separated from the main bursts in their lightcurves. These GRBs are GRB 060614-like or GRB 061006-like. They may be produced by mergers of compact stars (e.g., Zhang et al. 2007).  The detections of the EE tails and late X-ray flares indicate that the GRB central engine of Type I GRBs does not die rapidly. Different from the precursor emission, the EE tails are statistically softer than the main burst. Metzger et al. (2008) presented time-dependent models of the remnant accretion disks created during compact object mergers and showed that the late-time accretion can in principle provide sufficient energy to power the late time activities observed by {\em Swift}/BAT from some Type I GRBs. Liu et al. (2012) showed that the radial angular momentum transfer may significantly prolong the lifetime of the accretion process and multiple episodes may be switched by the magnetic barrier. Their numerical calculations based on the model of the neutrino-dominated accretion flows suggest that the disk mass is critical for producing the observed EE tails.

We appreciate helpful comments from the referee. This work made use of data supplied by the UK Swift Science Data Centre at the University of Leicester. It is supported by the National Basic Research Program (973 Programme) of China (Grant 2014CB845800), the National Natural Science Foundation of China (Grants 11025313, 11163001, 11363002), Guangxi Science Foundation (2013GXNSFFA019001), Key Laboratory for the Structure and Evolution of Celestial Objects of Chinese Academy of Sciences, and the Strategic Priority Research Program "The Emergence of Cosmological Structures" of the Chinese Academy of Sciences, Grant No. XDB09000000.


\begin{deluxetable}{cccccccccccc}
\tabletypesize{\tiny}
\tablecaption{Our analysis results for triggered and non-triggered precursors.}
\tablewidth{0pt}
\tablehead{
\colhead{GRB}&
\colhead{$T_{90}(s)$}&
\colhead{BG(s)\tablenotemark{a}}&
\colhead{Precursor(s)\tablenotemark{b}}&
\colhead{$C_{\rm p}(c/s)$\tablenotemark{c}}&
\colhead{Main burst(s)\tablenotemark{d}}&
\colhead{$C_{\rm m}(c/s)$\tablenotemark{e}}&
\colhead{$R_{\rm pm}$\tablenotemark{f}}&
\colhead{$\Gamma_{\rm p}$\tablenotemark{g}}&
\colhead{$\Gamma_{\rm m}$\tablenotemark{h}}
}
\startdata
Triggered&&&&&&&&&\\
\hline
050820A&26&-300$\sim$-200&-33.79$\sim$24.002&0.335&213.89$\sim$238.658&1.343&0.249&1.722$\pm$0.079&1.118$\pm$0.054\\
060322&221.5&400$\sim$600&-55.285$\sim$35.531&0.274&151.115$\sim$200.651&0.562&0.488&1.379$\pm$0.065&1.712$\pm$0.040\\
060904A&80.1&700$\sim$900&-29.646$\sim$36.402&0.41&36.402$\sim$94.194&1.324&0.309&1.647$\pm$0.048&1.577$\pm$0.023\\
061007&75.3&700$\sim$900&-5.845$\sim$10.667&2.981&18.923$\sim$93.227&7.231&0.412&1.061$\pm$0.033&0.998$\pm$0.013\\
061121&81.3&700$\sim$900&-4.103$\sim$4.217&0.804&58.297$\sim$99.897&5.646&0.142&1.464$\pm$0.058&1.431$\pm$0.015\\
061202&91.2&700$\sim$900&-5.845$\sim$10.667&0.108&43.691$\sim$151.019&0.823&0.131&2.357$\pm$0.329&1.523$\pm$0.038\\
061222A&71.4&400$\sim$500&-22.817$\sim$10.207&0.141&18.463$\sim$117.535&2.029&0.069&1.458$\pm$0.158&1.344$\pm$	 0.022\\
071021&225&500$\sim$700&-38.563$\sim$60.509&0.053&60.509$\sim$101.789&0.183&0.289&2.051$\pm$0.245&1.585$\pm$0.112\\
080229A&64&700$\sim$900&-12.392$\sim$20.632&0.596&20.632$\sim$70.168&2.757&0.216&1.484$\pm$0.085&1.952$\pm$0.029\\
080928&280&-250$\sim$-150&-13.795$\sim$111.013&0.093&193.573$\sim$259.621&0.484&0.192&1.781$\pm$0.182&1.832$\pm$0.064\\
081210&146&500$\sim$700&-14.569$\sim$14.807&0.319&14.807$\sim$21.335&1.101&0.289&1.462$\pm$0.081&1.376$\pm$0.062\\
081221&34&500$\sim$700&-5.74$\sim$10.771&1.464&10.771$\sim$35.54&6.596&0.222&1.775$\pm$0.042&1.816$\pm$0.015\\
090401A&112&500$\sim$700&-19.157$\sim$88.171&0.176&88.171$\sim$162.476&2.054&0.086&1.603$\pm$0.106&1.718$\pm$0.026\\
090621A&n/a&-200$\sim$-100&-46.819$\sim$43.997&0.12&217.373$\sim$233.885&0.677&0.177&1.629$\pm$0.108&1.680$\pm$0.059\\
090618&113.2&500$\sim$700&-5.071$\sim$44.465&2.673&44.465$\sim$120.059&16.271&0.164&1.434$\pm$0.016&1.628$\pm$0.011\\
090904A&122&500$\sim$700&27.485$\sim$77.021&0.131&118.301$\sim$184.349&0.631&0.207&2.806$\pm$0.223&1.860$\pm$0.049\\
100902A&428.8&700$\sim$900&-53.914$\sim$36.902&0.074&127.718$\sim$243.302&0.21&0.352&1.822$\pm$0.202&2.023$\pm$0.069\\
101023A&80.8&500$\sim$700&-14.488$\sim$18.536&1.235&35.048$\sim$92.84&4.926&0.251&1.634$\pm$0.072&1.580$\pm$0.016\\
110102A&264&500$\sim$700&-38.563$\sim$35.741&0.182&93.533$\sim$275.165&0.931&0.195&1.609$\pm$0.104&1.548$\pm$0.021\\
111228A&101.2&500$\sim$700&-19.221$\sim$30.315&0.285&30.315$\sim$112.875&2.018&0.141&2.490$\pm$0.178&2.266$\pm$0.035\\
120102A&38.7&500$\sim$700&-6.297$\sim$10.215&0.214&18.471$\sim$51.495&2.033&0.105&2.034$\pm$0.199&1.619$\pm$0.031\\
120327A&62.9&500$\sim$700&-22.728$\sim$26.808&0.283&26.808$\sim$76.344&1.12&0.253&1.548$\pm$0.071&1.627$\pm$0.044\\
\hline
Non&-triggered&&&&&&&&\\
\hline
041224&177.2&200$\sim$300&-115.568$\sim$-24.752&0.196&-24.752$\sim$57.808&0.943&0.208& 1.657$\pm$0.102&1.693$\pm$0.034\\
050315&95.6&-300$\sim$-200&-66.967$\sim$-9.175&0.105&-9.175$\sim$65.129&0.613&0.171&1.714$\pm$0.205&2.169$\pm$0.050\\
050319&152.5&200$\sim$300&-143.973$\sim$-124.389&0.262&-0.357$\sim$10.523&0.518& 0.506&1.934$\pm$0.174&2.037$\pm$0.098\\
050713A&124.7&-300$\sim$-200&-74.28$\sim$-49.512&0.123&-8.232$\sim$74.328&1.565&0.079&1.398$\pm$0.421&1.488$\pm$0.036\\
060115&139.6&200$\sim$300&-66.612$\sim$-17.076&0.052&-17.076$\sim$115.02&0.291&0.179& 1.795$\pm$0.308&1.773$\pm$0.075\\
060204B&139.4&200$\sim$300&-189.227$\sim$-164.459&0.063&-40.619$\sim$17.173&0.387&0.163&1.682$\pm$0.397&1.397$\pm$0.044\\
060418&103.1&500$\sim$700&-88.446$\sim$-55.422&0.122&-22.398$\sim$51.906&1.396&0.087& 2.657$\pm$0.453&1.666$\pm$0.029\\
060707&66.2&500$\sim$700&-53.194$\sim$-32.394&0.186&-7.434$\sim$30.006&0.39&0.477& 1.416$\pm$0.213&1.679$\pm$0.086\\
070306&209.5&500$\sim$700&-19.753$\sim$21.527&0.223&79.319$\sim$153.623&0.979&0.228& 1.597$\pm$0.142&1.663$\pm$0.033\\
070628&39.1&-200$\sim$-100&-39.466$\sim$-14.698&0.168&-6.442$\sim$18.326&1.835& 0.092&	2.651$\pm$0.344&1.901$\pm$0.045\\
070911&162&-300$\sim$-200&-73.796$\sim$-15.295&0.202&-15.295$\sim$149.825&0.854&0.236& 1.719$\pm$0.077&1.760$\pm$0.023\\
071010B&$>$35.7&-250$\sim$-150&-47.093$\sim$-5.813&0.13&-5.813$\sim$27.211&2.535& 0.051&1.844$\pm$0.134&2.051$\pm$0.026\\
080710&120&-250$\sim$-150&-99.921$\sim$-29.201&0.099&-4.241$\sim$24.879&0.194&0.510&0.922$\pm$0.216&1.954$\pm$0.176\\
080714&33&500$\sim$700&-103.587$\sim$-62.307&0.052&-12.771$\sim$45.021&0.858&0.0606& 1.063$\pm$0.720&1.509$\pm$0.054\\
080725&120&-250$\sim$-150&-113.472$\sim$-96.96&0.104&-22.656$\sim$51.648&0.711&0.146&1.803$\pm$0.477&1.482$\pm$0.049\\
080906&147&500$\sim$700&-88.526$\sim$-14.222&0.065&-14.222$\sim$93.106&0.452&0.144&1.600$\pm$0.246&1.553$\pm$0.047\\
081109A&190&500$\sim$700&-88.962$\sim$-22.914&0.042&-22.914$\sim$84.414&0.546&0.077&1.418$\pm$0.319&1.674$\pm$0.041\\
081128&100&500$\sim$700&-63.331$\sim$-5.539&0.096&-5.539$\sim$77.021&0.351&0.273&1.831$\pm$0.157&2.041$\pm$0.063\\
090123&131&500$\sim$700&-52.512$\sim$-15.072&0.19&-15.072$\sim$63.968&0.427&0.445&1.540$\pm$0.162&1.685$\pm$0.089\\
090410&165&300$\sim$400&-54.164$\sim$-4.628&0.138&-4.628$\sim$135.724&0.87&0.158&1.349$\pm$0.117&1.159$\pm$0.032\\
091221&68.5&600$\sim$700&-47.488$\sim$-6.208&0.147&-6.208$\sim$43.328&1.214&0.121& 2.517$\pm$0.276&1.600$\pm$0.029\\
100727A&84&500$\sim$700&-88.913$\sim$-14.609&0.05&-6.353$\sim$26.671&0.307&0.162&1.783$\pm$0.251&1.898$\pm$0.074\\
110213A&48&500$\sim$700&-41.137$\sim$-7.857&0.595&-7.857$\sim$25.423&1.975&0.301& 1.758$\pm$0.257&1.947$\pm$0.070\\
110305A&12&500$\sim$700&-103.49$\sim$-86.978&0.116&-12.674$\sim$12.094&0.319&0.364&3.167$\pm$1.529&1.781$\pm$0.162\\
110315A&77&400$\sim$600&-80.682$\sim$-35.598&0.133&-35.598$\sim$20.978&0.683&0.195&1.311$\pm$0.196&1.893$\pm$0.049\\
120116A&41&500$\sim$700&-31.154$\sim$-6.385&0.206&-6.385$\sim$26.639&1.397&0.147&2.830$\pm$0.257&2.853$\pm$0.046\\
120324A&118&500$\sim$700&-71.95$\sim$-12.545&0.106&-12.545$\sim$86.527&1.378&0.077&1.486$\pm$0.160&1.284$\pm$0.022\\
120326A&69.6&300$\sim$400&-71.901$\sim$-5.853&0.085&-5.853$\sim$18.915&1.401&0.061&2.039$\pm$0.169&2.074$\pm$0.037\\

\enddata
\tablenotetext{a}{Time interval for the background calculation.}
\tablenotetext{b}{Time interval of the precursor.}
\tablenotetext{c}{Average count rate of the precursor estimated with our BB analysis.}
\tablenotetext{d}{Time interval of the main burst.}
\tablenotetext{e}{Average count rate of the main burst estimated with our BB analysis.}
\tablenotetext{f}{The ratio of $C_{\rm p}/C_{\rm m}$.}
\tablenotetext{g}{$\gamma$-ray photo index of the precursor.}
\tablenotetext{h}{Photon index of the main burst.}
\end{deluxetable}

\begin{deluxetable}{cccccccccccc}
\tabletypesize{\tiny}
\tablecaption{The BAT Observations and our fitting results of the GRBs with a extended emission.}
\tablewidth{0pt}
\tablehead{
\colhead{GRB}&
\colhead{$T_{90}(s)$}&
\colhead{BG(s)\tablenotemark{a}}&
\colhead{Main burst(s)\tablenotemark{b}}&
\colhead{$C_{\rm m}(c/s)$\tablenotemark{c}}&
\colhead{EE tail(s)\tablenotemark{d}}&
\colhead{$C_{\rm e}(c/s)$\tablenotemark{e}}&
\colhead{$R_{\rm me}$\tablenotemark{f}}&
\colhead{$\Gamma_{\rm m}$\tablenotemark{g}}&
\colhead{$\Gamma_{\rm e}$\tablenotemark{h}}
}
\startdata
050717&85&-300$\sim$-200&-9.062$\sim$32.218&1.047&32.218$\sim$114.778&0.26&4.027 &1.221 $\pm$0.029 &1.656 $\pm$0.069 \\
050724&96&-200$\sim$-100&-1.547$\sim$2.805&1.024&2.805$\sim$108.341&0.089&11.506 &1.706$\pm$0.113 &2.020 $\pm$0.163 \\
050911&16.2&-500$\sim$-200&-2.329$\sim$19.431&0.2&28.135$\sim$156.519&0.076&2.632 &1.858$\pm$0.184 &2.142 $\pm$0.642 \\
051227&114.6&-300$\sim$-200&-1.53$\sim$3.91&0.343&3.91$\sim$145.35&0.077&4.455 &1.202$\pm$0.133 &1.534 $\pm$0.192 \\
060306&61.2&-300$\sim$-200&-6.89$\sim$1.43&1.228&22.23$\sim$63.83&0.249&4.932 &1.617 $\pm$0.070 &1.874 $\pm$0.091 \\
060428A&39.5&500$\sim$700&-1.173$\sim$18.411&0.873&24.939$\sim$46.699&0.252&3.464 &2.057 $\pm$0.068 &2.036 $\pm$0.124 \\
060507&183.3&500$\sim$700&-12.257$\sim$30.175&0.439&30.175$\sim$199.903&0.2&2.195 &1.558$\pm$0.063 &1.966 $\pm$0.074 \\
060607A&102.2&500$\sim$700&-22.648$\sim$26.888&0.388&26.888$\sim$117.704&0.069&5.623 &1.450 $\pm$0.042 &1.626 $\pm$0.120 \\
060614&108.7&-250$\sim$-150&-2.839$\sim$3.689&5.254&17.833$\sim$91.817&3.213&1.635 &1.631 $\pm$0.041 &2.115 $\pm$0.023 \\
060814&145.3&700$\sim$900&-14.214$\sim$84.858&1.823&84.858$\sim$159.162&0.289&6.308 &1.499 $\pm$0.019 &1.720 $\pm$0.045 \\
061006&129.9&600$\sim$800&-25.968$\sim$-23.791&2.224&-23.791$\sim$36.049&0.158&14.076 &2.471$\pm$2.600 &1.975 $\pm$0.131 \\
061210&85.3&600$\sim$800&-0.901$\sim$1.099&1.2&1.099$\sim$105.723&0.124&9.677 &0.824$\pm$0.130 &1.550 $\pm$0.206 \\
070223&88.5&500$\sim$700&-6.007$\sim$43.529&0.245&43.529$\sim$101.321&0.104&2.356 &1.725 $\pm$0.080 &2.213 $\pm$0.160 \\
070420&76.5&-250$\sim$-150&-6.08$\sim$19.264&2.822&19.264$\sim$72.064&0.463&6.095 &1.511 $\pm$0.037 &1.790 $\pm$0.080 \\
070714B&64&600$\sim$800&-1.7$\sim$1.564&1.47&1.564$\sim$109.276&0.065&22.615 &1.049$\pm$0.071 &1.946 $\pm$0.288 \\
071227&1.8&700$\sim$900&-1.479$\sim$1.785&0.683&1.785$\sim$164.985&0.044&15.523 &0.969$\pm$0.151 &2.148 $\pm$0.339 \\
080603B&60&400$\sim$500&-3.632$\sim$21.328&1.044&37.968$\sim$67.088&0.342&3.053 &1.643 $\pm$0.038 &2.114 $\pm$0.073\\
080905B&128&700$\sim$900&-8.149$\sim$8.491&0.431&58.411$\sim$104.171&0.189&2.280 &1.429 $\pm$0.117 &1.521 $\pm$0.084 \\
090530&48&-250$\sim$-150&-1.122$\sim$11.934&1.334&26.078$\sim$43.486&0.155&8.606 &1.544 $\pm$0.089 &2.183 $\pm$0.261 \\
090715A&63&500$\sim$700&-3.567$\sim$4.753&0.522&17.233$\sim$62.993&0.095&5.495 &0.912 $\pm$0.209 &1.791 $\pm$0.456 \\
090916&63.4&500$\sim$700&-1.173$\sim$5.355&0.489&5.355$\sim$82.603&0.081&6.037 &1.498$\pm$0.190 &1.492 $\pm$0.303 \\
091208A&29.1&-250$\sim$-150&-14.69$\sim$26.59&0.288&26.59$\sim$101.26&0.057&5.053 &0.838 $\pm$0.074 &1.727 $\pm$0.213 \\
100423A&n/a&-150$\sim$-50&-3.697$\sim$9.358&2.191&9.358$\sim$97.487&0.865&2.533 &0.823 $\pm$0.030 &1.313 $\pm$0.037 \\
100522A&35.3&500$\sim$700&-3.648$\sim$4.672&1.777&21.312$\sim$37.952&0.505&3.519 &1.601 $\pm$0.042 &2.577 $\pm$0.107 \\
100606A&480&500$\sim$700&-2.145$\sim$23.199&1.15&23.199$\sim$84.447&0.505&2.277 &1.047 $\pm$0.059 &1.382 $\pm$0.068 \\
100704A&197.5&400$\sim$600&-22.744$\sim$10.28&0.936&142.376$\sim$191.912&0.335&2.794 &1.326 $\pm$0.031 &2.371 $\pm$0.076 \\
100725A&141&500$\sim$700&-6.245$\sim$16.987&0.335&25.435$\sim$65.563&0.157&2.134 &0.904 $\pm$0.078 &1.171 $\pm$0.102 \\
100802A&487&800$\sim$900&-12.521$\sim$28.759&0.266&28.759$\sim$111.319&0.087&3.057 &1.342 $\pm$0.069 &1.824 $\pm$0.123 \\
100814A&174.5&500$\sim$700&-4.967$\sim$33.05&1.485&60.505$\sim$157.657&0.649&2.288 &1.170 $\pm$0.024 &1.751 $\pm$0.037 \\
100906A&114.4&500$\sim$700&-6.111$\sim$18.657&3.096&92.961$\sim$125.985&0.674&4.593 &1.541 $\pm$0.021 &2.532 $\pm$0.052 \\
101011A&71.5&500$\sim$700&-1.581$\sim$3.859&0.661&3.859$\sim$39.763&0.241&2.743 &1.121 $\pm$0.094 &1.303 $\pm$0.096 \\
101017A&70&500$\sim$700&-14.496$\sim$26.784&3.189&26.784$\sim$76.32&0.499&6.391 &1.382 $\pm$0.019 &1.898 $\pm$0.043 \\
110402A&60.9&-250$\sim$-150&-0.388$\sim$5.948&1.271&5.948$\sim$60.86&0.495&2.568 &1.078 $\pm$0.146 &1.796 $\pm$0.137 \\
111103B&167&400$\sim$500&-14.287$\sim$35.249&1.878&93.041$\sim$134.321&0.243&7.728 &1.384 $\pm$0.023 &1.911 $\pm$0.089 \\
111121A&119&500$\sim$700&-2.739$\sim$1.485&1.71&1.485$\sim$62.733&0.204&8.382 &1.136 $\pm$0.066 &2.017 $\pm$0.098 \\
120514A&164.4&500$\sim$700&-8.951$\sim$10.057&0.843&43.849$\sim$69.193&0.227&3.714 &1.229 $\pm$0.060 &1.927 $\pm$0.122 \\
\enddata
\tablenotetext{a}{Time interval for the background calculation.}
\tablenotetext{b}{Time interval of the main burst.}
\tablenotetext{c}{Average count rate of the main burst estimated with our BB analysis.}
\tablenotetext{d}{Time interval of the EE tail.}
\tablenotetext{e}{Average count rate of the EE tail estimated with our BB analysis.}
\tablenotetext{f}{The ratio of $C_{\rm m}/C_{\rm e}$.}
\tablenotetext{g}{$\gamma$-ray photo index of the main burst.}
\tablenotetext{h}{Photon index of the EE tail..}
\end{deluxetable}

\newpage

\newpage


\end{document}